\documentclass[a4paper,fleqn,usenatbib,useAMS]{mnras}
\usepackage[T1]{fontenc}
\usepackage{ae,aecompl}
\usepackage{graphicx}	
\usepackage{amsmath}	
\usepackage{amssymb}	
\usepackage{footnote}
\usepackage{enumitem}
\usepackage{tikz}
\usepackage{xcolor,colortbl}
\newcommand{\Msun}{$M_{\odot}$}

\newcommand{\Vmax}{$V_\mathrm{max}$}
\newcommand{\Vmaxi}{$V_{\mathrm{max}, i}$}
\newcommand{\zmax}{$z_\mathrm{max}$}

\newcommand{\zi}{$z_i$}

\newcommand{\zminsample}{0.02}
\newcommand{\zmaxsample}{0.06}
\newcommand{\Nes}{$\sim$110'000}
\newcommand{\mlim}{$m_\mathrm{lim} = 17.77$}
\newcommand{\Mstarintmax}{$10^{15}$}
\newcommand{\faintfrac}{20\%}
\newcommand{\masscompl}{95\%}
\newcommand{\zbinsize}{$\Delta z = 0.005$}
\newcommand{\Mstarentiresample}{$10.79 \pm 0.01$}
\newcommand{\firstphistarentiresample}{$-3.31 \pm 0.20$}
\newcommand{\firstalphaentiresample}{$-1.69 \pm 0.10$}
\newcommand{\secondphistarentiresample}{$-2.01 \pm 0.28$}
\newcommand{\secondalphaentiresample}{$-0.79 \pm 0.04$}
\newcommand{\pvalueET}{$p = 1.55\ 10^{-15}$}

\definecolor{grey}{RGB}{114, 122, 131}
\definecolor{orange}{RGB}{225, 165, 85}
\definecolor{red}{RGB}{195, 26, 0}
\definecolor{green}{RGB}{0, 166, 52}
\definecolor{blue}{RGB}{43, 112, 222}
\definecolor{purple}{RGB}{143, 104, 202}
\definecolor{pink}{RGB}{204, 95, 192}

\title[SMF: methods, systematics \& results for $z \sim 0$]{Stellar mass functions: methods, systematics and results for the local Universe}
\author[Anna K. Weigel et al.]{
Anna K. Weigel,$^{1}$\thanks{E-mail: anna.weigel@phys.ethz.ch}
Kevin Schawinski,$^{1}$
and Claudio Bruderer$^{1}$
\\
$^{1}$Institute for Astronomy, Department of Physics, ETH Zurich, Wolfgang-Pauli-Strasse 27, CH-8093 Zurich, Switzerland\\
}

\date{Accepted XXX. Received YYY; in original form ZZZ}

\pubyear{2016}

\begin{document}
\label{firstpage}
\pagerange{\pageref{firstpage}--\pageref{lastpage}}
\maketitle

\begin{abstract}
We present a comprehensive method for determining stellar mass functions, and apply it to samples in the local Universe. We combine the classical 1/$V_\mathrm{max}$ approach with STY, a parametric maximum likelihood method and SWML, a non-parametric maximum likelihood technique. In the parametric approach, we are assuming that the stellar mass function can be modelled by either a single or a double Schechter function and we use a likelihood ratio test to determine which model provides a better fit to the data. We discuss how the stellar mass completeness as a function of $z$ biases the three estimators and how it can affect, especially the low mass end of the stellar mass function. We apply our method to SDSS DR7 data in the redshift range from 0.02 to 0.06. We find that the entire galaxy sample is best described by a double Schechter function with the following parameters: $\log (M^{*}/M_\odot) = 10.79 \pm 0.01$, $\log (\Phi^{*}_1/\mathrm{h^3\ Mpc^{-3}}) = -3.31 \pm 0.20$, $\alpha_1 = -1.69 \pm 0.10$, $\log (\Phi^{*}_2/\mathrm{h^3\ Mpc^{-3}}) = -2.01 \pm 0.28$ and $\alpha_2 = -0.79 \pm 0.04$.
We also use morphological classifications from Galaxy Zoo and halo mass, overdensity, central/satellite, colour and sSFR measurements to split the galaxy sample into over 130 subsamples. We determine and present the stellar mass functions and the best fit Schechter function parameters for each of these subsamples.
\end{abstract}

\begin{keywords}
galaxies: luminosity function, mass function -- methods: data analysis-- galaxies: general -- galaxies: statistics
\end{keywords}


\section{Introduction}
Stellar mass functions describe the number density of galaxies as a function of their stellar mass. They represent a key measure for the properties of the galaxy population and allow us to trace the assembly of stellar mass and the evolution of the star formation rate (SFR)  through cosmic time. 
 
Today, large redshift surveys, for instance the the Sloan Digital Sky Survey (SDSS) \citep{York:2000aa, Abazajian:2009aa}  or the zCOSMOS survey \citep{Lilly:2007aa}, allow us to probe the galaxy population in great detail. Recent work on stellar mass functions in the local Universe has been published by \cite{Panter:2007aa}, \cite{Baldry:2008aa, Baldry:2012aa}, \cite{Perez-Gonzalez:2008aa}, \cite{Peng:2010aa}, \cite{Kelvin:2014aa}  \cite{Taylor:2014aa} and \cite{Moffett:2015aa}. Stellar mass functions at higher redshift have for example been estimated by  \citet{Marchesini:2009aa, Marchesini:2010aa}, \cite{Stark:2009aa}, \cite{Pozzetti:2010aa}, \cite{McLure:2011aa},\cite{Caputi:2011aa}, \cite{Lee:2012aa}, \cite{Muzzin:2013aa}, \cite{Ilbert:2013aa}, \cite{Duncan:2014aa}, \cite{Grazian:2015aa} and \cite{Mortlock:2015aa}. 

In recent years studies have shown that for star-forming galaxies the characteristic stellar mass $M^{*}$ and the low mass end slope $\alpha$  of the stellar mass function stay constant out to redshifts of at least $z\sim2$ \citep{Bell:2003aa, Peng:2010aa, Pozzetti:2010aa, Ilbert:2013aa}. The shape of the star-forming mass function thus stays the same even though individual galaxies gain significant amounts of mass over this redshift range. Only $\Phi^{*}$, the normalisation of the mass function is redshift dependent. The mass function of quiescent galaxies is commonly fit with a double Schechter function \citep{Schechter:1976aa}, whereas star forming galaxies are often described using a single Schechter function (e.g. \citealt{Li:2009aa, Peng:2010aa, Pozzetti:2010aa, Peng:2012aa, Baldry:2012aa,  Ilbert:2013aa,  Muzzin:2013aa}). The process of star formation quenching is therefore directly imprinted on the shape and the evolution of the stellar mass function. Stellar mass functions hence provide a powerful tool for trying to understand the physical processes that cause the bimodality in colour-mass and colour-magnitude space \citep{Bell:2003aa, Baldry:2004aa, Martin:2007aa, Faber:2007aa, Schawinski:2014aa}. 

For simulations, especially semi analytic models (SAMs), stellar mass functions represent both, an important tool to constrain parameters and a fundamental indicator to test the predictions of the model. A comparison between the stellar mass function and the halo mass function has proven the need for feedback. The assumption that stellar mass follows halo mass produces too many low- and too many high-mass galaxies. This motivates the introduction of processes such as supernovae and AGN feedback which reduce the gas cooling efficiency at the low and the high mass end of the galaxy population, respectively \citep{Kauffmann:1993aa, Benson:2001aa, Croton:2006aa, Bower:2006aa, Somerville:2008aa}.

Stellar mass functions in the local universe have also been used as the base of purely empirical models. \cite{Peng:2010aa,Peng:2012aa} have developed such a model that predicts the stellar mass function of red galaxies based on two quenching processes. They distinguish between the mass dependent, but environment independent mass quenching and the mass independent, but environment dependent environment or satellite quenching. Environment quenching affects galaxies on all mass scales and thus produces a single Schechter function with the same shape as blue galaxies. Environment quenching could be associated with ram pressure stripping \citep{Gunn:1972aa} or strangulation \citep{Larson:1980aa, Balogh:2000aa}. In their model mass quenching is star formation rate (SFR) dependent and causes massive galaxies to stop forming stars. The cause of mass quenching might be feedback from active galactic nuclei or supernovae. By combining environment and mass quenching, \cite{Peng:2010aa,Peng:2012aa} they then predict the mass function of red, quenched galaxies to be a double Schechter function. 

Recently, \cite{Taylor:2015aa} have argued that at low stellar masses ($\log M < 9.3$), red and blue galaxies become indistinguishable and considering these galaxies to be part of two distinct populations might be inappropriate. The low-mass end upturn that we see in the mass function of red, quenched galaxies and the fact that these galaxies are well described by a double Schechter function, might hence be caused by the simple cut in the colour-mass digram which is usually used to distinguish red from blue galaxies.
	
We do not follow the approach by \cite{Taylor:2015aa} who model the distribution of galaxies in the colour-mass diagram as two independent, but overlapping populations. It is however important to note that we are assuming that stellar mass functions can be modelled by either a single Schechter function or a double Schechter functions. In contrast to previous work, we are not making any a priori assumptions about which functional form provides a better fit. For all subsamples we estimate both, the likelihood of a single and a double Schechter solution without putting any constraints on the functional parameters. We then use the likelihood ratio to determine if the subsample in question is better described by a single or a double Schechter function.

We are taking advantage of the comprehensive data that is available for the local universe. Based on six basic galaxy properties, we are splitting the entire SDSS DR7 galaxy sample into over 130 subsamples and determine the stellar mass function for each of them. We are using Galaxy Zoo DR1 \citep{Lintott:2008aa,Lintott:2011aa} data to not only split the sample by colour, but also by morphology. Furthermore, we slice the sample by specific star formation rate (sSFR)  \citep{Brinchmann:2004aa,Kauffmann:2003ac}, overdensity, halo mass \citep{Yang:2007aa} and into centrals and satellites \cite{Yang:2007aa}. We use three classical stellar mass function estimators: the 1/\Vmax\ approach by \cite{Schmidt:1968aa}, the non-parametric maximum likelihood method by \cite{Efstathiou:1988aa} (SWML) and the parametric maximum likelihood technique by \cite{Sandage:1979aa} (STY). We refine, test and compare these estimators and present an extensive collection of stellar mass functions in the local universe.

This paper is organized as follows. In Sec. \ref{sec:data} we introduce our sample, illustrate our method to estimate the overdensity as an additional environment measure and introduce our colour definitions. This is followed by a discussion of our method to determine the stellar mass completeness in Sec. \ref{sec:mass_completeness}. Sec. \ref{sec:Schechter_function} introduced the Schechter function \citep{Schechter:1976aa} and in Sec. \ref{sec:construction} we concentrate on the methods used to determine our stellar mass functions and discuss them in great detail. We also test our code using mock catalogues, discuss the effects that the mass completeness can have on the low mass end of the mass function and compare our results to previously published work in Sec. \ref{sec:test_code}. In Sec. \ref{sec:mass_functions} we present all of our stellar mass functions and their best fit parameters. 

Throughout this paper we assume a $\Lambda$CDM cosmology with $h_0 = 0.7$, $\Omega_\mathrm{m} = 0.3$ and $\Omega_\Lambda = 0.7$. 

\section{Data and Sample}
\label{sec:data}
\begin{table*}
 \begin{centering}
 \caption{Overview of quantities and variables used in our analysis. We mainly rely on data from the NYU VAGC \citep{Blanton:2005aa,Padmanabhan:2008aa}, the MPA JHU \citep{Brinchmann:2004aa,Kauffmann:2003ab} and the \protect \cite{Yang:2007aa} catalogue and morphological classifications from Galaxy Zoo \citep{Lintott:2008aa,Lintott:2011aa}. }
 \label{tab:variable_ov}
 \begin{tabular}{llll}
 	\hline
 	{quantity/variable} & {symbol/value} & {reference} & {column name/comment}\\
 	\hline
 	{minimum sample redshift} & {$z_\mathrm{min}^\mathrm{s}$ = \zminsample} & {} & {}\\
 	{maximum sample redshift}& {$z_\mathrm{max}^\mathrm{s}$ = \zmaxsample} & {} & {}\\
 	{sample area} & {$\Omega^\mathrm{s} = 7748\ \mathrm{deg}^{2}$} & {\cite{Yang:2007aa}} & {}\\
 	{magnitude limit} & {\mlim} & {\cite{Strauss:2002aa}, } & {}\\
 	{} {($r$ band, Petrosian magnitude)} & {} & {\cite{Abazajian:2009aa}} & {}\\
 	{spectral type} & {spec\_type = `GALAXY'} & {MPA JHU} & {[SPECTROTYPE]}\\
 	{stellar mass} & {$M$} & {MPA JHU} & {[TOTGM\_AVG]}\\
 	{} {(total, not fibre)} & {} & {} & {}\\
 	{redshift} & {$z$} & {NYU VAGC} & {[Z]}\\
 	{} {(sepctroscopic)} & {} & {} & {}\\
 	{apparent magnitudes} & {$m$} & {NYU VAGC} & {[PETROFLUX]}\\
 	{spectral completeness} & {$w_\mathrm{spec}$} & {Yang et al.} & {[COMPL]}\\
 	{Morphology} & {ETs, LTs, Int} & {GZ1} & {[ELLIPTICAL], [SPIRAL],}\\
 	{} {Early types, Late types, indeterminates} & {} & {} & {[UNCERTAIN]}\\
 	{specific star formation rate} & {sSFR} & {MPA JHU} & {[TOTSSFR\_AVG]}\\
 	{colour} & {red, green, blue} & {see Sec. \ref{sec:colour_def}} & {defined in colour mass diagram}\\
 	{} {red, green, blue} & {} & {} & {}\\
 	{Halo mass} & {$M_\mathrm{h}$} & {\cite{Yang:2007aa}} & {[MGROUP]}\\
 	{centrals \& satellites} & {cent, sat} & {\cite{Yang:2007aa}} & {[RANK] = 1, [RANK] = 2}\\
 	{overdensity} & {$\delta$} & {see Sec. \ref{sec:overdensity}} & {5th nearest neighbour}\\
 	{characteristic stellar mass } & {$M^{*}$}  & {see Sec. \ref{sec:Schechter_function}} & {}\\
 	{} {Schechter function} & {} & {} & {}\\
 	{characteristic number density } & {$\Phi^{*}$}  & {see Sec. \ref{sec:Schechter_function}} & {}\\
 	{} {Schechter function} & {} & {} & {}\\
 	{powerlaw slope Schechter function} & {$\alpha$}  & {see Sec. \ref{sec:Schechter_function}} & {}\\
 	{stellar mass completeness function } & {$M_\mathrm{lim}(z)$}  & {see Sec. \ref{sec:mass_completeness} \& \ref{sec:compl}} & {based on \cite{Pozzetti:2010aa}}\\
 	{minimum stellar mass at which object $i$} & {$M_{\mathrm{min}, i}$}  & {see Sec. \ref{sec:mass_completeness} \& \ref{sec:compl}} & {}\\
 	{} {can be observed} & {}  & {} & {}\\
 	{maximum $z$ to which object $i$} & {$z_{\mathrm{max}, i}$}  & {see Sec. \ref{sec:mass_completeness} \& \ref{sec:compl}} & {}\\
 	{} {can be observed} & {}  & {} & {}\\
 	{likelihood for single/double Schechter function} & {$\mathcal{L}$} & {see Sec. \ref{sec:STY_tech}} & {according to \cite{Sandage:1979aa}}\\
 	\hline
 	\end{tabular}
 \end{centering}
\end{table*}

\begin{table}
 \begin{centering}
 \caption{\label{tab:sample}Our entire sample contains \Nes\ objects. Here we show how these are split within the morphology, colour and environment subsamples.}
 	\begin{tabular}{llllll}
 	\hline
 	\multicolumn{2}{c}{morphology} & \multicolumn{2}{c}{colour} & \multicolumn{2}{c}{environment}\\
 	\hline
 	{Early types} & {8.44\%} & {red} & {26.99\%} & {centrals} & {66.62\%}\\
	{Indeterminates} & {58.27\%} & {green} & {16.97\%} & {} & {}\\
	{Late types} & {33.29\%} & {blue} & {55.93\%} & {satellites} & {33.38\%}\\
 	\hline
 	\end{tabular}
 \end{centering}
\end{table}

We base our sample on the seventh data release (DR7) of the Sloan Digital Sky Survey (SDSS, \citealt{York:2000aa, Abazajian:2009aa}). We cross match the New York Value-Added Galaxy Catalog (NYU VAGC, \citealt{Blanton:2005aa, Padmanabhan:2008aa}) with the Max Planck Institute for Astrophysics \ John Hopkins University (MPA JHU, \citealt{Kauffmann:2003ab, Brinchmann:2004aa}) catalogue. Furthermore, we extract morphological classifications from the first Galaxy Zoo data release \citep{Lintott:2008aa, Lintott:2011aa}, include absorption and emission-line measurements by \cite{Oh:2011aa} (OSSY) and add halo masses and the classification into centrals and satellites from the catalogue by \cite{Yang:2007aa} for DR7. 

The SDSS DR7 provides photometry over 11'663 $\mathrm{deg}^{2}$ for the filters $u, g, r, i\ \mathrm{and}\ z$ and spectroscopy over 9'380 $\mathrm{deg}^{2}$ down to a Petrosian magnitude limit of \mlim\ \citep{Strauss:2002aa, Abazajian:2009aa}. We base our work on the spectroscopic sample and extract the spectroscopic redshift ($z$) and apparent magnitude ($m$) values from the NYU VAGC. For stellar mass ($M$), star formation rate ($SFR$) and specific star formation rate ($sSFR = SFR/ M$) values we use the values by \cite{Brinchmann:2004aa} given in the MPA JHU catalogue. To estimate stellar masses, \cite{Brinchmann:2004aa} perform fits to the photometry using model spectra by \cite{Bruzual:2003aa}. This method is different to the approach by \cite{Kauffmann:2003ab} who constrain the mass and age of a galaxy's stellar population based on the $4000 \AA$ break and the $\mathrm{H}\delta_\mathrm{A}$ Balmer absorption line, the estimated masses do however agree well\footnote{\url{http://wwwmpa.mpa-garching.mpg.de/SDSS/DR7/mass_comp.html}}. The $SFR$ values are determined through emission line modeling using models by \cite{Charlot:2001aa}. To exclude quasars from our sample, we only use objects of the MPA JHU spectral type `GALAXY'. Note that \cite{Brinchmann:2004aa} use a \cite{Kroupa:2001aa} IMF to model $M$, SFR and sSFR. 

All galaxies in our sample were visually classified by volunteer users of the Galaxy Zoo\footnote{\url{http://www.galaxyzoo.org}} website. Users were shown $g$, $r$ and $i$ colour composite SDSS images and asked to classify each galaxy into one of six categories (Elliptical galaxy, Clockwise spiral galaxy, Anti-clockwise spiral galaxy, other Spiral galaxy, Star/Do not know, Merger). Over 100'000 citizen scientists participated and the mean number of classification per galaxy is about $38$. \cite{Lintott:2008aa, Lintott:2011aa} apply a debiasing procedure and flag each galaxy as `Spiral', `Elliptical' or `Uncertain'. For a galaxy to be flagged as `Spiral' or `Elliptical', 80 percent of its votes have to be in that category. Objects that are classified as neither `Spiral' nor `Elliptical', are flagged as `Uncertain'. We refer to galaxies with the `Spiral' flag as spirals or Late type galaxies, to sources with the `Elliptical' flag as ellipticals or Early type galaxies and to all objects classified as `Uncertain' as indeterminates.     

To be able to correct for dust (see Section \ref{sec:colour_def}), we extract $E(B-V)$ values from the publicly available OSSY catalogue by \cite{Oh:2011aa}. \cite{Oh:2011aa} provide absorption and emission-line measurements for the DR7 of the SDSS that were determined by using the pixel-fitting method by \cite{Cappellari:2004aa} (\textsc{pPXF}) and the \textsc{gandalf} code by \cite{Sarzi:2006aa}.  

For halo mass ($M_h$) measurements and the classification into centrals and satellites, we rely on the work by \cite{Yang:2007aa}. We use their sample which is based on Petrosian magnitudes and mass, not luminosity, ranked halo masses. We thus refer to galaxies as being centrals if they are the most massive ones in their group. We also use the spectral completeness values ($w_{spec}$) that are provided by \cite{Yang:2007aa} for the construction of our mass functions (see Section \ref{sec:Vmax_tech}). The survey area we are analysing here hence corresponds to the area considered by \cite{Yang:2007aa} which is $\Omega^{s} = 7748\ \mathrm{deg}^{2}$.      

The quantities mentioned here allow us to construct stellar mass functions. We are also able to split the sample by morphology, $sSFR$, halo mass and into centrals and satellites. In sections \ref{sec:overdensity} and \ref{sec:colour_def} we illustrate how we estimate an additional environment measure and split the sample according to colour.   

We limit our sample to the redshift range between $z_{min}^{s} =$ \zminsample\ and $z_{max}^s = $ \zmaxsample. After excluding objects with invalid stellar masses, our main sample contains \Nes\ objects. Table \ref{tab:variable_ov} gives an overview of the quantities and variables that are used in the following analysis.

\subsection{Overdensity}
\label{sec:overdensity}
\begin{figure}
\includegraphics[width=.5\textwidth]{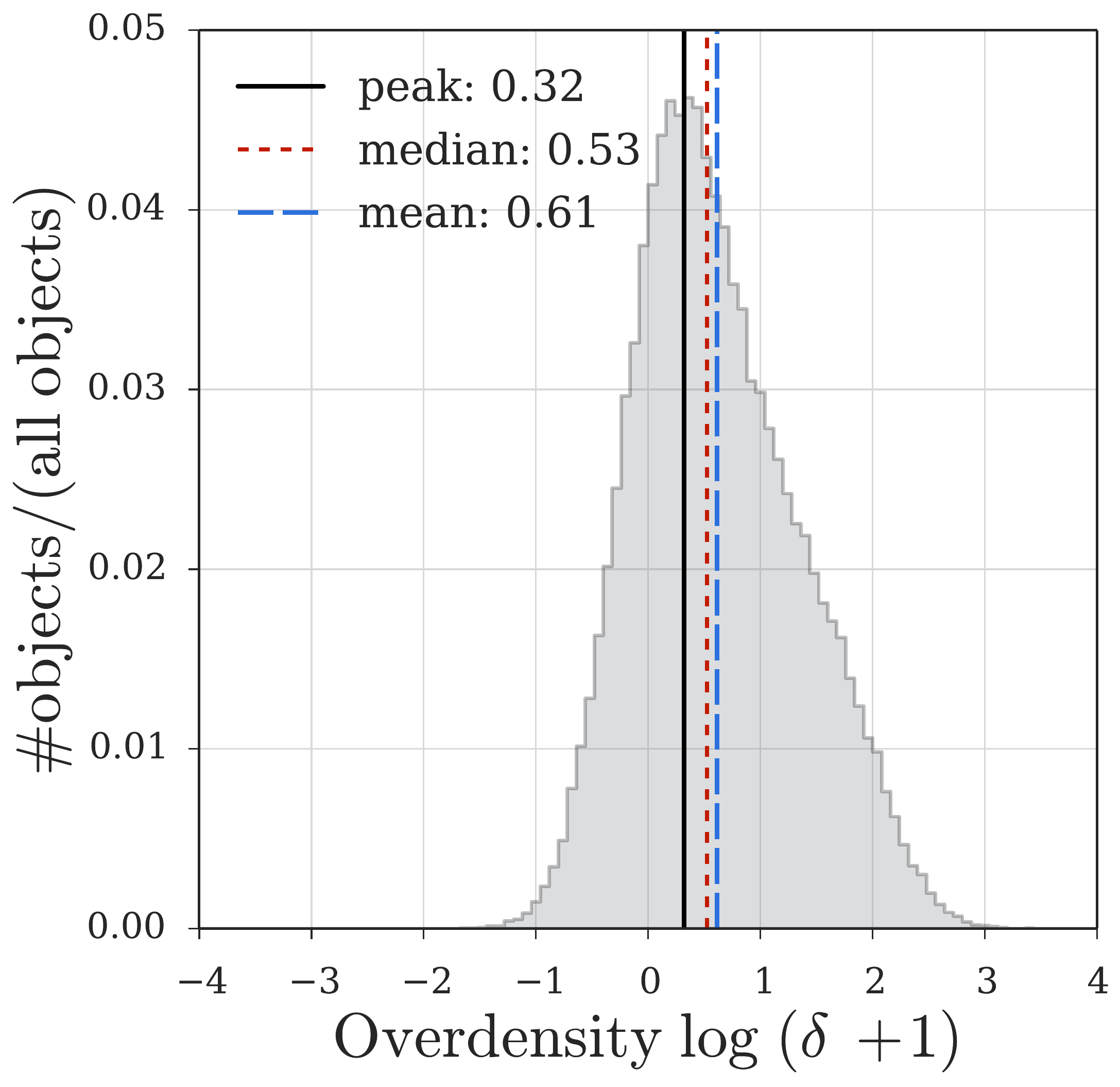}
\caption{\label{fig:ovdens}Distribution of overdensity values for the entire sample. We use a 5th nearest neighbour approach to estimate the environmental density for our sample sources (see Sec. \ref{sec:overdensity}).}
\end{figure}

We use a nearest neighbour approach to determine an environment measure for our sample sources. For each galaxy with redshift $z$, we determine the 5th nearest neighbour that has a mass above $M_\mathrm{cut} = 10^{9}$ \Msun\ and lies within the allowed recession velocity range of $\pm 1000\ \mathrm{km}/\mathrm{s}$ around the recession velocity of the galaxy. Using the projected comoving distance to the 5th nearest neighbour $r_{5\mathrm{NN}}$ we then compute the surface number density $\rho_{5\mathrm{NN}}$ defined as:
\begin{equation}
\rho_{5\mathrm{NN}} = \frac{5}{\pi\ r_{5\mathrm{NN}}^{2}}
\end{equation}

The average surface number density at redshift $z$ is given by:
\begin{equation}
\rho_\mathrm{avg} = \frac{N_{z_{\mathrm{range}}}}{A_{z}} = \frac{N_{z_\mathrm{range}}}{d_c^2\ \frac{\pi^2}{180^2}\ \Omega} 
\end{equation}
with $N_{z_{range}}$ corresponding to the total number of objects in the allowed recession velocity range with a stellar mass $> M_{cut}$, $d_c$ being the comoving distance to redshift $z$ and $\Omega$ being equivalent to the sample area in $\mathrm{deg}^{2}$.

The overdensity $\delta$ for this galaxy is then given by:
\begin{equation}
\delta = \frac{\rho_{5\mathrm{NN}}}{\rho_\mathrm{avg}}  - 1 
\end{equation}
Figure \ref{fig:ovdens} shows the distribution of $\delta$ values for the entire SDSS DR7 in the redshift range $\zminsample \leq z \leq \zmaxsample$.

\subsection{Colour definition}
\label{sec:colour_def}
\begin{figure*}
\includegraphics[width=\textwidth]{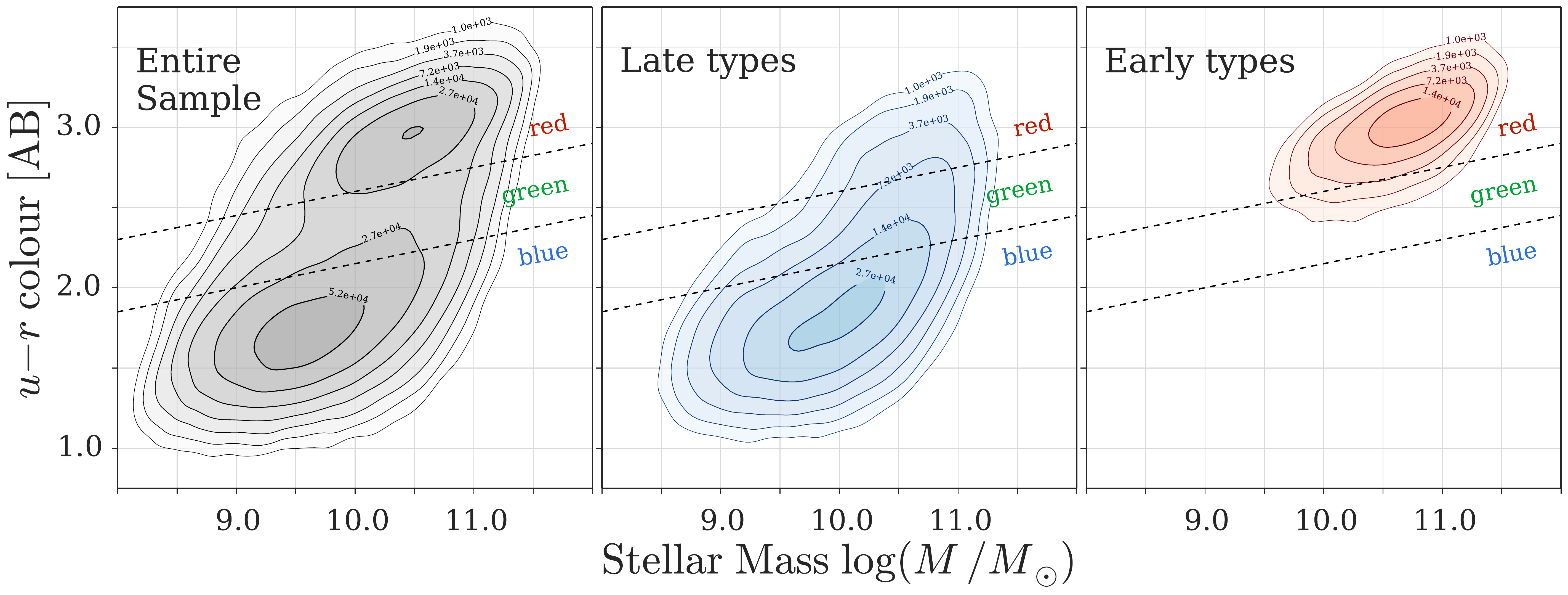}
\caption{\label{fig:cm}Colour-mass diagram for the entire sample, Early types and Late types. The dashed lines show how we separate the sample into red, green and blue galaxies (see equation \ref{eq:cut_red}, \ref{eq:cut_blue}). This figure illustrates that a cut in colour is not the same as splitting by morphology. All colours are dust and k-corrected (see Section \ref{sec:colour_def}). Out of the \Nes\ galaxies in our main sample, 8.44\% are Early and 33.29\% are Late type galaxies (see Table \ref{tab:sample}). The contours correlate with the number of objects and are linearly spaced in log-space.}
\end{figure*}

We split our sample into red, green and blue galaxies using the colour-mass diagram. To determine the colour of our sources we use dust and k-corrected Petrosian flux values from the NYU VAGC \citep{Blanton:2005aa,Padmanabhan:2008aa}. We calculate the K-correction values using the \textsc{kcorrect} \textsc{idl} package (version 4.2) by \cite{Blanton:2007aa} and correct the magnitudes to redshift 0. For an object at redshift $z$ with an absolute magnitude $M_Q$ in the emitted-frame filter $Q$, the apparent magnitude $m_R$ in the observed filter $R$ is given by:

\begin{equation}
m_R = M_Q + DM(z) + K_{QR}(z)
\end{equation}
where $K_{QR}$ is the K-correction value and $DM$ is the distance modulus for redshift $z$ \citep{Oke:1968aa, Hogg:2002aa, Blanton:2007aa}. We correct for dust by applying the Calzetti law \citep{Calzetti:2000aa} with $E(B - V)$ values from OSSY ([EBV\_STAR], \citealt{Oh:2011aa}).

We define the colour of our galaxies in the dust-corrected $u - r$ colour-mass diagram which is shown in Figure \ref{fig:cm}. Similar to \cite{Schawinski:2014aa}, we refer to galaxies lying above
\begin{equation}\label{eq:cut_red}
u- r= 0.6 + 0.15 \times \log M
\end{equation} 
as `red', where as objects below 
\begin{equation}\label{eq:cut_blue}
u - r  = 0.15 + 0.15 \times \log M
\end{equation}
are defined to be `blue'. Objects lying between relations \ref{eq:cut_red} and \ref{eq:cut_blue} in the colour-mass diagram are part of the green valley \cite{Bell:2004aa, Martin:2007aa, Fang:2012aa, Schawinski:2014aa} and are thus referred to as being `green'. 

Figure \ref{fig:cm} shows the colour-mass diagram for the entire sample and for Late and Early types only. This figure illustrates that a simple colour cut is not equivalent to splitting the sample by morphology. It is thus important to remember that not all Late types are blue  \citep{Masters:2010aa} and not all Early types are red \citep{Schawinski:2009aa} and to split the sample by both, morphology and colour \citep{Schawinski:2014aa}.  

\section{Stellar Mass Completeness}
\label{sec:mass_completeness}
\begin{figure*}
	\includegraphics[width=\textwidth]{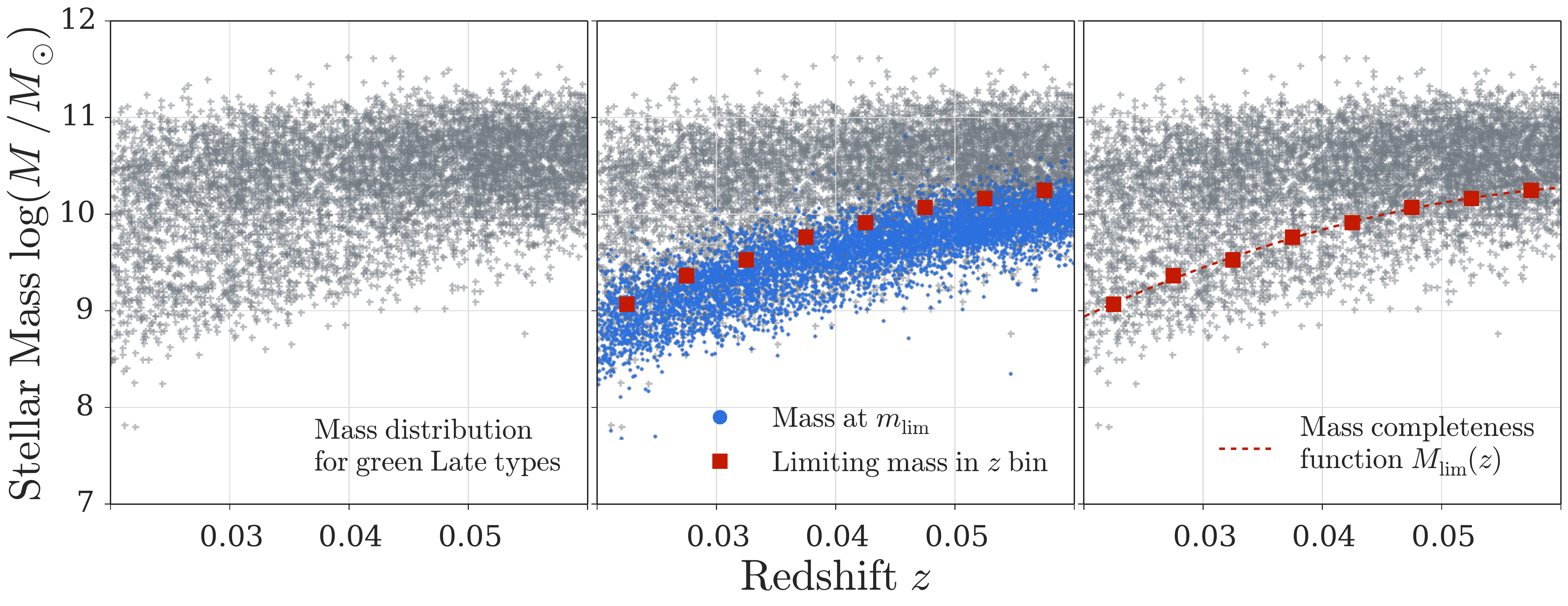}
	\caption{\label{fig:Pozzetti} Determination of the stellar mass completeness as a function of redshift.  Due to the range in M/L, constraining the mass completeness of a flux limited sample is not straight forward. We are following the approach by \protect\cite{Pozzetti:2010aa} which is illustrated here. In the left panel we are showing the distribution of green Late type galaxies in the stellar mass redshift plane as an example. By keeping the M/L ratio and the redshift of each source constant, we can determine the stellar masses that these objects would have if their flux was equal to the flux limit. These limiting mass values are shown as blue dots in the middle panel. We bin in redshift, sort the sources in each redshift bin according to their magnitude, select the faintest \faintfrac, and determine the stellar mass below which lie \masscompl\ of these faint objects. We repeat this procedure for each bin and obtain the mass values that are shown as red squares in the middle panel. The mass completeness function $M_\mathrm{lim}(z)$ is then estimated by fitting a second order polynomial to the limiting mass value in each redshift bin. $M_\mathrm{lim}(z)$ is shown as a red dahsed line in the right panel.} 
\end{figure*}

\begin{figure*}
	\includegraphics[width=\textwidth]{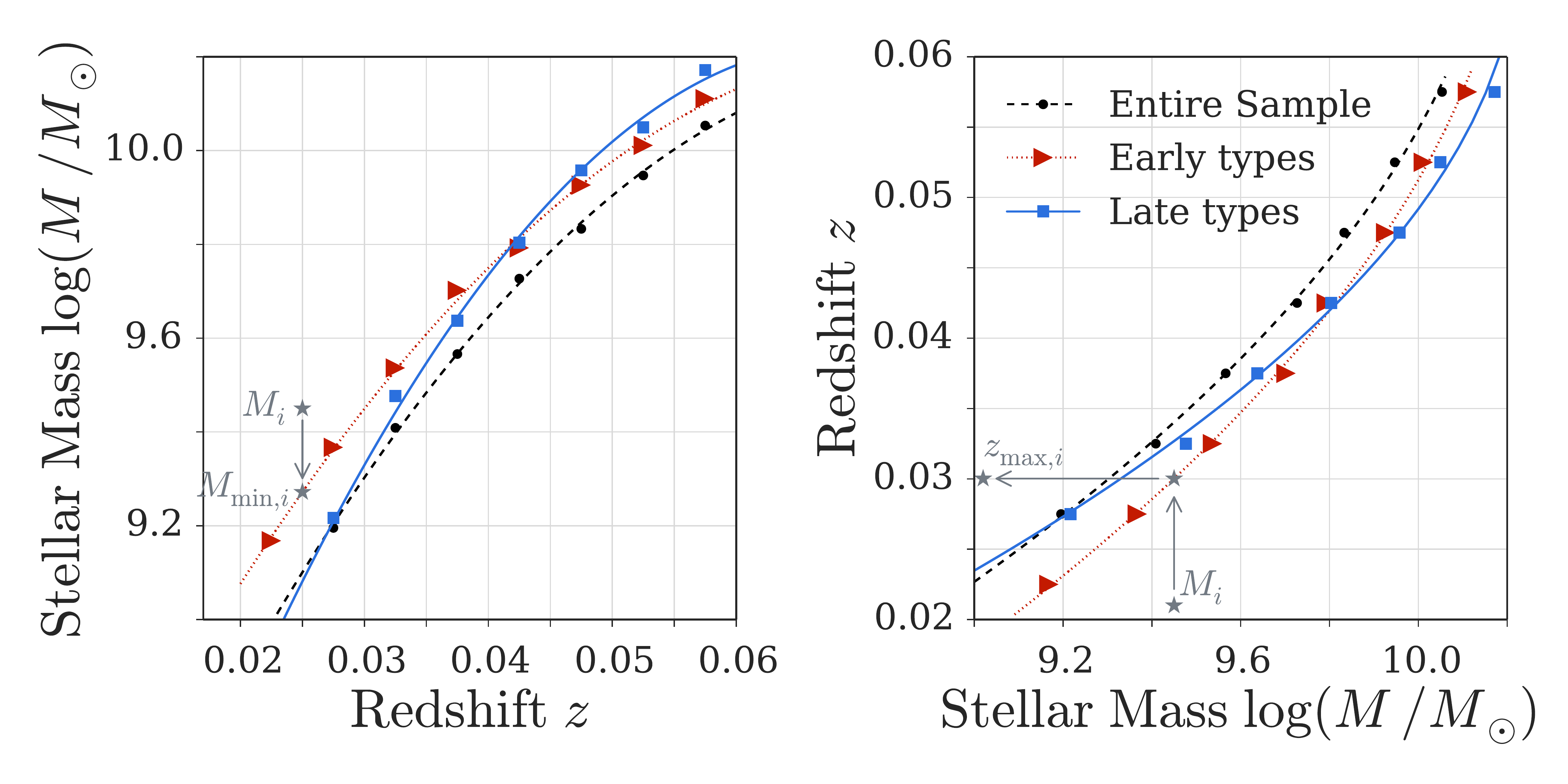}
	\caption{\label{fig:Mlim}Stellar mass completeness function $M_\mathrm{lim}(z)$. The left panel shows the stellar mass completeness function as a function of redshift for the entire sample (dashed line), Early type (dotted line) and Late type galaxies (solid line). We estimate $M_\mathrm{lim}(z)$ following the approach by \protect\cite{Pozzetti:2010aa}. We use $M_\mathrm{lim}(z)$ to determine $M_{\mathrm{min}, i}$, the limiting mass that source $i$ would have at the magnitude limit assuming a constant $\mathrm{M}_i/\mathrm{L}_i$. Inverting $M_\mathrm{lim}(z)$ allows us to estimate $z_{\mathrm{max}, i}$, the redshift at which source  $i$ would no longer be part of the sample, given the sample's stellar mass completeness. This is illustrated in the right panel.} 
\end{figure*}

To estimate the number density $\Phi$ we have to derive $M_{\mathrm{lim}}(z)$, the completeness in stellar mass as a function of redshift for our flux-limited sample. The galaxies in our sample show a range of mass-to-light ratios (M/L) which has to be taken into account when determining $M_{\mathrm{lim}}(z)$. By simply using the completeness as a function of magnitude and not mass, we would on one hand allow even relatively low mass objects to be part of the sample if they are luminous enough, whereas on the other hand, massive, but faint sources that lie below the magnitude cut would be excluded even if their masses where high enough to lie above $M_\mathrm{lim}(z)$. Objects with high M/Ls (high luminosity compared to their stellar mass) could thus be falsely included, whereas sources with low M/Ls (low luminosity compared to their stellar mass) could be falsely excluded \citep{Marchesini:2009aa}. $M_\mathrm{lim}(z)$ is dependent on the mass and M/L distribution of the sample and therefore has to be redetermined for each of the subsamples. Before determining the shape of the stellar mass function for a certain subsample, we therefore estimate its stellar mass completeness limit as a function of redshift. 

For the 1/\Vmax\ technique, we need to know $z_{\mathrm{max}, i}$, the maximum redshift at which source $i$ with stellar mass $M_i$ would no longer be part of the subsample. STY and SWML require $M_{\mathrm{min}, i}$, the mass at which object $i$ at redshift \zi\ falls below the stellar mass completeness (see Sec. \ref{sec:construction}).  

In the literature, different techniques are used to estimate $M_\mathrm{lim}(z)$. \cite{Dickinson:2003aa}, \cite{Fontana:2006aa}, \cite{Perez-Gonzalez:2008aa} and others use an approach based on a single stellar population (SSP). One generates the spectrum of a passively evolving SSP that formed at high redshift. By scaling the flux of this model spectrum, the maximum stellar mass that a galaxy could have if its magnitude corresponds to $m_\mathrm{lim}$ can be estimated. \cite{Marchesini:2009aa} introduce a technique in which they take advantage of the availability of deeper survey data. They scale the flux and mass of objects from the deeper surveys up to match the magnitude cut of their sample. The most massive objects of this scaled mass distribution, represent the sources with the lowest M/Ls that might be missed by a simple magnitude cut (see \citealt{Marchesini:2009aa} for a more detailed discussion of their technique and caveats that might affect the SSP approach). \cite{2012ApJ...744...88Q}  and \cite{Tomczak:2014aa} slightly modify the approach by \cite{Marchesini:2009aa}. Instead of employing deeper survey data, they scale the masses and flux values of objects above the flux completeness down to the magnitude limit. The upper envelope of the scaled down masses is then used as a redshift dependent mass completeness limit.  

We are following the approach by \cite{Pozzetti:2010aa}, illustrated in Figure \ref{fig:Pozzetti}. In a first step, we determine $M_{\mathrm{lim}, i}$ for all sources in the sample. $M_{\mathrm{lim}, i}$ corresponds to the mass that source $i$ would have if its magnitude was equal to the SDSS magnitude limit (\mlim, \citealt{Strauss:2002aa, Abazajian:2009aa}). $M_{\mathrm{lim}, i}$ is M/L dependent and thus has to be determined for each source individually. For each galaxy we keep $M_i/L_i$ constant and can therefore also assume that, for a constant $z_i$, the ratio between stellar mass and flux stays constant:
\begin{equation}\label{eq:ML_const}
\frac{M_i}{L_i} \propto \frac{M_i}{F_i(z_i)} = \frac{M_{\mathrm{lim}, i}}{F_{\mathrm{lim}, i} (z_i)}.
\end{equation} 
$L_i$ is the luminosity of object $i$. $F_{i}(z_i)$ is the corresponding flux at $z_i$, $F_{\mathrm{lim}, i} (z_i)$ is the limiting flux at the same redshift $z_i$ and $M_{\mathrm{lim}, i}$ is the limiting mass at this redshift.
 $M_{\mathrm{lim}, i}$ can hence be computed in the following way:
\begin{equation}\label{eq:Mlim}
\log(M_{\mathrm{lim}, i}) = \log(M_{i}) + 0.4 \times (m_i - m_\mathrm{lim})
\end{equation}
Using equation \ref{eq:Mlim}, these limiting mass values can easily be determined for each object individually. 

 After computing $M_\mathrm{lim}$ for all galaxies in the subsample, we sort all sources by magnitude and select the faintest \faintfrac. We bin in redshift-space (\zbinsize, mean error on $z = 1.26\ 10^{-5}$ ) and in each bin determine the mass below which lie \masscompl\ of these faint objects. The function $M_\mathrm{lim}(z)$ is then given by fitting a second order polynomial to the estimated mass limits in each bin. The left panel of Fig. \ref{fig:Mlim} shows $M_{\mathrm{lim}}(z)$ for the entire sample and for Early and Late types only.
 
 Once we have determined $M_\mathrm{lim}(z)$, we eliminate all sources with stellar masses below the mass completeness limit. Furthermore, we compute $M_{\mathrm{min}, i} = M_\mathrm{lim}(z_i)$ for all of the remaining sources since this limiting mass value is needed as input for the STY and SWML methods. As can be seen in the right panel of Figure \ref{fig:Mlim}, we also invert the $M_\mathrm{lim}(z)$ function and determine $z_{\mathrm{max}, i}$, the maximum redshift out to which source $i$ at \zi\ with stellar mass $M_{i}$ can be detected, given our mass completeness limits. $z_{\mathrm{max}, i}$ is the most important input parameter for the 1/\Vmax\ technique.

The shape of the mass completeness function has significant impact on the low mass end of the stellar mass function. We discuss this effect in Sec. \ref{sec:test_code}. Furthermore we show, which influence the choice of redshift bin size, fitting function and completeness level have on the results. 

\section{The Schechter function}
\label{sec:Schechter_function}
Luminosity and stellar mass functions are well described by \cite{Schechter:1976aa} functions. For mass functions,  the Schechter function $\Phi(M)$ parametrises the number density of galaxies as a function of their stellar mass. The number density of galaxies in a mass bin $dM$ is then given by: 

\begin{equation}
\begin{aligned}
n_\mathrm{galaxies}  & =   \Phi(M) dM \\
								  &   = \Phi^{*} e^{-M/M^{*}}\Big(\frac{M}{M^{*}}\Big)^{\alpha} dM
\end{aligned}
\end{equation}
$M^{*}$ is referred to as the `knee' of the Schechter function. It corresponds to the stellar mass at which the Schechter function transitions from a simple power law with slope $\alpha$ at lower masses into an exponential function at higher masses. The normalisation $\Phi^{*}$ corresponds to the number density at $M^{*}$.  

For stellar mass functions it is more convenient to work in $\log M$ space and express the Schechter function in the following way: 
\begin{equation} \label{eq:single_Schechter}
\begin{aligned}
\Phi\ d\log M = & \ln(10) \Phi^{*} e^{-10^{\log M - \log M^{*}}} \\
						  & \cdot \Big(10^{\log M - \log M^{*}}\Big)^{\alpha + 1} d\log M
\end{aligned}
\end{equation}

Recent studies have shown that the galaxy population at low redshift is better described by a double, or two component, Schechter function \citep{Li:2009aa, Peng:2010aa, Pozzetti:2010aa, Peng:2012aa, Baldry:2012aa,  Ilbert:2013aa,  Muzzin:2013aa}. A double Schechter  simply corresponds to the sum of two single Schechter functions with the same $M^{*}$:
\begin{equation} \label{eq:double_Schechter}
\begin{aligned}
\Phi\ d\log M = & \ln(10) e^{-10^{\log M - \log M^{*}}} \Big[ \Phi^{*}_1 (10^{\log M - \log M^{*}})^{\alpha_1 + 1} \\
						  & +  \Phi^{*}_2 (10^{\log M - \log M^{*}})^{\alpha_2 + 1}]\ d\log M
\end{aligned}
\end{equation}

The STY maximum likelihood approach allows us to determine if the data is better fit with a single or a double Schechter function (see Sec. \ref{sec:single_double}). We report the best fit parameters for all of our subsamples in Tables \ref{tab:results_1} and \ref{tab:results_2}. 

\section{Stellar Mass Function Construction}
\label{sec:construction}
We use three different approaches to construct our stellar mass functions: the classical 1/\Vmax\ method by \cite{Schmidt:1968aa} (Section \ref{sec:Vmax_tech}), the parametric technique developed by \cite{Sandage:1979aa} (STY, Section \ref{sec:STY_tech}) and the non-parametric step-wise-maximum likelihood method (SWML, Section \ref{sec:SWML_tech}) by \cite{Efstathiou:1988aa}.

The 1/\Vmax\ technique is popular due to its simplicity. We correct for the Malmquist bias \citep{Malmquist:1920aa,Malmquist:1922aa} by weighing each object by the maximum volume it can be detected in.  We do not need to assume a  functional form and the method directly provides the normalisation of the mass function. The 1/\Vmax\ technique does however implicitly assume that all sources follow a uniform spatial distribution which can lead to a distortion in the case of over- or under-dense regions \citep{Efstathiou:1988aa}. We are treating the SDSS survey area as one complete sample with constant depth and follow the simple approach by \cite{Schmidt:1968aa}. 

STY and SWML are maximum likelihood methods which, unlike 1/\Vmax\, do not require any initial assumptions on the spatial distribution of objects and are therefore unbiased with respect to density inhomogeneities. For both we do however have to determine the normalisation $\Phi^{*}$ independently. SWML, a non-parametric approach, requires binning in stellar mass. For STY, a parametric approach, we have to assume a functional form for the stellar mass function. STY alone does not allow us to constrain the goodness of fit to the data which is why STY and SWML are usually combined.

Other methods to estimate stellar mass and luminosity functions include for example the non-parametric maximum likelihood $C^{-}$ \citep{Lynden-Bell:1971aa} and $C^{+}$ \citep{Zucca:1997aa} estimators. \cite{Willmer:1997aa} and \cite{Takeuchi:2000aa} compare various estimators using Monte-Carlo simulations, \cite{Ilbert:2004aa} analyse the biases that might affect them and \cite{Binggeli:1988aa} and \cite{Johnston:2011aa} summarise the different approaches. 
\subsection{The 1/\Vmax\ technique}
\label{sec:Vmax_tech}
\subsubsection*{The basic principle}
In the 1/\Vmax\ method \citep{Schmidt:1968aa} we weigh each object by the maximum volume it could be detected in, given the redshift range and the mass completeness of the sample. This corrects for the Malmquist bias, i.e. the fact that faint, low mass objects can only be detected in a small volume, whereas bright, massive sources can be detected in the entire sample volume.  

\Vmaxi\ thus corresponds to the maximum volume in which object $i$ at redshift \zi\ with stellar mass $M_i$ could be detected in. To estimate the number density $\Phi$ we bin in stellar mass. The number density in mass bin $j$ is then given by the sum over all $N_\mathrm{bin}$ objects in this bin:

\begin{equation}\label{eq:Vmax_phi}
\Phi_{j} d\log M = \sum_{i}^{N_\mathrm{bin}} \frac{w_{\mathrm{spec}, i}}{V_{\mathrm{max},i}}.
\end{equation}
$w_{\mathrm{spec}, i}$ is the spectroscopic completeness of source $i$ which we extract from the catalogue by \cite{Yang:2007aa}.

\subsubsection*{The method}
In a first step, we bin in stellar mass in the mass range $10^9 - 10^{12} M_\odot$ using a bin size of $\Delta \log M = 0.2$. 
We then determine \Vmaxi, the maximum volume in which object $i$ with redshift \zi\ and stellar mass $M_i$ could be detected for each source in the subsample. In a flat universe the comoving volume \Vmaxi\ is given by \citep{Hogg:1999aa}:
\begin{equation}\label{eq:Vmax}
V_{\mathrm{max}, i} = \frac{4\pi}{3}\ \frac{\Omega^\mathrm{s}}{\Omega^\mathrm{sky}}\ (d_c(z_{\mathrm{max}, i})^{3} - d_c(z_{\mathrm{min}, i})^{3})
\end{equation}
where $\Omega^\mathrm{sky} = 41'253\ \mathrm{deg}^2$ is the surface area of the entire sky, $\Omega^\mathrm{s}$ is the surface area covered by our sample and $d_c(z)$ corresponds to the comoving distance to redshift $z$. 
We are not considering the lower magnitude limit of SDSS \citep{Blanton:2001aa}. $z_{\mathrm{min}, i}$ is thus simply given by the lower redshift limit of our main sample $z_\mathrm{min}^{s}$ = \zminsample. To determine $z_{\mathrm{max}, i}$, we compare the maximum redshift of our main sample $z_\mathrm{max}^\mathrm{s}$ = \zmaxsample\ and $z_{\mathrm{max}, i}^\mathrm{mass}$, the maximum redshift that we determine for object $i$ based on its stellar mass $M_i$ and the mass completeness of the subsample (see Sec. \ref{sec:mass_completeness}). $z_{\mathrm{max}, i}$ is given by the smaller of those two values:
\begin{equation}
z_{\mathrm{max}, i} = \mathrm{min}\left(z_\mathrm{max}^\mathrm{s},\ z_{\mathrm{max}, i}^\mathrm{mass}\right)
\end{equation}
Once we have determined the \Vmaxi\ values for all objects in our subsample, we calculate the number density in each stellar mass bin using equation \ref{eq:Vmax_phi}.

\subsubsection*{Error calculation}
To estimate the random errors on $\Phi$ we follow the approach by \cite{Zhu:2009aa} and \cite{Gilbank:2010ab}. For each mass bin $j$ we determine the effective weight which is defined as: 

\begin{equation}
W_{\mathrm{eff}, j} = \frac{\sum_i^{N_\mathrm{bin}} \frac{w_{\mathrm{spec}, i}^2}{V_{\mathrm{max},i}^2}} {\sum_i^{N_\mathrm{bin}} \frac{w_{\mathrm{spec}, i}}{V_{\mathrm{max},i}}}
\end{equation}
and the effective number given by:
\begin{equation}
N_{\mathrm{eff}, j} = \frac{\sum_i \frac{w_{\mathrm{spec}, i}}{V_{\mathrm{max},i}}}{W_{\mathrm{eff}, j}} = \frac{(\sum_i \frac{w_{\mathrm{spec}, i}}{V_{\mathrm{max},i}})^2}{\sum_i \frac{w_{\mathrm{spec}, i}^2}{V_{\mathrm{max},i}^2}}
\end{equation}

We use the results by \cite{Gehrels:1986aa} to calculate $\lambda_\mathrm{up}$ and $\lambda_\mathrm{low}$, the upper and lower limits on $N_\mathrm{eff}$ ($S = 1$, equations (7) and (11)). The $1\sigma$ upper and lower errors on $\Phi_j$ are then given by:
\begin{equation}
\begin{aligned}
\sigma_{\Phi_j, \mathrm{up}} = - \Phi_j + W_{\mathrm{eff}, j} \times \lambda_{\mathrm{up}, j}(N_{\mathrm{eff}, j})\\
\sigma_{\Phi_j, \mathrm{low}} = \Phi_j - W_{\mathrm{eff}, j} \times \lambda_{\mathrm{low}, j}(N_{\mathrm{eff}, j})
\end{aligned}
\end{equation} 

For large $N$, $\sigma_{\Phi, \mathrm{up}}$ and $\sigma_{\Phi, \mathrm{low}}$ both approach the limit:
\begin{equation}
W_\mathrm{eff} \ \times \sqrt{N_\mathrm{eff}} = \sqrt{\sum_i^{N_\mathrm{bin}} \frac{w_{\mathrm{spec}, i}^2}{V_{\mathrm{max}, i}^2}}  
\end{equation}
which is commonly used in the literature \citep{Marshall:1985aa, Boyle:1988aa}.

For stellar mass bins with $N = 0$ we use a similar approach to calculate $1\sigma$ upper limits. In this case $V_\mathrm{max}$ is given by the comoving volume of our entire sample $V^{s}(z_\mathrm{min}^\mathrm{s}, z_\mathrm{max}^\mathrm{s})$. Following  \cite{Gehrels:1986aa} (Table 1) we set $\lambda_\mathrm{up}(N_\mathrm{eff} = 0) = 1.841$. The upper limit on $\Phi$ is then given by:
\begin{equation}\label{eq:Vmax_up_lim}
\begin{aligned}
\sigma_{\Phi, \mathrm{limit}} &= -\Phi_\mathrm{limit} + W_{\mathrm{eff}, \mathrm{limit}} \times \lambda_\mathrm{up}(N_\mathrm{eff} = 0) \\
											    &= -\frac{1}{V^\mathrm{s}} + 1.841 \times \frac{1}{V^\mathrm{s}} \\
											    &= 0.841 \times \frac{1}{V^\mathrm{s}}.
\end{aligned}
\end{equation}

\subsection{The STY technique}
\label{sec:STY_tech}
\begin{figure*}
 	\includegraphics[width=\textwidth]{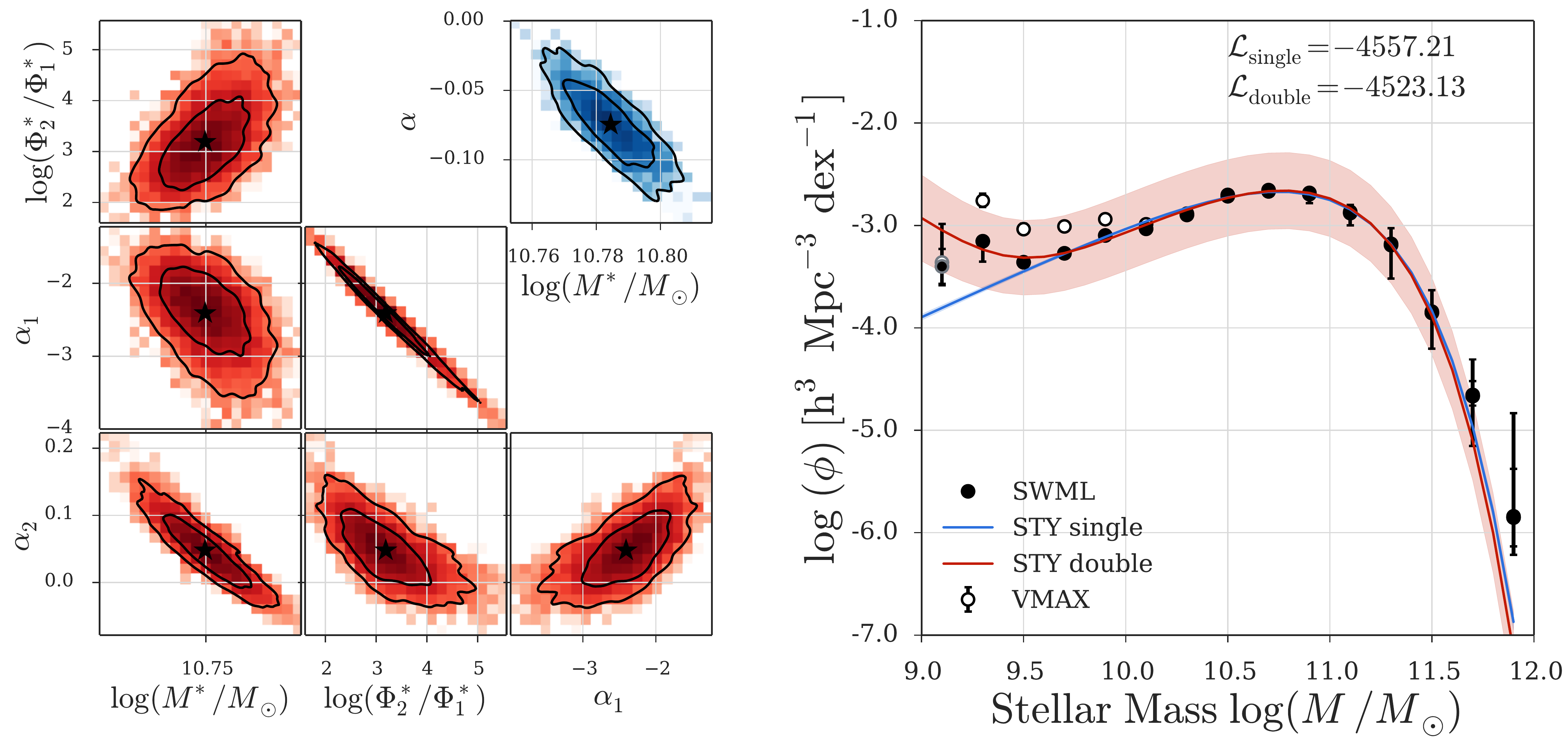}
	\caption{\label{fig:MCMC_results}Results for Early type galaxies. On the left we are showing the STY MCMC results for Early type galaxies  for the single Schechter function in blue in the top right panel and for the double Schechter function in red on the left and at the bottom. For the single and double Schechter function we are varying two ($M^{*}$, $\alpha$) and four ($M^{*}$, $\log (\Phi^{*}_2/\Phi^{*}_1)$, $\alpha_1$, $\alpha_2$) free parameters, respectively. The $1\sigma$ and $2\sigma$ contours are shown in black. The mean is marked with a star. Illustrated on the right is the stellar mass function for Early type galaxies. The 1/\Vmax\ points are shown with open, the SWML values with filled symbols. The best fit single Schechter function according to STY is shown with a blue, dashed line. The red solid line is illustrating the best fit STY double Schechter function. The shaded regions show the $1\sigma$ uncertainties on the STY results. Also given in the right panel are the STY likelihoods values for the single and double Schechter. For this subsample we measure $\mathcal{L}_\mathrm{double} > \mathcal{L}_\mathrm{single}$ and with the likelihood ratio test (see Sec. \ref{sec:single_double}) we derive a p-value of \pvalueET. We thus conclude that this subsample is better described by a double Schechter function.}
 \end{figure*}

\subsubsection*{The basic principle}
The STY technique is a parametric maximum likelihood method that was originally developed by \cite{Sandage:1979aa}. STY relies on the fact that the probability of object $i$ being part of a subsample can be expressed as \citep{Efstathiou:1988aa, Willmer:1997aa}:
\begin{equation}\label{eq:prob_STY}
p_i(M_i) \propto \frac{\Phi(M_i)}{\int_{M_{\mathrm{min}, i}}^{M^{\mathrm{max}, i}} \Phi(M) dM}.
\end{equation}
$M_{\mathrm{min},i}$ is the mass at which source $i$ drops below the mass completeness limit (see Sec. \ref{sec:mass_completeness}) of the subsample. $M_{\mathrm{max},i}$ is the upper stellar mass limit which we set to \Mstarintmax \Msun\ for all objects.

The likelihood of all galaxies being part of the same subsample is then given by:
\begin{equation}\label{eq:likeli_intro_STY}
\mathcal{L} \propto \prod_{i = 1}^{N_\mathrm{g}} p_i(M_i)
\end{equation}
where $N_\mathrm{g}$ is the total number of galaxies in the subsample.
We can now assume a Schechter function for the functional form of $\Phi$ and maximise the likelihood $\mathcal{L}$ relative to the function's parameters. We are are maximizing $\mathcal{L}$ for both, a single and a double Schechter function.   

\subsubsection*{The method}
Similar to the 1/\Vmax\ method, we are also including the spectroscopic completeness $w_{\mathrm{spec}, i}$ here. $\ln \mathcal{L}$ can thus be written as:
\begin{equation}\label{eq:likelihood_STY}
\begin{aligned}
\ln \mathcal{L} = & \sum_{i = 1}^{N_\mathrm{g}} w_{\mathrm{spec}, i} \Big[ \ln(\Phi(M))\\
							& - \ln \Big (\int_{\log M_{\mathrm{min},i}}^{\log M_{\mathrm{max},i}} \Phi(M) d\log M \Big) \Big]
\end{aligned}
\end{equation}

\subsubsection*{Single Schechter}
In the single Schechter function case the two free parameters are $\log M^{*}$ and $\alpha$. In equation \ref{eq:likelihood_STY} the normalisation $\Phi^{*}$ is eliminated and can thus not be constrained. We assume the following Schechter function:
\begin{equation} \label{eq:single_Schechter_STY}
\tilde{\Phi}\ d\log M = e^{-10^{\log M - \log M^{*}}} (10^{\log M - \log M^{*}})^{\alpha + 1}\ d\log M
\end{equation}
Given this functional form, equation \ref{eq:likelihood_STY} can be rewritten in the following way:
\begin{equation} \label{eq:STY_ln_likeli_single}
\begin{aligned}
\ln \mathcal{L}  = & - \frac{1}{M^{*}} \sum^{N_g}_{i} (w_{\mathrm{spec}, i} M_i) + (\alpha + 1)\ \sum^{N_\mathrm{g}}_{i} w_{\mathrm{spec}, i} \ln ( M_i / M^{*}) \\
						     &-\sum^{N_g}_{i} w_{\mathrm{spec}, i} \ln \Big(\int_{\log M_{\mathrm{min}, i}}^{\log M_{\mathrm{max}, i}} \tilde{\Phi} d\log M\Big)
\end{aligned}
\end{equation}

\subsubsection*{Double Schechter}
For the double Schechter function we consider four free parameters: $\log M^{*}$, $\alpha_1$, $\alpha_2$ and $\log(\Phi^{*}_2 / \Phi^{*}_1)$. We assume the following Schechter function:
\begin{equation} \label{eq:double_Schechter_STY}
\begin{aligned}
\tilde{\Phi} d\log M = &  e^{-10^{\log M - \log M^{*}}} \Big[\Big(10^{\log M - \log M^{*}}\Big)^{\alpha_1 + 1} \\
								   &+ 10^{\log(\frac{\Phi^{*}_2}{\Phi^{*}_1})} \Big(10^{\log M - \log M^{*}}\Big)^{\alpha_2 + 1} \Big] d\log M
\end{aligned}
\end{equation}
Equation \ref{eq:likelihood_STY} then has the following form for the double Schechter function:
\begin{equation} \label{eq:STY_ln_likeli_double}
\begin{aligned}
\ln \mathcal{L} = & - \frac{1}{M^{*}} \sum^{N_\mathrm{g}}_{i} (w_{\mathrm{spec}, i} M_i) \\
							&+ \sum^{N_g}_{i} w_{\mathrm{spec}, i} \ln \Big[\Big(10^{\log M - \log M^{*}}\Big)^{\alpha_1 + 1} \\
							&+ \frac{\Phi^{*}_2}{\Phi^{*}_1} \Big(10^{\log M - \log M^{*}}\Big)^{\alpha_2 + 1}\Big] \\
							&+ \sum^{N_g}_{i} w_{\mathrm{spec}, i} \ln \Big(\int^{\log M_{\mathrm{max}, i}}_{\log M_{\mathrm{min}, i}} \tilde{\Phi} d\log M  \Big)
\end{aligned}
\end{equation}

\subsubsection*{MCMC}
\begin{table}
	\caption{\label{tab:MCMC_boundaries}Parameter boundaries for the MCMC run. The normalisation $\Phi^{*}$ can not be constrained by STY and is thus not listed here. For the double Schechter function run we are constraining $\log(\Phi^{*}_2/\Phi^{*}_1)$ to be positive and $\alpha_2$ to be $< \alpha_1$. This keeps the MCMC sampler from jumping between the two allowed, but equivalent solutions.}
	\centering
	\begin{tabular}{lll}
		\hline
		{} & {min} & {max}\\
		\hline 
		{$\log M^{*}$} & {9.5} & {11.5}\\
		{$\alpha$} & {-4.0} & {4.0}\\
		{} & {} & {}\\
		{$\log M^{*}$} & {9.5} & {11.5}\\
		{$\log(\Phi^{*}_2/\Phi^{*}_1)$} & {0.0} & {4.0}\\
		{$\alpha_1$} & {-4.0} & {-1.0}\\
		{$\alpha_2$} & {-1.0} & {4.0}\\
	\end{tabular}
\end{table}

We maximise $\ln \mathcal{L}$ using a Markov Chain Monte Carlo (MCMC) sampler. Specifically, we are using the \textsc{python} package \textsc{cosmohammer}\footnote{\url{http://cosmohammer.readthedocs.org/}} by \cite{Akeret:2013aa}. \textsc{cosmohammer} is based on \textsc{emcee} by \cite{Foreman-Mackey:2013aa} which uses the MCMC ensemble sampler by \cite{Goodman:2010aa}. In contrast to the classical Metropolis-Hastings algorithm \citep{Metropolis:1953aa, Hastings:1970aa} which only uses one walker, the method by \cite{Goodman:2010aa} uses an ensemble of walkers that sample the parameter space in parallel. The \cite{Goodman:2010aa} algorithm, does not require a covariance matrix as input. The efficiency of the sampling process is thus not sensitive to how well we can describe the target distribution. Furthermore, the use of multiple walkers allows the sampling to be parallelised. In addition to \textsc{cosmohammer}, we are also using \textsc{Hope}\footnote{\url{http://pythonhosted.org/hope/}} \citep{Akeret:2015aa}, a just-in-time \textsc{python} to \textsc{C++} compiler, to speed up the sampling process.

We are using 40 parallel walkers which results in a walkers ratio (number of walkers/free parameters) of 20 for the single and 10 for the double Schechter function, respectively. Furthermore, we are restricting the parameters to the values given in Table \ref{tab:MCMC_boundaries}. For the double Schechter function we are enforcing $\Phi^{*}_2 > \Phi^{*}_1$ and $\alpha_1 < \alpha_2$ to keep the sampler from jumping between the two allowed, but equivalent solutions. 

As an example we show the results of the MCMC run for the Early type galaxies subsample in the left panel of Figure \ref{fig:MCMC_results}. 

\subsubsection*{Normalisation}
The STY method does not constrain the normalisation $\Phi^{*}$ of the stellar mass function. We thus have to separately constrain $\Phi^{*}$ for the single and the double Schechter function. We are following the approach by \cite{Efstathiou:1988aa} that has for example been used by \cite{Chen:2003aa} and \cite{Ilbert:2005aa}.  

The number density of objects in the local Universe is given by:
\begin{equation}
n = \Phi^{*} \int^{\infty}_{-\infty} \tilde{\Phi} d\log M. 
\end{equation}

The probability of the single object $i$ being included in our subsample is therefore given by the selection function $s(z_i)$: 
\begin{equation}
 s(z_i) = \frac{\int^{\log M_{\mathrm{max}, i}}_{\log M_\mathrm{min, i}} \tilde{\Phi} d\log M}{\int^{\infty}_{-\infty} \tilde{\Phi} d\log M}
\end{equation}

Following \cite{Efstathiou:1988aa}, the total number of objects can be estimated by summing over all $N_\mathrm{g}$ sources in our subsample and weighing each source by $s(z_i)$, the object's probability of being included in the sample. The number density $n$ is then given by dividing this sum by the total volume we are sampling:
\begin{equation}
n = \frac{1}{V_\mathrm{tot}} \sum_i^{N_\mathrm{g}} \frac{w_{\mathrm{spec}, i}}{s(z_i)}
\end{equation}
$V_\mathrm{tot}$ is the comoving volume between $z_\mathrm{min}^\mathrm{s}$ and $z_\mathrm{max}^\mathrm{s}$. The normalisation $\Phi^{*}$ of the stellar mass function is thus given by:
\begin{equation} \label{eq:phistar_norm}
\Phi^{*} = \frac{1}{V_\mathrm{tot}} \sum_{i}^{N_\mathrm{g}} \frac{w_{\mathrm{spec}, i}}{\int^{\log M_{\mathrm{max}, i}}_{\log M_{\mathrm{min}, i}} \tilde{\Phi} d\log M}
\end{equation}
For the single Schechter function we assume equation. \ref{eq:single_Schechter_STY} multiplied by $\ln(10)$ for the $\tilde{\Phi}$ given here. We include the factor of $\ln(10)$ so that $\Phi^{*}$ is conform with our Schechter function definition in Sec. \ref{sec:Schechter_function}. For the double Schechter function we assume the functional form given in equation \ref{eq:double_Schechter_STY} multiplied by $\ln(10)$. The $\Phi^{*}$ value that we determine in equation \ref{eq:phistar_norm} then corresponds to $\Phi^{*}_1$ in equation \ref{eq:double_Schechter}. $\Phi^{*}_2$ is easily determined by multiplying the $\Phi^{*}_2 / \Phi^{*}_1$ value that we determine through the MCMC with $\Phi^{*}_1$. 

\subsubsection*{Error calculation}
The STY method has the advantage that the errors on the Schechter function parameters can easily be determined from the MCMC chain. We use the standard deviation of the parameter values tested by the MCMC after the burn-in to determine the $1\sigma$ errors on $M^{*}$ and $\alpha$ for the single Schechter function and on $M^{*}$, $\log(\Phi^{*}_2/\Phi^{*}_2)$, $\alpha_1$ and $\alpha_2$ for the double Schechter function. 
 
We also determine the random error on $\Phi$ as a function of stellar mass $M$ based on the covariance matrix. We compute the covariance matrix $\Sigma$ from the MCMC chain. For the single Schechter function $\Sigma$ has the following form:
\begin{equation}
\Sigma = 
	\begin{pmatrix}
	\Sigma_{M^{*}M^{*}} & \Sigma_{M^{*} \alpha}\\
	\Sigma_{M^{*} \alpha} & \Sigma_{\alpha \alpha}\\
	\end{pmatrix}.
\end{equation}
$\sqrt{\Sigma_{M^{*}M^{*}}}$ and  $\sqrt{\Sigma_{\alpha \alpha}}$ correspond to the marginalized $1\sigma$ errors on $M^{*}$ and $\alpha$, respectively. The standard deviation on $\Phi$ (see equation \ref{eq:single_Schechter}) as a function of stellar mass is then given by:
\begin{equation}
\begin{aligned}
\sigma^{2}_{\Phi} (\log M) = & \Big[\Big(\frac{\partial \Phi}{\partial \log M} \Big)^2 \Sigma_{M^{*}M^{*}} + \Big(\frac{\partial \Phi}{\partial \alpha}\Big)^2 \Sigma_{\alpha \alpha} \\
											 &+2 \Big(\frac{\partial \Phi} {\partial \log M}\Big) \Big(\frac{\partial \Phi}{\partial \alpha}\Big) \Sigma_{M^{*} \alpha}\Big]^{1/2}.
\end{aligned}
\end{equation}

$\Phi^{*}$ cannot be constrained by the STY method and needs to be determined separately. It is thus neither part of the MCMC chain, nor the covariance matrix $\Sigma$. We nonetheless want to constrain $\Phi^{*}$ and thus use 
\begin{equation}\label{eq:error_phi}
\sigma_{\log \Phi^{*}} = \sqrt{ \sigma^{2}_{\Phi} (\log M^{*})}/(\ln(10) \Phi(\log M^{*}))
\end{equation}
 as the error on $\Phi^{*}$.

We use the same approach to derive $\sigma_{\Phi}$ for the double Schechter function. Since the double Schechter function has four free parameters, $\Sigma$ takes the following forms:
\begin{equation} \label{eq:STY_single_error}
\Sigma = 
	\begin{pmatrix}
	\Sigma_{M^{*}M^{*}} & \Sigma_{M^{*}R} & \Sigma_{M^{*} \alpha_1} & \Sigma_{M^{*} \alpha_2}\\
	... & \Sigma_{RR} & \Sigma_{R \alpha_1} & \Sigma_{R \alpha_2}\\
	... & ... & \Sigma_{\alpha_1 \alpha_1} & \Sigma_{\alpha_1 \alpha_2}\\
	\text{symmetric} & ... & ... & \Sigma_{\alpha_2 \alpha_2}\\
	\end{pmatrix}
\end{equation}
with $R = \log(\Phi^{*}_2/\Phi^{*}_1)$. $\sigma_{\Phi}$ with $\Phi$ being given by equation \ref{eq:double_Schechter} can then be calculated in analogy to equation \ref{eq:STY_single_error}.

We also determine the error on $\log \Phi^{*}_1$ with equation \ref{eq:error_phi}. As $\log \Phi^{*}_2 = R + \log \Phi^{*}_1$, we compute the error on $\log \Phi^{*}_2$ by using $\sigma^2_{\log \Phi^{*}_2} = \sigma^2_R + \sigma^2_{\log \Phi^{*}_1}$.

\subsection{The SWML technique}
\label{sec:SWML_tech}
\subsubsection*{The basic principle}
The SWML technique is a non-parametric maximum likelihood approach to determine the shape of the stellar mass function. It was originally developed by \cite{Efstathiou:1988aa}. Instead of assuming a functional form for the stellar mass function like in the STY method, we bin in stellar mass, similar to the 1/\Vmax\ approach. By rewriting the likelihood in equation \ref{eq:likeli_intro_STY} we then derive an equation that allows us to determine the $\Phi$ values in each mass bin iteratively. 

\subsubsection*{The method}
To derive an equation for $\Phi$ \citep{Takeuchi:2000aa}, we bin in $\log M$ and rewrite the stellar mass function as:
\begin{equation}
\Phi(M) = \sum_{j}^{N_\mathrm{bins}} \Phi_j W(\log M_j - \log M)
\end{equation} 
where $W(\log M_j - \log M)$ is a step function defined in the following way:
\begin{equation}
\begin{aligned}
W(x) = 
   \begin{cases}
      1 & \text{for}\ -\frac{\Delta \log M}{2} \leq x \leq \frac{\Delta \log M}{2}\\
      0 & \text{otherwise}
   \end{cases}
\end{aligned}
\end{equation}
We choose $\Delta \log M = 0.2$, which is the same bin width as in the 1/\Vmax\ approach (see Sec. \ref{sec:Vmax_tech}).

The likelihood $\mathcal{L}$ in equation \ref{eq:likeli_intro_STY} can then be rewritten as:
\begin{equation} \label{eq:likeli_prod_SWML}
\begin{aligned}
\mathcal{L} = \prod_i^{N_\mathrm{g}} & \Bigg  [\sum_j^{N_\mathrm{bins}} W(\log M_j - \log M_i) \Phi_j \Bigg]\\
															  &\cdot  \Bigg [\sum_j^{N_\mathrm{bins}} \Phi_j H(\log M_{\mathrm{max}, i} - \log M_j)  \\
															  &\cdot H(\log M_j - \log M_{\mathrm{min, i}}) \Delta\log M \Bigg]^{-1}.
\end{aligned}
\end{equation}
$N_\mathrm{g}$ is the total number of objects in our subsample, $\log M_j$ is the central mass of the stellar mass bin $j$, $M_{\mathrm{max}, i}$ is the upper stellar mass limit, which we set to \Mstarintmax \Msun\ for all objects, and $M_{\mathrm{min},i}$ is the mass at which source $i$ drops below the mass completeness limit (see Sec. \ref{sec:mass_completeness}). $H(x)$ is a step function which is defined in the following way:
\begin{equation}\label{eq:window_H}
H(x) = \\
   \begin{cases}
     0 & \text{for}\ x < -\frac{\Delta \log M}{2}\\
     \frac{x}{\Delta \log M} + \frac{1}{2} & \text{for}\ -\frac{\Delta \log M}{2} \leq x \leq \frac{\Delta \log M}{2}\\
     1 & \text{for}\ x > \frac{\Delta \log M}{2} 
   \end{cases}
\end{equation}

Given these definitions, we can write $\ln \mathcal{L}$ as:
\begin{equation} \label{eq:SWML_likeli}
\begin{aligned}
\ln \mathcal{L} = \sum_i^{N_\mathrm{g}} & \Big[ \sum_j^{N_\mathrm{bins}} \Big( W(\log M_j - \log M_i) \ln \Phi_j \Big)  \\
																   &- \ln  \Big(\sum_j^{N_\mathrm{bins}}  \Phi_j  H(\log M_{\mathrm{max}, i} - \log M_j)\\
																   &\cdot  H(\log M_j - \log M_{\mathrm{min, i}}) \Delta \log M\Big)\Big]
\end{aligned}
\end{equation}
Note that in the first term, the $\ln$ can be pulled into the sum over $j$ due to the window function $W(x)$. Our stellar mass bins are not overlapping and so $\sum_j^{N_\mathrm{bins}} W(\log M_j - \log M_i) = 1$ for fixed $i$. 
 
To maximise $\ln \mathcal{L}$, we set $\frac{d\ln \mathcal{L}}{d\Phi_k}  = 0$ and solve the equation for $\Phi_k d\log M$:

\begin{equation}\label{eq:SWML_iter}
\begin{aligned}
\Phi_k d\log M = & \sum_i^{N_\mathrm{g}} w_{\mathrm{spec}, i} W(\log M_k - \log M_i) \\
							&\cdot \left[\sum_i^{N_\mathrm{g}}\frac{w_{\mathrm{spec}, i} H_{1, k} H_{2, k}} {\sum_j^{N_\mathrm{bins}} \Phi_j H_{1, j} H_{2, j} d\log M} \right]^{-1}.
\end{aligned}
\end{equation}

$H_{1, k} = H(\log M_{\mathrm{max}, i} - \log M_k)$, $H_{2, k} = H(\log M_k - \log M_{\mathrm{min, i}})$, $H_{1,j} = H(\log M_{\mathrm{max}, i} - \log M_j)$ and $H_{2,j} = H(\log M_j - \log M_{\mathrm{min, i}})$. 

We also include the spectroscopic completeness values according to \cite{Ilbert:2005aa}. equation \ref{eq:SWML_iter} now allows us to determine the $\Phi_k$ value in each of the $N_\mathrm{bins}$ iteratively. 
Following \cite{Efstathiou:1988aa}, we also include an additional constraint which fixes the normalisation of the $\Phi_k$ values: 
\begin{equation}
g = \sum_j^{N_\mathrm{bins}} \Phi_j \Delta \log M (\log M_j - \log M_f)^{\beta} - 1 = 0.
\end{equation}
$M_f$ is a fiducial mass, which we set to $10^7$ \Msun. $\beta$ is a constant which we choose to be equal to 1.5 \citep{Efstathiou:1988aa}. We add this condition to the likelihood equation by using a Lagrangian multiplier, $\ln \mathcal{L}' = \ln \mathcal{L} + \lambda g(\Phi_k)$, and maximise $\ln \mathcal{L}'$ with respect to $\Phi_k$ and $\lambda$:
\begin{equation}
\begin{aligned}
\frac{\partial \ln \mathcal{L}'}{\partial \Phi_k} &= \frac{\partial \ln \mathcal{L}'}{\partial \Phi_k} + \lambda \frac{\partial g(\Phi_k)}{\partial \Phi_k} = 0\\
\frac{\partial \ln \mathcal{L}'}{\partial \lambda} &= g(\Phi_k) = 0.
\end{aligned}
\end{equation}

We test the convergence by determining $\epsilon = |\ln \mathcal{L}(\Phi_\mathrm{previous}) - \ln \mathcal{L}(\Phi_\mathrm{new})|$ and stop the iteration once $\epsilon < 10^{-5}$.

\subsubsection*{Error calculation}
To estimate the random errors on $\Phi$ for the SWML method, we compute the covariance matrix $C$, which is the inverse of the Fisher-information matrix $I$, according to \cite{Efstathiou:1988aa}. $C$ has the following form:

\begin{equation}
\footnotesize
\arraycolsep=3pt
\medmuskip = 1mu
\begin{aligned}
&C(\Phi) = I^{-1}(\Phi) = \\
&\begin{pmatrix}
\frac{\partial^2 \ln \mathcal{L}}{\partial \Phi_1^2} + \Big(\frac{\partial g}{\partial \Phi_1}\Big)^2 & 
... & 
\frac{\partial^2 \ln \mathcal{L}}{\partial \Phi_1 \partial_\mathrm{N_\mathrm{bins}}} + \frac{\partial g}{\partial \Phi_1} \frac{\partial g}{\partial \Phi_\mathrm{N_\mathrm{bins}}} & 
\frac{\partial g}{\partial \Phi_1}\\

... & ... & ... & ...\\

\frac{\partial^2 \ln \mathcal{L}}{\partial \Phi_\mathrm{N_\mathrm{bins}}  \partial \Phi_1} + \frac{\partial g}{\partial \Phi_\mathrm{N_\mathrm{bins}}} \frac{\partial g}{\partial \Phi_1} &
... & 
\frac{\partial^2 \ln \mathcal{L}}{\partial \Phi_\mathrm{N_\mathrm{bins}}^2} + \Big( \frac{\partial g}{\partial \Phi_\mathrm{N_\mathrm{bins}}} \Big)^2 &
\frac{\partial g}{\partial \Phi_\mathrm{N_\mathrm{bins}}}\\

\frac{\partial g }{\partial \Phi_1} & 
... & 
\frac{\partial g }{\partial \Phi_\mathrm{N_\mathrm{bins}}} &
0\\
\end{pmatrix}^{-1}.
\end{aligned}
\end{equation}
$\ln \mathcal{L}$ is given by equation \ref{eq:SWML_likeli}. $\frac{\partial \ln \mathcal{L}}{\partial \Phi_i \Phi_j}$ is given by \citep{Ilbert:2005aa}:

\begin{equation}
\begin{aligned}
\frac{\partial \ln \mathcal{L}}{\partial \Phi_i \partial \Phi_j}  = &  - \sum_l^{N_{\mathrm{g}}} \frac{w_{\mathrm{spec}, l} \delta_{ij} W(\log M_l - \log M_j) (\Delta \log M)^2} {(\Phi_j \Delta \log M/(g + 1))^2}\\
&+ \sum_l^{N_\mathrm{g}} \frac{w_{\mathrm{spec}, l} (\Delta \log M)^2 H_1}{\Big(\sum_k^{N_\mathrm{bin}} \Phi_k \Delta \log M H_2/(g + 1)\Big)^2}
\end{aligned}
\end{equation}

with $H_1 = H(\log M_i - \log M_{\mathrm{min}, i}) H(\log M_{\mathrm{max}, i} - \log M_i ) H(\log M_j - \log M_{\mathrm{min}, j}) H(\log M_{\mathrm{max}, j} - \log M_j)$ and $H_2 = H(\log M_k - \log M_{\mathrm{min}, l}) H(\log M_{\mathrm{max}, l} - \log M_k)$. To determine the random error on $\Phi$ we take the square root of the diagonal elements of $C$.

For the SWML $\Phi$ values we also account for the systematic errors due to stellar mass uncertainties in addition to the random errors. We redetermine the SWML $\Phi$ values for the subsample using the 16th ($-1\sigma$) and 84th ($+1\sigma$) percentile values for $\log M$.  We calculate the upper and lower $1\sigma$ systematic error on $\Phi$ by measuring $\Phi_{84} - \Phi$ and $\Phi - \Phi_{16}$, respectively. 
We derive the total error on $\Phi$ by adding the random and systematic errors in quadrature. 
 
 \subsection{Single vs. double Schechter}
 \label{sec:single_double}
 For each of our subsamples, we want to know if the stellar mass function shape is better fit by a single or by a double Schechter function. The single and the double Schechter functions are however non-linear, which implies that their number of degrees of freedom cannot be estimated. We can thus not simply compare their reduced chi-squared values \citep{Andrae:2010aa}. Instead we use a likelihood ratio test to determine the better fitting model. 
 
 We use the best-fit parameters that we have determined trough the MCMC and equation \ref{eq:STY_ln_likeli_single} and equation \ref{eq:STY_ln_likeli_double} to estimate the maximised $\ln \mathcal{L}_\mathrm{single}$ and $\ln \mathcal{L}_\mathrm{double}$ values, respectively. The double Schechter function can easily be reduced to a single Schechter function by setting $\alpha_1 = \alpha_2$ and $\Phi^{*}_1 = \Phi^{*}_2$. The double Schechter function hence represents the alternative model whereas, the single Schechter function corresponds to the null model. 
 
 In the case of $\ln \mathcal{L}_\mathrm{single} \geq \ln \mathcal{L}_\mathrm{double}$, the simple null model describes the data better. If  $\ln \mathcal{L}_\mathrm{single} < \ln \mathcal{L}_\mathrm{double}$, we need to test if the alternative model is significantly better than the null model. To do so, we compute the likelihood ratio:
 \begin{equation}\label{eq:likeli_ratio}
 R = -2 \ln \mathcal{L}_\mathrm{single} + 2 \ln \mathcal{L}_\mathrm{double}
 \end{equation}
 The test statistic is approximately $\chi^2$ distributed with $d = d_\mathrm{double} - d_\mathrm{single} = 4 - 2 = 2$ degrees of freedom. We estimate the $p$-value with the null hypothesis that the data is better fit with the alternative model, the double Schechter function. We accept this null hypothesis if $p < 5$\%. 
 
 In our stellar mass function figures we hence show the single Schechter function that we determined with the STY method, if $\ln \mathcal{L}_\mathrm{single} \geq \ln \mathcal{L}_\mathrm{double}$ or if $p \geq 5$\% and we plot the double Schechter function if $p < 5$\%.
 
 In Figure \ref{fig:MCMC_results} we show the results for the Early type galaxies subsample. For this set of sources $\mathcal{L}_\mathrm{single} < \mathcal{L}_\mathrm{double}$ with \pvalueET\ and so we conclude that the data is better described by a double Schechter function. 
 
Alternative techniques to determine if the data is better described by a single or by a double Schechter function include the Bayesian information criterion (BIC, \citealt{Schwarz1978}) and the Akaike information criterion (AIC, \citealt{Akaike:1974aa}). Similar to the likelihood ratio test, the BIC and the AIC are based on comparing likelihoods and introducing penalties which depend on the number of free parameters: 
\begin{equation}
\begin{aligned}
\mathrm{BIC} &= -2 \ln \mathcal{L} + d \ln{n}\\
\mathrm{AIC} &= 2d - 2\ln \mathcal{L}.
\end{aligned}
\end{equation}
 $d$ and $n$ correspond to the number of free model parameters and the sample size, i.e. the number of objects above the mass completeness limit, respectively. The single Schechter model is preferred by the BIC (AIC) if $\mathrm{BIC}_\mathrm{single} \leq \mathrm{BIC}_\mathrm{double}$ ($\mathrm{AIC}_\mathrm{single} \leq \mathrm{AIC}_\mathrm{double}$). The double Schechter function describes the data better if $\mathrm{BIC}_\mathrm{single} > \mathrm{BIC}_\mathrm{double}$ ($\mathrm{AIC}_\mathrm{single} > \mathrm{AIC}_\mathrm{double}$). We can thus use the following equations to distinguish between single and double Schechter:
\begin{equation}
\begin{aligned}
R_\mathrm{BIC} &=  \mathrm{BIC}_\mathrm{single} - \mathrm{BIC}_\mathrm{double}\\
						  & = -2 \ln \mathcal{L}_\mathrm{single}  + 2 \ln n + 2 \ln \mathcal{L}_\mathrm{double} - 4 \ln n\\
						  & = R - 2 \ln n\\
R_\mathrm{AIC} & =  \mathrm{AIC}_\mathrm{single} - \mathrm{AIC}_\mathrm{double}\\
						  & = 4 - 2\ln \mathcal{L}_\mathrm{single} - 8 + 2\ln \mathcal{L}_\mathrm{double}\\
						  & = R - 4
\end{aligned}
\end{equation} Similar to the likelihood ratio test, a single Schechter fit is preferred if $R_\mathrm{BIC} \leq 0$ ($R_\mathrm{AIC} \leq 0$) and a double Schechter fit is favoured if $R_\mathrm{BIC} > 0$ ($R_\mathrm{AIC} > 0$).

We compare the results based on the likelihood ratio test, the BIC and the AIC. We find discrepancies between the preferred model according to the likelihood ratio test and the BIC for ten out of the 135 subsamples (red, $\log(\delta + 1) \leq 0.05$, Late types \&  $-12 \leq \log(\text{sSFR}) < -11$, Late types \& $-11 \leq \log(\text{sSFR}) < -10 $, indeterminate \& red, indeterminate \& green, blue \& $-12 \leq \log(\text{sSFR}) < -11$,  $\log(\delta + 1) > 0.05$ \& $-12 \leq \log(\text{sSFR}) < -11$, red \& $\log(\delta + 1) > 0.05$, satellites \& $\log(\delta + 1) \leq 0.05$). For each of these subsamples, the BIC prefers a single Schechter fit whereas the likelihood ratio favours a double Schechter fit. The likelihood ratio test and the AIC show disagreement for two of the 135 subsamples (indeterminate \& $\log(\delta + 1) \leq 0.05$, $\log(\text{sSFR}) < -12$ \& $12 \leq \log(M_h) < 13.5 $, $\log(\delta + 1) > 0.05$ \& $-12 \leq \log(\text{sSFR}) < -11$).  In both cases, the AIC favours a double Schechter fit, whereas the likelihood ratio test prefers a single Schechter fit.

For the ten subsamples mentioned above, BIC favours a single Schechter fit, i.e. the model with fewer free parameters. For the three subsamples where we find discrepancies between the AIC and the likelihood ratio test, the AIC suggests that the data is better described by a double Schechter fit. For those two subsamples $\mathcal{L}_\mathrm{single}$ is indeed smaller than $\mathcal{L}_\mathrm{double}$. The difference between $\mathcal{L}_\mathrm{single}$ and $\mathcal{L}_\mathrm{double}$ is however not significant ($p \geq 5$\%) and the likelihood ratio hence favours a single Schechter fit.

Besides the differences in how free parameters are penalized, the likelihood ratio test, the BIC and the AIC show good overall agreement. In the few cases where we find discrepancies, the difference between $\mathcal{L}_\mathrm{single}$ and $\mathcal{L}_\mathrm{double}$ is small and single and double Schechter fits result in similar reduced $\chi^2$ values. We chose to rely on the likelihood ratio test and show the better fitting model according to this measure in our plots and in our result tables.
 
 \subsection{The procedure}
 \label{sec:procedure}
 \begin{figure}
 	\includegraphics[width=.5\textwidth]{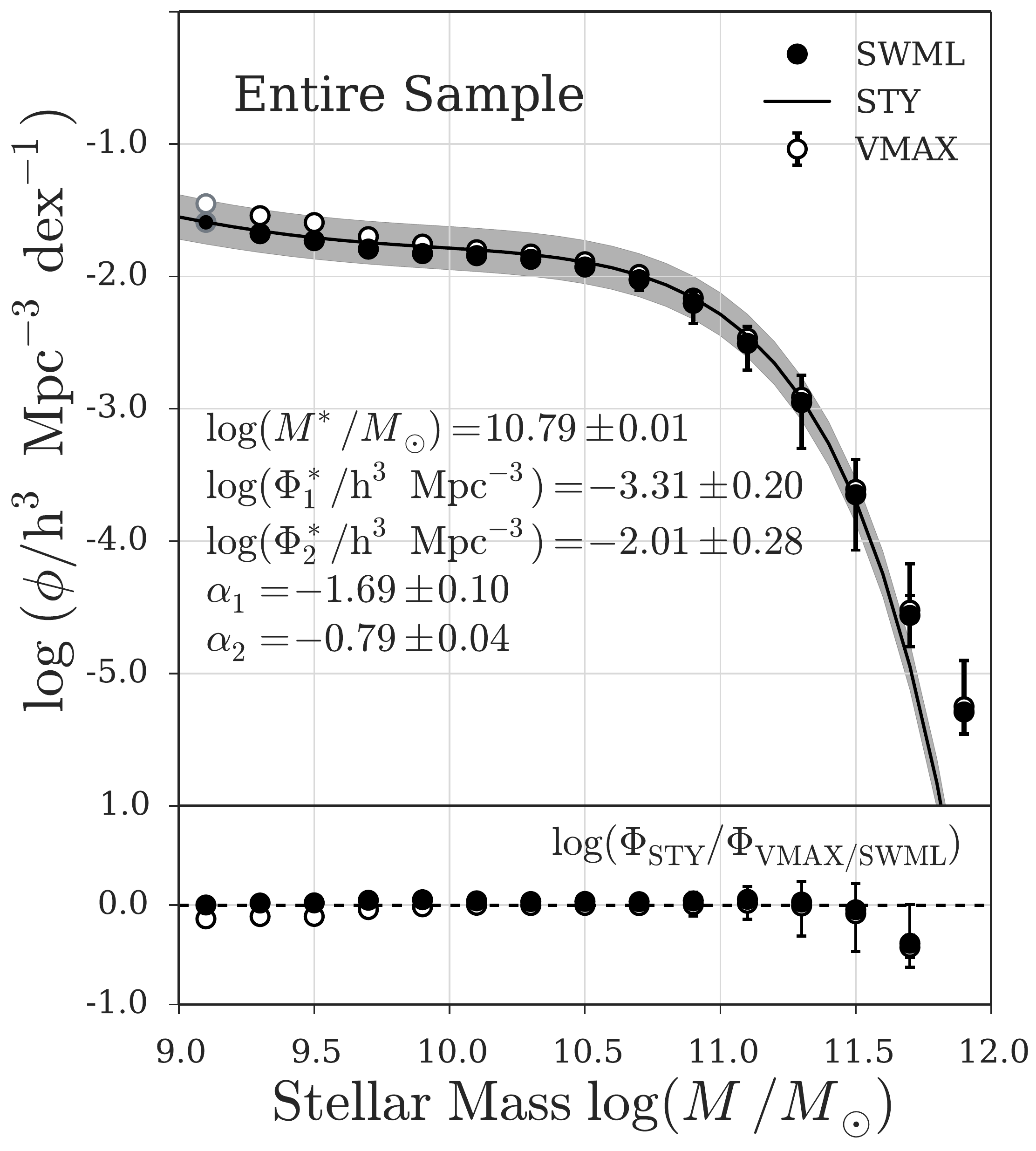}
 	\caption{\label{fig:mass_fct_entire_sample} Mass function for our $\sim 110'000$ main sample galaxies with redshifts in the range \zminsample $< z < $ \zmaxsample. In the top panel we show the results from the classical 1/\Vmax\ approach \protect\citep{Schmidt:1968aa} as open symbols, the $\Phi$ values estimated with the non-parametric maximum likelihood method by \protect\cite{Efstathiou:1988aa} (SWML) as filled symbols and the best fitting Schechter function determined with the parametric maximum likelihood method by \protect\cite{Sandage:1979aa} (STY) as a solid line. The shaded region shows the $1\sigma$ uncertainty according to the STY method. For the 1/\Vmax\ points we show random errors and for the SWML points we show the combination of random errors and the systematic error due to uncertainties in stellar mass. The best fit parameters according to STY are shown within the figure. The bottom panel shows the offset between STY and 1/\Vmax\ as open and the offset between STY and SWML as filled symbols, respectively. For a comparison to previously published mass functions see Sec. \ref{sec:previous_work} and Figure \ref{fig:comparison}.}
 \end{figure}
 
 Let us now summarise how we combine the methods discussed above to determine the stellar mass functions for each of our over 130 subsamples. Our main sample consists of about $ 110'000$ sources between $0.02 < z < 0.06$ which have been classified as galaxies in the MPA JHU catalogue \citep{Brinchmann:2004aa,Kauffmann:2003ab} and for which spectroscopic completeness values are available from the \cite{Yang:2007aa} catalogue. The mass function for the entire sample is shown in Figure  \ref{fig:mass_fct_entire_sample}. According to the STY method, our sample is best described by a double Schechter function with $\log (M^{*}/M_\odot) = $ \Mstarentiresample, $\log (\Phi^{*}_1/\mathrm{h^3\ Mpc^{-3}}) = $ \firstphistarentiresample, $\log (\Phi^{*}_2/\mathrm{h^3\ Mpc^{-3}}) = $ \secondphistarentiresample, $\alpha_1 = $ \firstalphaentiresample\ and $\alpha_2 = $ \secondalphaentiresample. For a comparison to previously published mass functions see Sec. \ref{sec:previous_work}.
 
In a first step we define our subsample based on, for example, colour and halo mass.

 We then use the approach by \cite{Pozzetti:2010aa} to determine the mass completeness function $M_\mathrm{lim}(z)$ (see Sec. \ref{sec:mass_completeness}). This step has to be taken every time since the range of M/L ratios is different for each of the subsamples. Once $M_\mathrm{lim}(z)$ has been derived, we eliminate all sources below the stellar mass completeness. For each of the subsample objects, we also estimate $M_{\mathrm{min}, i}$ and $z_{\mathrm{max}, i}$, which are input parameters for the maximum likelihood and the 1/\Vmax\ methods, respectively (see e.g. Figure \ref{fig:Mlim}). 
 
After binning in stellar mass, we determine $\Phi(M)$ in each bin using the 1/\Vmax\ method \citep{Schmidt:1968aa} (see Sec. \ref{sec:Vmax_tech}).  In our plots, the 1/\Vmax\ points are shown with open symbols. We also estimate random errors on $\Phi$ following the approach by \cite{Zhu:2009aa} and \cite{Gilbank:2010aa}.

We fit a single Schechter function to the 1/\Vmax\ points and use this as an initial guess for the non-parametric maximum likelihood method by \cite{Efstathiou:1988aa}, SWML (see Sec. \ref{sec:SWML_tech}). For SWML we estimate random errors and the systematic error due to stellar mass uncertainties and indicate the $\Phi$ values using filled symbols.

When plotting the stellar mass functions, we highlight the SWML and 1/\Vmax\ $\Phi$ value of the first bin above the lower stellar mass completeness cut, i.e. the first bin for which $\log M \geq M_\mathrm{lim}(z_\mathrm{min}^{s})$, with a gray border. Bins below the mass completeness function are not shown. For the 1/\Vmax\ method we compute upper limits with equation \ref{eq:Vmax_up_lim}. For the SWML upper limits we also take the systematic error on the stellar mass measurements into account and thus find higher upper limits compared to the 1/\Vmax\ values.

Furthermore, we determine the best fit single and double Schechter parameters using the parametric maximum likelihood approach by \cite{Sandage:1979aa} (see Sec. \ref{sec:STY_tech}). Analogous to SWML, STY is unable to constrain the normalisation of the stellar mass function. We hence have to constrain $\Phi^{*}$ separately. To maximise the likelihood we are varying two ($M^{*}$, $\alpha$) and four ($M^{*}$, $\log (\Phi^{*}_1/\Phi^{*}_2)$, $\alpha_1$, $\alpha_2$) free parameters for the single and the double Schechter functions, respectively. The errors on the Schechter function parameters can be derived directly from the MCMC chain. In our plots we show the $1\sigma$ random error on the Schechter function, which we estimate using the MCMC covariance matrix, as a shaded region. 

To conclude if the subsample is better fit by a single or by a double Schechter function, we use the STY likelihoods to calculate the likelihood (see Sec. \ref{sec:single_double} and Figure \ref{fig:MCMC_results}). In our figures we show the better fitting model as a solid line. 

For the better fitting model, we also compare the STY and SWML results by estimating the reduced $\chi^2$ value. We estimate $\chi^2_\mathrm{red}$ in log-space and use the mean of the upper and the lower SWML error on $\log \Phi$ to derive the residuals.

\section{Testing the code}
\label{sec:test_code}
\subsection{Simulation}
\label{sec:simulation}
\begin{figure*}
	\includegraphics[width=\textwidth]{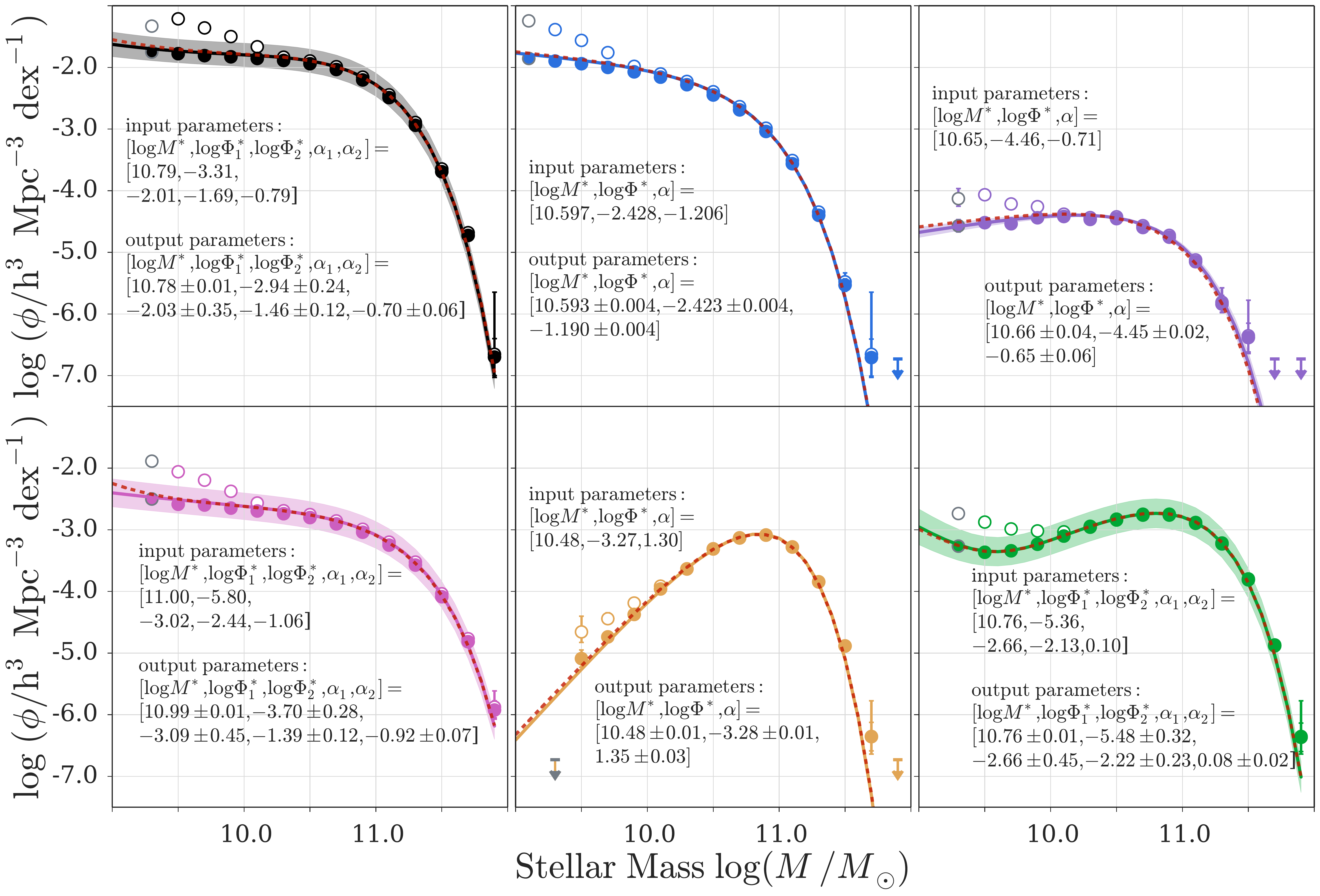}
	\caption{\label{fig:sim_comp} Stellar mass function simulation results. To test our stellar mass function code and compare the different techniques to estimate $\Phi$ we create a simulation that generates a mock catalogue based on an input Schechter function. The red dashed lines show the input Schechter functions. The correspodining input parameters are given within the panels. We use the mock catalogues as input for our stellar mass function code and show the 1/\Vmax\ with open and the SWML results with filled symbols, respectively. The best fit Schechter functions according to STY are illustrated with solid lines, the corresponding parameters are given within the panels. This figure illustrates that for 1/\Vmax\ we measure a significant discrepancy between input and output at the low mass end. This offset could be due to our simulation not correctly reproducing the underlying stellar mass - redshift distribution. This would cause a distortion of the stellar mass completeness function and could thus explain the overprediction of low mass galaxies and the lowest mass bins not containing any sources. The discrepancy could also be caused by an intrinsic bias in the 1/\Vmax\ method itself or a combination of those two effects. The STY and SWML methods retrieve the input Schechter functions more reliably even though they are based on the same stellar mass completeness function as 1/\Vmax. SWML marginally underpredicts the $\Phi$ values at the low mass end for samples with $\alpha + 1 < 0$.  STY slightly underpredicts the low mass end slope of the mass functions shown in the left column and overpredicts the low mass end slope for the mass function in the bottom right panel. Note, that the error on $\log \Phi^{*}$, which is particullary small for the mass functions shown in the top middle and top left panel, cannot be constrained through STY and the MCMC chain. Instead, we calcualte the error on $\log \Phi^{*}$ by estimating the uncertainty in $\Phi$ at $\log M^{*}$ (see Sec. \ref{sec:STY_tech}).}
\end{figure*}

\begin{figure*}
	\includegraphics[width=\textwidth]{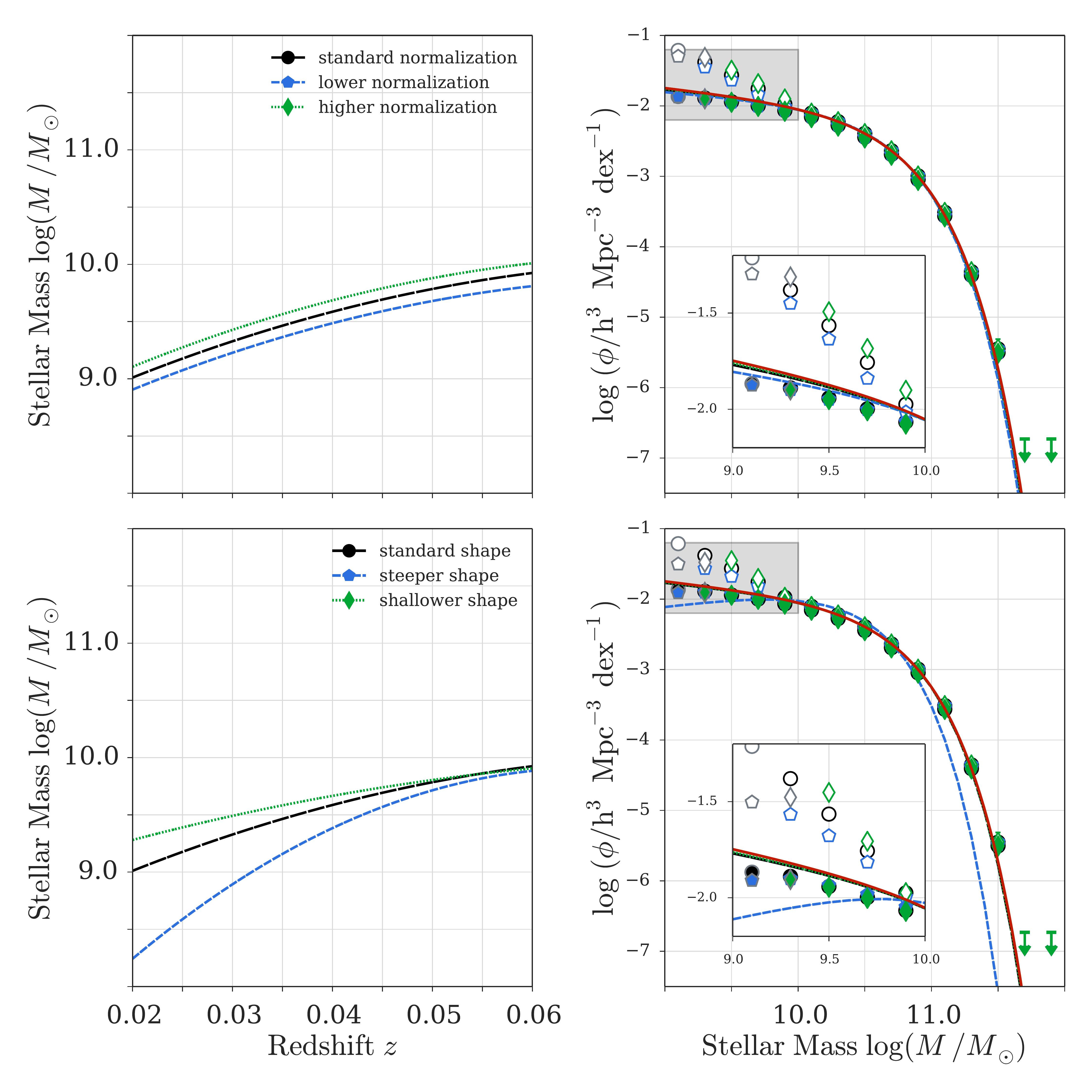}
	\caption{\label{fig:sim_comp_tests} Stellar mass function simulation results for different stellar mass completness functions. We are showing the effects of different completeness functions on the low mass end of the mass function. To generate these mass functions we again used our simulation and always assumed the same input Schechter function, which is shown with a red solid line on the right. Illustrated in the left panels are the $M_\mathrm{lim}(z)$ functions that we use to determine the stellar mass functions shown on the right. The insets in the left panels show zoom-ins of the grey shaded regions. This figure illustrates that the 1/\Vmax\ points are significantly offset from the input function. They do however behave as expected: the higher the normalisation of $M_\mathrm{lim}(z)$, the lower the $z_{\mathrm{max}, i}$ value that source $i$ with mass $M_i$ will be assigned and the higher the resulting $\Phi$ value. The shape of $M_\mathrm{lim}(z)$ only determines which bins are strongly affected by this effect. Compared to the 1/\Vmax\ results, the maximum likelihood methods determine $\Phi$ values that follow the input Schechter function more closely. SWML biases low at the low mass end. The discrepancy is however not as significant as for the 1/\Vmax\ results. STY correctly reproduces the input Schechter function. It only under-predicts $M^{*}$ for the steep completeness function which indeed produces significantly more low mass galaxies than the standard completeness function. As in Figure  \ref{fig:sim_comp}, the 1/\Vmax\  and SWML values are illustrated with open and filled symbols, respectively. For clarity we are not showing the 1$\sigma$ uncertainties on the STY results which are marked with dotted and dashed lines.}
\end{figure*}

To test our method of determining stellar mass functions, we generate mock catalogues. Assuming a Schechter function and a M/L ratio we create catalogues containing stellar mass, redshift and magnitude values. By running our mass function code on these mock catalogues we can determine if our method correctly reproduces the input Schechter function and investigate potential biases. Our results are shown in Figure \ref{fig:sim_comp}. 

We generate a mock catalogue by:
\begin{itemize}
	\item{constructing the cumulative distribution function (CDF) based on the input Schechter function. The CDF is defined in the following way:
		\begin{equation}
		\mathrm{CDF}(\log M) = \frac{\int_{\log M_{\mathrm{min}, i}}^{\log M} \Phi\ d\log M}{\int_{\log M_{\mathrm{min}, i}}^{\log M_{\mathrm{max}, i}} \Phi\  d\log M}
		\end{equation}
		and we choose $\log M_{\mathrm{min}} = 8$ and $\log M_\mathrm{max} = 12$.}
	\item{drawing $N$ stellar mass values from CDF. $N$ depends on the shape of the input mass function and the simulated volume: 
		\begin{equation}
		N = V \int_{\log M_{\mathrm{min}}}^{\log M_{\mathrm{max}}} \Phi\ d\log M.
		\end{equation}
		$V$ is the comoving volume between $z_\mathrm{min}^\mathrm{s}$ = \zminsample\ and $z_\mathrm{max}^{s}$ = \zmaxsample\ for the surface area of the entire sky (see equation \ref{eq:Vmax})}. 
	\item{assigning each of the $N$ objects a luminosity value using the input M/L relation and drawing from a normal distribution}
	\item{assigning redshifts to each of the $N$ objects, drawing the values from a uniform distribution in the range $z_\mathrm{min}^\mathrm{s} \leq z \leq z_\mathrm{max}^\mathrm{s}$}
	\item{converting the luminosity into magnitude values and eliminating all objects below the magnitude limit \mlim}
\end{itemize}

After having created the mock catalogue,  we run our mass function code on the simulated sample and compare our initial Schechter function input parameters to the output values. 

To determine the luminosity of each object, we are using a constant M/L which is based on real data. In the second panel of Figure \ref{fig:sim_comp} we are for example showing the results of the simulation for which we use the Schechter function of blue galaxies as input. To calibrate the M/L for this simulation run, we measure the mean M/L and its standard deviation for all blue galaxies in our sample. In the simulation we then draw the luminosity of object $i$ from a normal distribution around $\log M_i - \log(M/L)$.  Note that in Figure \ref{fig:sim_comp} we are only showing random and no systematic errors on the SWML $\Phi$ values since we are not determining errors on the stellar mass values in our simulation. 

Going from the top left to bottom right, the simulation results shown in Figure \ref{fig:sim_comp} are based on the mass functions of the entire sample, blue galaxies, blue Early type galaxies, galaxies in halos with $13.5 \leq \log M_h < 15$, satellite galaxies with $\log \mathrm{sSFR} < -12$ and elliptical galaxies in overdense regions. We used the following M/L values as input: $(M/L)_\mathrm{tot} = 2.00\ M_\odot/L_\odot$, $\sigma_\mathrm{tot} = 1.52\ M_\odot/L_\odot$, $(M/L)_\mathrm{blue} = 1.20\ M_\odot/L_\odot$, $\sigma_\mathrm{blue} = 0.71\ M_\odot/L_\odot$, $(M/L)_\mathrm{blue\ ET} = 2.21\ M_\odot/L_\odot$, $\sigma_\mathrm{blue\ ET} = 0.85\ M_\odot/L_\odot$, $(M/L)_\mathrm{hm3} = 3.11\ M_\odot/L_\odot$, $\sigma_\mathrm{hm3} = 0.96\ M_\odot/L_\odot$,  $(M/L)_\mathrm{sat\ sSFR1} = 3.79\ M_\odot/L_\odot$, $\sigma_\mathrm{sat\ sSFR1} = 0.49\ M_\odot/L_\odot$ and $(M/L)_\mathrm{ET\ overd} = 3.51\ M_\odot/L_\odot$, $\sigma_\mathrm{ET\ overd} = 0.66\ M_\odot/L_\odot$.

Within figure \ref{fig:sim_comp} we give the parameters of the Schechter functions that we are drawing from to create the mock catalogues and the STY parameters that we determine by running our code on the mock data. Going from top left to bottom right, we are drawing from a double, a single, a single, a double, a single and a double Schechter function. For the simulation shown in the top left panel, we are for example creating a mock catalogue from a double Schechter function. We treat this mock data set as if it was observed data and use it as input for our mass function code. Based on the likleihood ratio test, the code then correctly returns a double and not a single Schechter function.

Going from top left to bottom right the likelihood values (see equation \ref{eq:likeli_ratio}) are $R = 271.45$ ($p = 0$), $R = -17.31$, $R = -0.66$, $R = 11.39$ ($p = 0.003$), $R = -0.02$, $R = 246.67$ ($p = 0$). As we have discussed in Sec. \ref{sec:single_double}, $R \leq 0$ shows that the data is well fit with a single Schechter function and $R > 0$ indicates that a double Schechter function provides a better fit. Comparing the Schechter functions that we were drawing from and the shape of the mass function determined by our code shows, that our method reliably recovers the input mass function shape.

This is even true for the mass function shown in the bottom left panel. To create the mock catalogue for this mass function, we used the M/L of galaxies in halos with $13.5 \leq \log M_h < 15$ as input. In the data, the single and double Schechter likelihoods for this subsample have similar values ($R = -19.19$). The real subsample does thus not show a prominent double Schechter shape and is then fit with a single Schechter function. For the construction of the mock catalogue representing this subsample we did however choose to draw from the double, and not the single Schechter function fit to the data.  Figure \ref{fig:sim_comp} shows that the mock data is then indeed fit with a double Schechter function. This illustrates that our method is sensitive to even small changes in the input mass function and correctly retrieves the mass function that we were sampling from, even if it is only a weak double Schechter function.

Figure \ref{fig:sim_comp} illustrates that the $\Phi$ values determined with 1/\Vmax\ show a significant discrepancy at the low mass end. In our simulation, the 1/\Vmax\ technique is thus either over-predicting the number of low mass objects or under-estimating the maximum volume $V_\mathrm{max}$. For the blue galaxies for instance, which are shown in the middle panel at the top, this low mass end based on 1/\Vmax\ implies more than $50\%$ more objects between $10^9 M_\odot$ and $M^{12} M_\odot$ compared to the maximum likelihood methods. 
  
The results of the maximum likelihood methods show a less significant discrepancy to  the input Schechter function. SWML biases low at the low mass end for the samples with $\alpha + 1 < 0$. STY slightly under-predicts the low mass end of the mass functions shown in the left column and over-predicts the low mass end of the mass function shown in the bottom right panel. 
 
The over-prediction of the low mass end by the 1/\Vmax\ method could have two possible explanations. 

First, our simple assumption of a constant M/L value that we use as input could distort the shape of the mass completeness function $M_\mathrm{lim}(z)$ which we have introduced in Sec. \ref{sec:mass_completeness}. To calibrate M/L we use the real SDSS subsamples, as we have discussed above. The sources in our SDSS subsamples do not reflect the true, underlying M/L distribution though. Faint objects are not included since they fall below the magnitude limit and we are thus missing galaxies with low M/L (low luminosity compared to their stellar mass). Including these very low M/L sources would increase the mean $\log(M/L)$ value that we use as input for our simulation. A higher mean $\log(M/L)$ value would lead to low mass objects being more likely to be assigned a low luminosity value since we draw those values from a normal distribution around $\log M_i - \log(M/L)$. This would cause more low mass galaxies to fall below the magnitude limit and would result in fewer low mass galaxies, i.e. lower 1/\Vmax\ $\Phi$ values. 

Furthermore, the real, magnitude limited SDSS subsamples contain more galaxies at high than at low stellar masses. The mean M/L will thus be dominated by the many bright objects at high stellar masses and we might not be reproducing the M/L of faint, low mass galaxies correctly. This affects the shape of the stellar mass completeness function at low redshifts and therefore also the number of galaxies in the lowest stellar mass bin. This effect can be seen in Figure \ref{fig:sim_comp}. For four of our six testsamples galaxies below $10^{9.3}\ M_\odot$ are not part of the mass function determination since they lie below the mass completeness function. Except for galaxies with $\log \mathrm{sSFR} < -12$, we do not see the same trend in the real subsamples which we use to calibrate our simulations.  This is an additional indication towards our assumption of a constant M/L value not reproducing the true, underlying stellar mass - redshift distribution. For completeness we are showing the mass-redshift distributions of the mock catalogues and of the real data sets which our simulations are based on in Figure \ref{fig:mass_z_dis_mock} in the appendix.

The question remains why the maximum likelihood methods are retrieving the input mass function more reliably if our mock catalogues indeed represent a distorted mass - redshift distribution. The second reason for the discrepancy which we see in Figure \ref{fig:sim_comp} could thus be a bias in the 1/\Vmax\ method itself. The tendency of 1/\Vmax\ to over-predict the number of low mass sources has previously been discussed by \cite{Efstathiou:1988aa} and \cite{Willmer:1997aa}. 

For luminosity functions the various estimators have been compared in great detail (see \cite{Johnston:2011aa} for an overview). To construct luminosity functions we need to know the completeness in terms of magnitude, and not stellar mass. For a flux limited sample, this is easier to determine since M/L do not have to be considered. Nonetheless, STY and 1/\Vmax\ have been found to be affected by biases when being used to determine luminosity functions. 

\cite{Willmer:1997aa} use Monte Carlo simulations to test estimators by \cite{Choloniewski:1986aa} (SWML), \cite{Choloniewski:1987aa} ($C^-$), \cite{Turner:1979aa} ($\phi/\Phi$) and also STY \citep{Sandage:1979aa}, SWML \citep{Efstathiou:1988aa} and 1/\Vmax \citep{Schmidt:1968aa}, the methods that we use here. They conclude that STY and $C^-$ are the best estimators to construct luminosity functions. STY does however tend to slightly underestimate the faint end which has previously been concluded by \cite{Efstathiou:1988aa}. They show that 1/\Vmax\ tends to overestimate the faint end slope even for spatially homogeneous samples. 

\cite{Takeuchi:2000aa} test four non-parametric estimators, including SWML and 1/\Vmax, using simulated luminosity functions, a mock catalogue for the Two Degree Field redshift survey \citep{Cole:1998aa} and photometric Hubble Deep Field data by \cite{Fernandez-Soto:1999aa}. For spatially homogeneous data sets they find agreement among all tested estimators. 1/\Vmax\ only produces different results if the objects are distributed inhomogeneously. \cite{Takeuchi:2000aa} thus do not find the bias that was previously found by \cite{Willmer:1997aa}. Yet, it is important to note, that \cite{Takeuchi:2000aa} use a modified version of the classical 1/\Vmax\ estimator which was developed by \cite{Eales:1993aa}. This approach also takes evolution with redshift into account.

\cite{Saunders:1990aa} (also see \citealt{Baldry:2012aa}) show that the 1/\Vmax\ method and the SWML approach are equivalent if corrections for large scale structure density variations are included in the 1/\Vmax\ estimator. \cite{Cole:2011aa} derives the same density corrected 1/\Vmax\ equation as \cite{Saunders:1990aa} by deriving a joint luminosity function and galaxy overdensity estimator. He points out that the maximum likelihood approach and the classical 1/\Vmax\ technique are equivalent if one assumes that there are no radial galaxy density fluctuations.

\cite{Ilbert:2004aa} compare 1/\Vmax, $C^+$ \citep{Zucca:1997aa}, STY and SWML. Their analysis focuses on the effects that arise if the wavelengths of selection and reference filter, for which the luminosity function is determined, are significantly different. At high redshift this bias needs to be considered if the sample consists of galaxies with different spectral energy distribution shapes. As we are working at low $z$, we do not need to take this issue into account.

In our simulation the discrepancy of the 1/\Vmax\ points could thus be explained by a combination of two effects. The steep low mass slope recovered by 1/\Vmax\ could be caused by an intrinsic bias of the 1/\Vmax\ technique in combination with us not correctly reproducing the underlying M/L distribution. When constructing the mock catalogues we are drawing the redshift values from a uniform distribution and are thus not considering density fluctuations. According to \cite{Willmer:1997aa} these results could still be affected by a bias of the 1/\Vmax\ method itself.  Based on the results of \cite{Saunders:1990aa} and \cite{Cole:2011aa}, the SWML and the 1/\Vmax\ technique should however lead to equivalent results. The input M/L determines the shape of the mass completeness function $M_\mathrm{lim}(z)$ and out of our three techniques, 1/\Vmax\ is the one most strongly dependent on both shape and normalisation of $M_\mathrm{lim}(z)$.  We illustrate this effect in Figure \ref{fig:sim_comp_tests} and will also discuss it in more detail in Sec. \ref{sec:compl}. 

Figure \ref{fig:sim_comp_tests}  shows the impact that the stellar mass completeness has on the shape of our resulting mass function. The top panels of this figure show how a change in the normalisation of $M_\mathrm{lim}(z)$ affects the low mass end. The bottom panels show the results for $M_\mathrm{lim}(z)$ functions with different shapes. In analogy to Figure \ref{fig:sim_comp}, we create this figure by generating a mock catalogue using our simulation. In the right panels, the input Schechter function is represented by the red solid line. 

This figure illustrates that the SWML points are more robust against changes in the normalisation and changes of the shape of $M_\mathrm{lim}(z)$. 
The STY technique correctly retrieves the input Schechter function in most cases. It only fails for the steep $M_\mathrm{lim}(z)$ function which is shown with a blue dashed line in the bottom panels of Figure \ref{fig:sim_comp_tests}. For this completeness function STY predicts a lower $M^{*}$ value than was put in since such a completeness cut causes the sample to contain significantly more low mass galaxies.

The values determined through 1/\Vmax\ are dependent on both, normalisation and shape, of the mass completeness function. A change in the normalisation of $M_\mathrm{lim}(z)$ has the expected effect. For a given stellar mass, a galaxy is assigned a lower $z_\mathrm{max}$ value if the normalisation is higher. A smaller $z_\mathrm{max}$ implies a smaller volume $V_\mathrm{max}$ and thus a higher $\Phi$ value. The same result can be seen in the bottom panels. The different shapes of $M_\mathrm{lim}(z)$ simply determine which bins will be affected most by this effect.

In conclusion, 1/\Vmax\ is the method that is most strongly affected by the shape of the stellar mass completeness function. The technique itself could also be biased towards over predicting the low mass end slope. The shape and normalisation of $M_\mathrm{lim}(z)$ have less impact on the maximum likelihood methods. Compared to 1/\Vmax, STY and SWML recover the input mass function more reliably.  
 
\subsection{Stellar mass completeness}
\label{sec:compl}
\begin{figure*}
	\includegraphics[scale=.35]{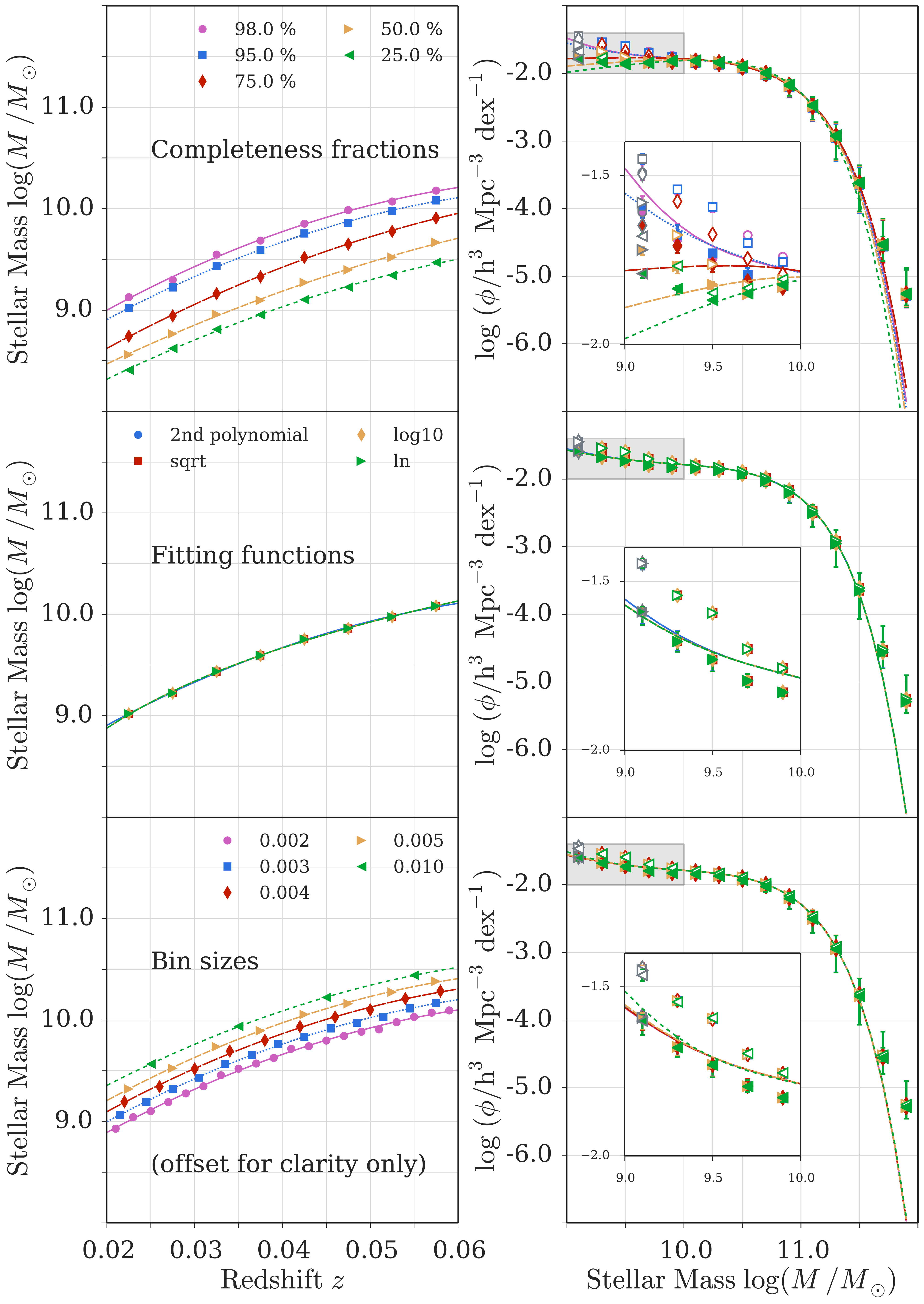}
	\caption{\label{fig:compl_frac} Shape and normalisation of the stellar mass completeness function. Following the approach by \protect\cite{Pozzetti:2010aa}, the shape and the normalisation of our stellar mass completeness functions are determined by three parameters: the level of completeness, the fitting function and the redshift bin size. Here, we are showing how these three parameters affect the low mass end of the stellar mass function. From top to bottom we are varying the completeness fraction, the fitting function and the redshift bin size. We are showing the stellar mass completeness functions and the corresponding Schechter functions on the left and on the right, respectively. The insets in the right panels show zoom-ins of the grey shaded regions. This figure clearly illustrates that neither our choice of fitting function, nor our choice of bin size change the shape of the resulting mass function. What does affect the low mass end of $\Phi(M)$ is the choice of completeness fraction shown in the top panels. For our mass functions we are using a completeness fraction of \masscompl. This ensures that at the magnitude limit, galaxies with high M/L, i.e. objects that are bright enough to lie above the flux limit but not massive enough to lie above the mass completeness limit, are not included in our sample. The 1/\Vmax\  and SWML values are marked with open and filled symbols, respectively. For clarity we are not showing the 1$\sigma$ uncertainties on the STY results which are illustrated with solid, dotted and dashed lines.}
\end{figure*}

In the previous section we have seen that the 1/\Vmax\ method strongly depends on the shape and the normalisation of the stellar mass completeness function $M_\mathrm{lim}(z)$. STY and SWML are more robust towards changes of $M_\mathrm{lim}(z)$. Here, we are discussing how the different methods depend on the stellar mass completeness and what makes STY and SWML more reliable.  For the maximum likelihood methods we determine $M_{\mathrm{min}, i} = M_\mathrm{lim}(z_i)$, the limiting mass of object $i$ with stellar mass $i$. The 1/\Vmax\ technique requires $z_{\mathrm{max}, i}$ as input and we thus need to invert $M_\mathrm{lim}(z)$ as is shown in Figure \ref{fig:Mlim}. It is also important to note that the limiting mass values which we determine for the maximum likelihood methods are purely redshift dependent. A massive $10^{12} M_\odot$ galaxy will have the same $M_{\mathrm{min}, i}$ value as a $10^{9} M_\odot$ galaxy, if they are at the same redshift.

In STY the probability of an object being part of the sample, given certain Schechter function parameters, depends on the mass range in which it could be found (see equation \ref{eq:prob_STY}). This is similar to the volume weighing used for 1/\Vmax. A change in the normalisation of $M_\mathrm{lim}(z)$ does increase or decrease the mass range over which we integrate by the same amount for all sources in the sample. This is significant for low mass galaxies which can only be found at low redshifts and thus have a low $M_{\mathrm{min}, i}$ value. For these sources the mass range that we are considering in equation \ref{eq:prob_STY} is small. A minor change in the normalisation of $M_\mathrm{lim}(z)$ can thus significantly change the probability $p_i$ of finding these galaxies in our sample. 

Yet, STY has the big advantage that it is a parametric approach. $\Phi(M)$ will thus always follow a Schechter function. On the one hand, this makes STY robust against uncertainties at low masses that might be due to the stellar mass completeness. On the other hand, this implies that STY is not sensitive to deviations from the traditional Schechter function.

For SWML we have a closer look at equation \ref{eq:SWML_iter} and the step function $H(x)$ defined in equation \ref{eq:window_H}. $H_{1,k}$ will always return 1 since we have chosen $M_{\mathrm{max}, i}  = $ \Mstarintmax $M_\odot$ and we are only considering sources up to $10^{12} M_\odot$. $H_{2,k}$ returns a value $>0$ for sources for whose limiting mass is smaller than the bin mass $M_k$ or lies within bin $k$. We can therefore think of equation \ref{eq:SWML_iter} in the following way:
\begin{equation}\label{eq:SWML_simple}
\begin{aligned}
& \text{Nr.\ density\ in\ bin}\ k \propto  \\
& \text{Nr. of objects in bin}\ k\ \\
& \cdot \Big(\sum \frac{\text{all objects with } M_\mathrm{min} \leq \text{bin mass }}{\text{Nr. density of all bins with mass}\ \leq M_\mathrm{min}}\Big)^{-1}
\end{aligned}
\end{equation}

We need to correct for the fact that due to the stellar mass completeness cut, mass bins at higher masses tend to include more objects than at lower masses. The completeness cut also implies that higher mass bins include objects over a wider redshift range. Furthermore, we know that the higher the redshift of a source, the higher its  $M_{\mathrm{min}, i}$ value. For a bin at, for example, $10^{10} M_\odot$ there are thus more objects with a limiting mass smaller than $10^{10} M_\odot$ compared to a bin at $10^{9.5} M_\odot$. These additional sources are however at higher $z$, therefore have higher  $M_{\mathrm{min}, i}$ values and are thus weighted by a larger number density. The second term in eq. \ref{eq:SWML_simple} is thus slightly larger for the $10^{10} M_\odot$ bin than for the $10^{9.5} M_\odot$ bin which compensates for the fact that the higher mass bin tends to contain more objects than the lower mass bin.

The SWML results do therefore depend on the overall shape and normalisation of $M_\mathrm{lim}(z)$. Compared to 1/\Vmax, SWML does however have the big advantage that we are determining $\Phi$ iteratively. Similar to STY, the $\Phi$ values in each bin are hence dependent on each other which makes SWML more robust against changes in $M_\mathrm{lim}(z)$. In addition, SWML has the benefit that it is a non-parametric approach. Instead of imposing the resulting mass function to follow a Schechter function as we do for STY, we are binning in stellar mass. The SWML results do thus however dependent on the stellar mass bin size.

In the 1/\Vmax\ method the only sources affected by the stellar mass completeness are the ones with $M_i < M_\mathrm{lim}(z_i)$ since for higher masses $z_{\mathrm{max}, i} = z_{\mathrm{max}}^\mathrm{s}$ = \zmaxsample. The shape and normalisation of $M_\mathrm{lim}(z)$ do hence change the low mass end of the mass function. In contrast to STY and SWML, the 1/\Vmax\ $\Phi$ bin values are determined independently from each other and the $M_\mathrm{lim}(z)$ - $\Phi$ relation is straight forward. Increasing the normalisation of $M_\mathrm{lim}(z)$ will cause low mass objects of a given mass to be assigned a smaller $z_\mathrm{max}$ value which results in a higher number density $\Phi$. This effect can clearly be seen in the top panels of Figure \ref{fig:sim_comp_tests}. In the bottom panels of this figure we show that a change in the shape of the stellar mass completeness function will affect some bins more strongly than others. In our case only $\Phi$ of the lowest mass bins is altered. 

In our analysis, the normalisation of the completeness function is determined by the fraction of faint objects (here: \faintfrac) that we consider to estimate $M_\mathrm{lim}(z)$ and the completeness level (here: \masscompl) that we chose. The shape of $M_\mathrm{lim}(z)$ depends on the redshift bin size (here: \zbinsize) and the function used to fit the relation (here: second order polynomial).

The top panels of Figure \ref{fig:compl_frac} show the effects of the completeness level on the stellar mass function. Note, that in comparison to Figure \ref{fig:sim_comp_tests} we are changing the normalisation of $M_\mathrm{lim}(z)$ by a larger factor and thus the SWML $\Phi$ values also vary. We want the completeness level to be as low as possible to maximise the number of objects in our sample. Furthermore, we want to ensure that at the magnitude limit galaxies with high M/L (high luminosity compared to their stellar mass) are excluded and galaxies with low M/L (low luminosity compared to their stellar mass) are included. As we have seen in equation \ref{eq:ML_const}, $M_{\mathrm{lim}, i} \propto F_{\mathrm{lim}, i} \frac{M_i}{L_i}$ and we thus also want to select a completeness level that is as high as possible. For our analysis we chose to use a completeness level of \masscompl. While a completeness level of \masscompl\ causes a steeper low mass end slope than lower completeness levels, it limits the number of galaxies with high M/L at the magnitude limit. Compared to lower completeness levels, it also  produces better agreement amongst our three mass function estimators. We did not chose a higher completeness level, for example 98\%, since this does not increase the agreement amongst the three estimators significantly. It does however lead to an even steeper low mass end slope and even fewer galaxies being included in the sample in general.

The panels in the middle of Figure \ref{fig:compl_frac} illustrate that our results are not affected by our choice of redshift bin size. For the chosen redshift bin size (\zbinsize), our results are also independent of our choice of completeness fitting function as is demonstrated in the bottom panels of Figure \ref{fig:compl_frac}. 

Here, we have discussed how our three mass function estimators are affected by the stellar mass completeness function. 1/\Vmax\ is directly dependent on the stellar mass completeness function and a change in the shape or normalisation of $M_\mathrm{lim}(z)$  can significantly alter the low mass end of the mass function. STY and SWML have the advantage that the $\Phi$ values at higher masses also affect the low mass end. The $\Phi$ values in each stellar mass bin are thus not independent from each other. STY and SWML are hence more robust and reliable towards changes in $M_\mathrm{lim}(z)$. Figure \ref{fig:compl_frac} justifies our choice of parameters for the $M_\mathrm{lim}(z)$ determination.

\subsection{Comparison to existing work}
\label{sec:previous_work}
\begin{figure*}
	\includegraphics[width=\textwidth]{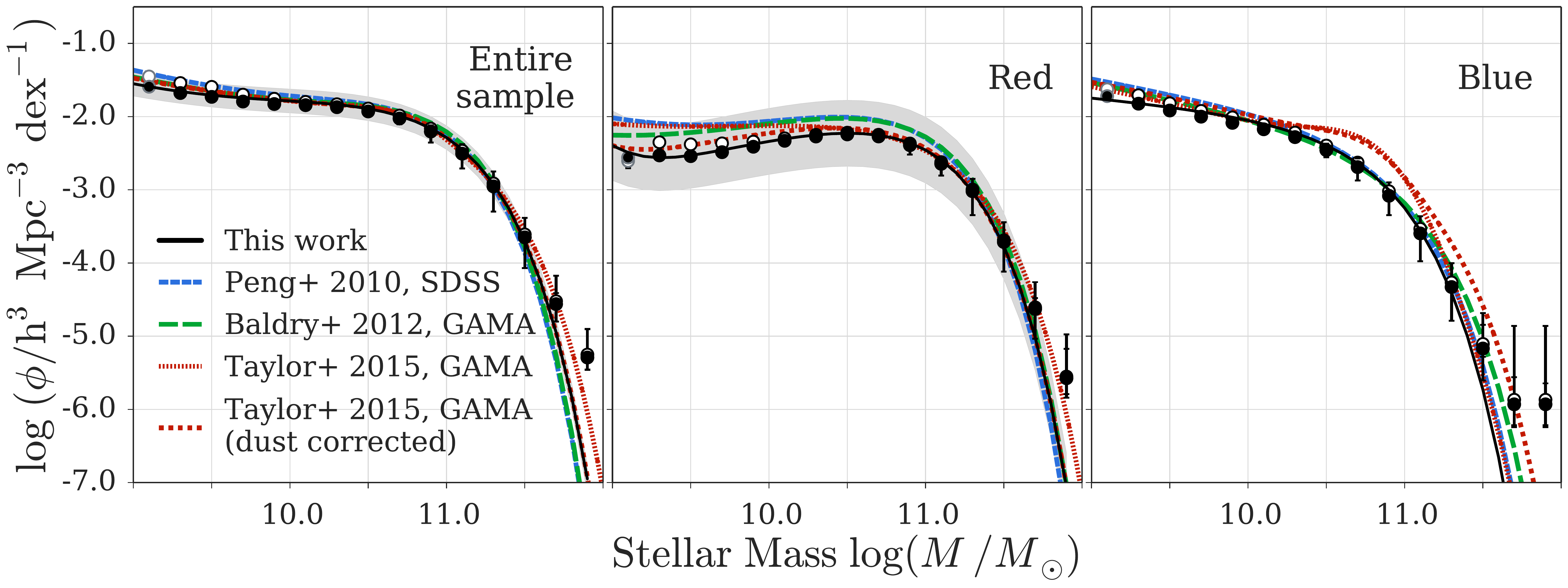}
	\caption{\label{fig:comparison} Comparison to published mass functions. Shown from left to right is a comparison between our work, \protect \cite{Peng:2010aa}, \protect \cite{Baldry:2012aa} and \protect\cite{Taylor:2015aa} for the entire sample, red, and blue galaxies. We show our stellar mass function estimates based on 1/\Vmax\ with open symbols, the SWML $\Phi$ values with filled markers and our best fit STY Schechter functions with solid lines. The shape of the stellar mass function for red and blue galaxies does depend on the definition of `red' and `blue'. \protect \cite{Peng:2010aa} and \protect \cite{Baldry:2012aa} use cuts in the $U - B$ and $u - r$ colour magnitude diagrams to split their sample into red and blue galaxies. We introduce a third colour by also defining a green valley transition zone (see Sec. \ref{sec:colour_def}). \protect\cite{Taylor:2015aa} model the distribution of galaxies in the colour-mass diagram as two independent, but overlapping populations. They argue that at $\log M < 9.3$ red and blue galaxies become indistinguishable and that the upturn we see at the low mass end of the red mass function is introduced artifically by using a simple cut in the colour mass diagram. \protect \cite{Peng:2010aa} and \protect \cite{Baldry:2012aa} fit the mass functions of red and blue galaxies with a double and single Schechter function, respectively. For the red galaxy sample, \protect \cite{Peng:2010aa} fix $\alpha$ of the low mass end slope to the low mass end slope of the entire sample and \protect \cite{Baldry:2012aa} assume $M^{*}_\mathrm{red} = M^{*}_\mathrm{blue}$, $\alpha_2 = \alpha_\mathrm{blue}$ and $\alpha_1 = \alpha_2 + 1$. We do not put any constraints on fitting parameters and determine $\log (M^{*}_\mathrm{red}/M_\odot) = 10.77 \pm 0.01$, $\alpha_{1, \mathrm{red}} = -2.46 \pm 0.33$, $\alpha_{2, \mathrm{red}} = -0.45 \pm 0.02$, $\log (M^{*}_\mathrm{blue}/M_\odot) = 10.60 \pm 0.01$ and $\alpha_\mathrm{blue} = -1.21 \pm 0.01$. Furthermore, we do not make the a priori assumption that the red and blue population should be fit with a double and a single Schechter function, respectively. Instead we use the results of the likelihood ratio test (see Sec. \ref{sec:single_double}) to determine if a single or a double Schechter function provides a better fit to the data. \protect\cite{Taylor:2015aa} describe both, the red and the blue mass function, as the sum of two Schechter functions. They derive their mass functions in the rest-frame $(g - i)$ and the intrinsic, dust corrected $(g^{*} - i^{*})$ colour-mass diagrams. In this figure we show both of their results.}
\end{figure*}

We show a comparison between our results and those by \cite{Peng:2010aa}, \cite{Baldry:2012aa} and \cite{Taylor:2015aa} in Figure \ref{fig:comparison}. For the \cite{Baldry:2012aa} mass functions we extract the best fit values from their Figures 13 and 15 and for the \cite{Peng:2010aa} mass functions we take the values from their Table 3. For the \cite{Taylor:2015aa} results we use their fits to the `R-type' and `B-type' mass functions which are shown in their figures 12 and 13. \cite{Peng:2010aa}, \cite{Baldry:2012aa} and \cite{Taylor:2015aa} use a \cite{Chabrier:2003aa} IMF, we thus convert their $M^{*}$ values to a \cite{Kroupa:2001aa} IMF \footnote{$\log(M_{\mathrm{Kroupa}}) = \log(M_{\mathrm{Chabrier}}) + 0.04$, $\log(M^{*}_{\mathrm{Kroupa}}) = \log(M^{*}_{\mathrm{Chabrier}}) + 0.04$}. 

To determine the stellar mass function shape, \cite{Peng:2010aa} apply the standard 1/\Vmax\ method,  using \Vmax\ values from \textsc{kcorrect} by \cite{Blanton:2007aa}. Note that these \Vmax\ values are only magnitude and not stellar mass dependent. \cite{Peng:2010aa} hence do not account for the range of M/L ratios (see Sec. \ref{sec:mass_completeness}). 

\cite{Baldry:2012aa} first determine the galaxy luminosity function shape using a density corrected 1/\Vmax\ approach \citep{Baldry:2006aa, Baldry:2008aa}, which, as we have seen above, should be equivalent to the SWML method  \citep{Saunders:1990aa, Cole:2011aa}. Their \Vmax\ values originate from the NYU-VAGC catalog \citep{Blanton:2005aa}  and are also magnitude and not stellar mass dependent. To convert from luminosity to stellar mass function, \cite{Baldry:2012aa} use a colour-based M/L.

\cite{Taylor:2015aa} model the distribution of galaxies in colour-mass space as two independent, but overlapping populations. The Schechter function parameters for the red and the blue stellar mass functions, are part of their 40 parameter model. They correct for incompleteness using magnitude based 1/\Vmax\ values and also discuss the impact of using a density corrected 1/\Vmax\ method. 

In the left panel of Figure \ref{fig:comparison} we compare our mass function of the entire sample to previously published mass functions. At $M = 10^{9} M_\odot$ our STY results lie about $0.2$, $ 0.1$ and $0.08$ dex below the mass functions of \cite{Peng:2010aa}, \cite{Baldry:2012aa} and \cite{Taylor:2015aa}, respectively. At $M = 10^{11.75} M_\odot$ the rest-frame mass function by \cite{Taylor:2015aa} lies about $0.46$ dex above ours. The \cite{Peng:2010aa} and \cite{Baldry:2012aa} mass functions have a $\Phi$ value that is about $0.4$ dex lower than ours. At our $M^{*} = 10^{10.79} M_\odot$ we find good agreement with offsets of about $+0.06$, $+0.07$ and less than $10^{-3}$ dex relative to our mass function for \cite{Peng:2010aa},  \cite{Baldry:2012aa} and \cite{Taylor:2015aa}, respectively. The mass functions considered here, thus show the biggest discrepancies at the high mass end which is presumably due to there generally being fewer high mass objects than for example at $M \sim M^{*}$. We can can also see  this affecting our own mass functions. For the entire sample for example, our 1/\Vmax\ and SWML $\Phi$ values in the mass bins at $10^{11.7} M_\odot$ and $10^{11.9} M_\odot$ show a significant offset relative to the STY solution. These bins contain 33 and 4 objects, respectively which is less than $0.4 \%$ of the number of objects at $M^{*}$. By containing fewer objects the $\Phi$ values at these high masses are hence easily affected by galaxies with uncertain stellar mass measurements that are being scattered into these bins. 

In the middle and right panels of Figure \ref{fig:comparison} we compare our red and blue  mass functions to the ones determined by \cite{Peng:2010aa}, \cite{Baldry:2012aa} and \cite{Taylor:2015aa}. The shape of the red and the shape of the blue mass functions depend on the colour definition. 
	
\cite{Peng:2010aa} separate their sample into red and blue galaxies based on a linear cut in the $U - B $ colour mass diagram. When fitting the blue population with a single and the red population with a double Schechter function, they fix the low mass end slope of the red sample to the low mass end $\alpha$ that was measured for the entire sample. 

\cite{Baldry:2012aa} use a linear cut in the $u - r$ colour magnitude diagram to define their red and blue populations. They then fit a single Schechter function to the blue and a double Schechter function to the red sample. Note, that when fitting they are enforcing $M^{*}_\mathrm{red} = M^{*}_\mathrm{blue}$, $\alpha_2 = \alpha_\mathrm{blue}$ and $\alpha_1 = \alpha_2 + 1$. 

Both, our red and our blue stellar mass function lie below the corresponding ones by \cite{Peng:2010aa} and  \cite{Baldry:2012aa}. This lower number density of red and blue galaxies might be due to us not only splitting the sample into red and blue galaxies, but also considering green galaxies separately (see Sec. \ref{sec:colour_def}).

When comparing our mass functions to the results by \cite{Baldry:2012aa}, we also need to be aware of the fact that \cite{Baldry:2012aa} are using data from the first three years of the Galaxy And Mass Assembly survey (GAMA, \citealt{Baldry:2010aa, Robotham:2010aa, Driver:2011aa}) survey. With $r$-band magnitude limits of 19.4 over two thirds and 19.8 over one third of the survey area, GAMA is significantly deeper than SDSS (\mlim) which adds additional completeness at the low mass end.

\cite{Taylor:2015aa} compare the definitions of `red' and `blue' galaxies by \cite{Bell:2003aa}, \cite{Baldry:2004aa} and \cite{Peng:2010aa} and illustrate how these different cuts in colour-mass and colour-magnitude diagrams can significantly change the shape of the resulting mass function. Different definitions could thus also be the reason why the mass functions in the middle and right panels of Figure \ref{fig:comparison} do not have the same shapes. Furthermore, \cite{Taylor:2015aa}  argue that at $M < 10^{9.3} M_\odot$ red and blue galaxies are indistinguishable and separating them into two categories might not be appropriate. Using their definition of `red' and `blue', the upturn at the low mass end of the red mass function, which we also see in our red mass function and the one by \cite{Peng:2010aa}, is not reproduced.  \cite{Taylor:2015aa} state that this upturn, which often motivates the fit of the red mass function with a double Schechter function, could be introduced artificially by using a simple cut in the colour-mass digaram to separate red from blue galaxies. 

We do not follow the approach by \cite{Taylor:2015aa}, but rely on a classical cut in the colour-mass diagram to define `red', `green' and `blue' galaxies (see Sec. \ref{sec:colour_def}). Yet, it is important to note, that in contrast to \cite{Peng:2010aa} and \cite{Baldry:2012aa} we do not make the a priori assumption that the red and the blue mass functions are well desribed by a double and a single Schechter function, respectively. Instead we determine the likelihoods of a single and a double Schechter solution for the red and the likelihoods of a single and a double Schechter solution to the blue mass function. We use the likelihood ratio test (see Sec. \ref{sec:single_double}) to determine which models provide better fits. When determining the single and double Schechter likelihoods with the STY approach, we put constraints on the allowed parameter range (see Table \ref{tab:MCMC_boundaries}), but we neither constrain $M^{*}_\mathrm{red}$ to be the same as $M^{*}_\mathrm{blue}$ nor enforce $\alpha_1 = \alpha_2 + 1$. Compared to \cite{Peng:2010aa} and \cite{Baldry:2012aa} we thus make fewer assumptions, but reach similar results.

Further stellar mass functions for starforming and quiescent galaxies have for example been published for the GAMA survey (\citealt{Baldry:2010aa, Robotham:2010aa, Driver:2011aa}) by \cite{Kelvin:2014aa}, \cite{Gunawardhana:2015aa} and \cite{Moffett:2015aa}. \cite{Bell:2003aa} combine SDSS Early Data Release Data \citep{Stoughton:2002aa} with the 2MASS extended source catalog \citep{Jarrett:2000aa}. \cite{Thanjavur:2016aa} use SDSS DR7 data and \cite{Moustakas:2013aa} base their analysis on a combination of SDSS DR7 and PRism MUlti-object Survey (PRIMUS, \citealt{Coil:2011aa, Cool:2013aa}) data. 

In comparison to the $\Phi^{*}$ values that we determine for our Early and Late type sample, some of these works report significantly higher number densities. We find that the Early and Late type mass functions are well described by double Schechter functions and determine $\log \Phi^{*}_{2}$ values of $-2.59$ and $-2.47\ \mathrm{h}^3\ \mathrm{Mpc}^{-3}$, respectively. For example, \cite{Kelvin:2014aa}, \cite{Moffett:2015aa} and \cite{Thanjavur:2016aa} split their sample into spheroid- and disk-dominated galaxies and find the following $\Phi^{*}$ values: \cite{Kelvin:2014aa}: $\log \Phi^{*}_\mathrm{sph} = -1.93$ and $\log \Phi^{*}_\mathrm{disk} = -2.54\  \mathrm{h}^3\ \mathrm{Mpc}^{-3}$, \cite{Moffett:2015aa}: $\log \Phi^{*}_\mathrm{sph} = -1.97$ and $\log \Phi^{*}_\mathrm{disk} = -2.60\  \mathrm{h}^3\ \mathrm{Mpc}^{-3}$, \cite{Thanjavur:2016aa}: $\log \Phi^{*}_\mathrm{sph} = -2.38$ and $\log \Phi^{*}_\mathrm{disk} = -2.34\  \mathrm{h}^3\ \mathrm{Mpc}^{-3}$. These discrepancies of up to $0.6$ dex in comparison to our results might be caused by the fact that more than $50\%$ of our sample is assigned to the indeterminate category (see Table \ref{tab:sample}). For a galaxy to be, for example, classified as an elliptical galaxy, we require more than $80\%$ of its Galaxy Zoo votes to be in this category. Our Early type and Late type galaxy samples are thus incomplete, but pure which explains the large offsets that we see in the $\Phi^{*}$ values in comparison to previous work.

At higher redshifts, \cite{Drory:2009aa} use COSMOS \citep{Scoville:2007aa} data and  \cite{Pozzetti:2010aa} use the zCOSMOS - 10k-bright spectroscopic sample \citep{Lilly:2007aa,Lilly:2009aa}. \cite{Muzzin:2013aa} and \cite{Ilbert:2013aa} use $K_S$-band selected galaxies from the COSMOS/UltraVISTA \citep{McCracken:2012aa} field to study stellar mass functions at $z < 4$. 

Recent publications regarding luminosity functions split by colour or morphology include work by \cite{Marchesini:2007aa}, \cite{Brown:2007aa}, \cite{Faber:2007aa}, \cite{Montero-Dorta:2009aa}, \cite{Loveday:2012aa}, \cite{Tomczak:2014aa} and \cite{Bonne:2015aa}. 

\section{Stellar Mass Functions}
\label{sec:mass_functions}
\begin{table*}
	\begin{center}
		\caption{\label{tab:Fig_overview}Overview over all stellar mass function figures. To split our main sample into subsamples we use morphological classifications from Galaxy Zoo \citep{Lintott:2008aa,Lintott:2011aa}, sSFR measurements from MPA JHU \citep{Brinchmann:2004aa,Kauffmann:2003ab}, halo masses and the central and satellite classification from \protect \cite{Yang:2007aa}. We define `red', `green' and `blue' in the $u - r$ colour mass digaram (see Sec. \ref{sec:colour_def}) and measure the environmental overdensity using a 5th nearest neighbour approach (see Sec. \ref{sec:overdensity}).}
		\begin{tabular}{llllllll}
			\hline
			{\cellcolor{grey!25}entire sample} & {\cellcolor{grey!25}} & {\cellcolor{blue!25}morphology} & {\cellcolor{purple!25}sSFR} & {\cellcolor{pink!25}colour} & {\cellcolor{red!25}halo mass} & {\cellcolor{orange!25}centrals \& satellites} & {\cellcolor{green!25}overdensity} \\
			
			{\cellcolor{grey!25}Fig. \ref{fig:mass_fct_entire_sample}} & {\cellcolor{grey!25}} & {\cellcolor{blue!25}Fig. \ref{fig:mass_fct_morphology}} & {\cellcolor{purple!25}Fig. \ref{fig:mass_fct_sSFR}} & {\cellcolor{pink!25}Fig.\ref{fig:mass_fct_color}} & {\cellcolor{red!25}Fig. \ref{fig:mass_fct_hm}} & {\cellcolor{orange!25}Fig. \ref{fig:mass_fct_cent_sat}} & {\cellcolor{green!25}Fig. \ref{fig:mass_fct_dens}} \\
			
			\hline
			{\cellcolor{blue!25}morphology (Galaxy Zoo)} & {\cellcolor{blue!25} Fig. \ref{fig:mass_fct_morphology}} & {\cellcolor{grey!25}} & {Fig. \ref{fig:mass_fct_morph_sSFR}} & {Fig. \ref{fig:mass_fct_morph_color}} & {Fig. \ref{fig:mass_fct_morph_hm}} & {Fig. \ref{fig:mass_fct_morph_cent_sat}} & {Fig. \ref{fig:mass_fct_morph_dens}}\\
			
			{\cellcolor{purple!25}sSFR (MPA JHU)} & {\cellcolor{purple!25} Fig. \ref{fig:mass_fct_sSFR}} & {Fig. \ref{fig:mass_fct_morph_sSFR}} & {\cellcolor{grey!25}} & {Fig. \ref{fig:mass_fct_sSFR_color}} & {Fig. \ref{fig:mass_fct_sSFR_hm}} & {Fig. \ref{fig:mass_fct_sSFR_cent_sat}} & {Fig. \ref{fig:mass_fct_sSFR_dens}}\\
			
			{\cellcolor{pink!25}colour (see Sec. \ref{sec:colour_def})} & {\cellcolor{pink!25} Fig. \ref{fig:mass_fct_color}} & {Fig. \ref{fig:mass_fct_morph_color}} & {Fig. \ref{fig:mass_fct_sSFR_color}} & {\cellcolor{grey!25}} & {Fig. \ref{fig:mass_fct_color_hm}} & {Fig. \ref{fig:mass_fct_color_cent_sat}} & {Fig. \ref{fig:mass_fct_color_dens}}\\
			
			{\cellcolor{red!25}halo mass (\cite{Yang:2007aa})} & {\cellcolor{red!25} Fig. \ref{fig:mass_fct_hm}} & {Fig. \ref{fig:mass_fct_morph_hm}} & {Fig. \ref{fig:mass_fct_sSFR_hm}} & {Fig. \ref{fig:mass_fct_color_hm}} & {\cellcolor{grey!25}} & {Fig. \ref{fig:mass_fct_hm_cent_sat}} & {Fig. \ref{fig:mass_fct_hm_dens}}\\
			
			{\cellcolor{orange!25}centrals \& satellites (\cite{Yang:2007aa})} & {\cellcolor{orange!25} Fig. \ref{fig:mass_fct_cent_sat}} & {Fig. \ref{fig:mass_fct_morph_cent_sat}} & {Fig. \ref{fig:mass_fct_sSFR_cent_sat}} & {Fig. \ref{fig:mass_fct_color_cent_sat}} & {Fig. \ref{fig:mass_fct_hm_cent_sat}} & {\cellcolor{grey!25}} & {Fig. \ref{fig:mass_fct_cent_sat_dens}}\\
			
			{\cellcolor{green!25}overdensity (see Sec. \ref{sec:overdensity})} & {\cellcolor{green!25} Fig. \ref{fig:mass_fct_dens}} & {Fig. \ref{fig:mass_fct_morph_dens}} & {Fig. \ref{fig:mass_fct_sSFR_dens}} & {Fig. \ref{fig:mass_fct_color_dens}} & {Fig. \ref{fig:mass_fct_hm_dens}} & {Fig. \ref{fig:mass_fct_cent_sat_dens}} & {\cellcolor{grey!25}}\\
			
			\hline
		\end{tabular}
	\end{center}
\end{table*}

\begin{figure*}
\includegraphics[width=\textwidth]{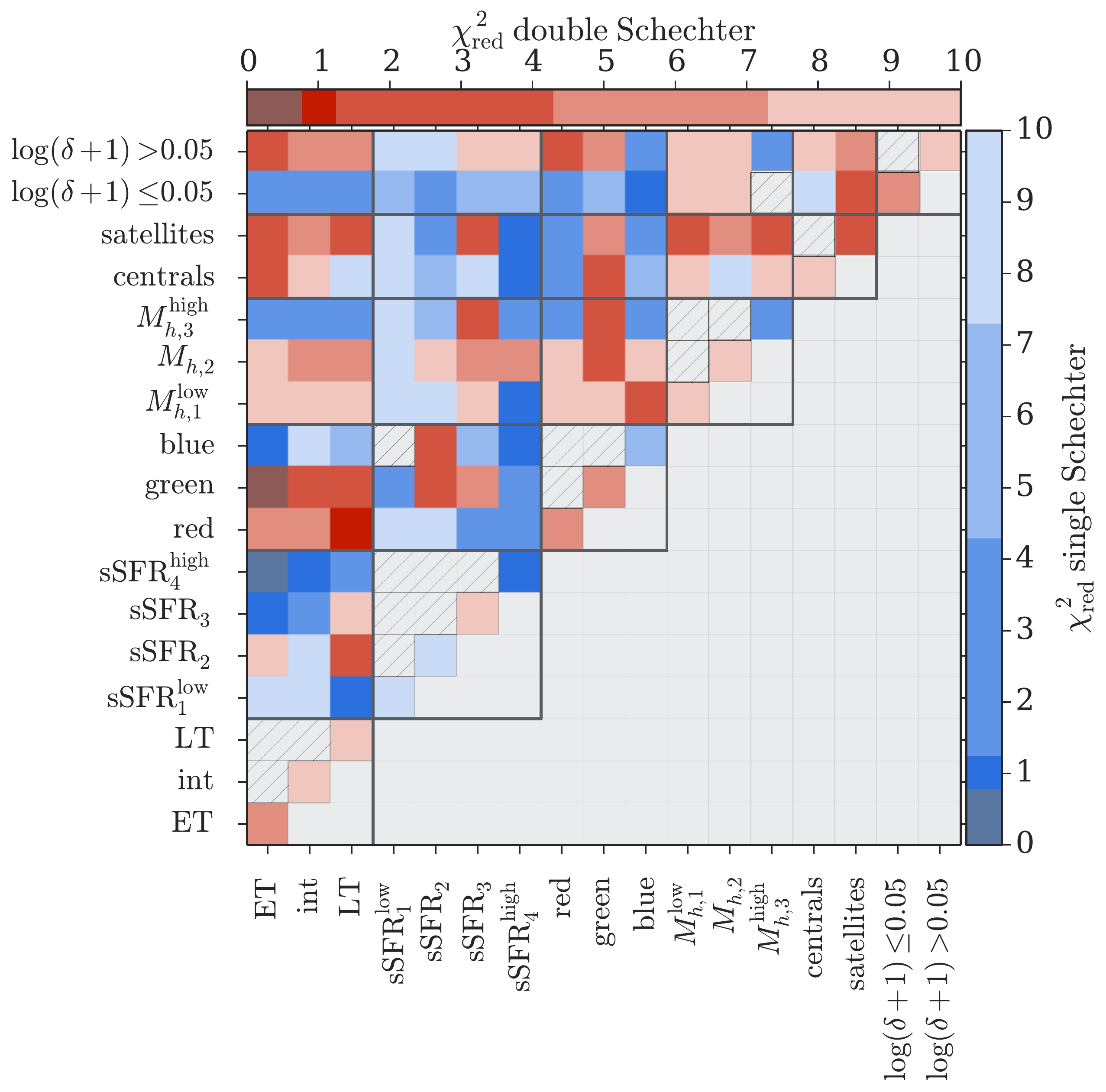}
\caption{\label{fig:single_vs_double}Summary of the over 130 stellar mass functions and their best fitting models. We use a likelihood-ratio test to determine if a subsample is better fit by a single or by a double Schechter function (see Sec. \ref{sec:single_double}). In this figure, single and double Schechter function subsamples are shown in blue and red, respectively. They are colour coded by their corresponding reduced $\chi^2$ value. Hatched cells show subsamples for which we cannot determine the stellar mass function. This is the case if the subsample contains too few objects (blue \& $\mathrm{sSFFR}_1^\mathrm{low}$, $\log(\delta + 1) \leq 0.05$ \& $M_{h, 3}^\mathrm{high}$) or if the properties that we would have to split by exclude each other. $\mathrm{sSFFR}_1^\mathrm{low}$, $\mathrm{sSFFR}_2$, $\mathrm{sSFFR}_3$ and $\text{sSFR}_4^\mathrm{high}$ correspond to the following cuts in sSFR: $\log \mathrm{sSFR} < -12$, $-12 \leq \log \mathrm{sSFR} < -11$, $-11 \leq \log \mathrm{sSFR} < -10$ and $-10 \leq \log \mathrm{sSFR} < -9$. $M_{h, 1}^\mathrm{low}$, $M_{h, 2}$ and $M_{h, 3}^\mathrm{high}$ refer to the following halo mass bins: $10.5 \leq \log M_h < 12$, $12 \leq \log M_h < 13.5$, $13.5 \leq \log M_h < 15$. }
\end{figure*}

We repeat the procedure outlined in Sec. \ref{sec:procedure} for $\sim 130$ subsamples. First, we slice our main sample by morphology, sSFR, colour, halo mass, centrals and satellites and overdensity. We then also split our sample by the combination of two properties, for example morphology and sSFR. Figure \ref{fig:single_vs_double} gives on overview of our results by showing which of these various subsamples are better fit by a single, and which are fit better by a double Schechter function. The individual stellar mass functions are shown in Figure \ref{fig:mass_fct_morphology} to Figure \ref{fig:mass_fct_cent_sat_dens}. Table \ref{tab:Fig_overview} illustrates which mass function can be found in which figure. 

The best fit Schechter function parameters, which we determine with the STY approach, are given in Tables \ref{tab:results_1} and \ref{tab:results_2}. The errors reported in these tables correspond to the $1\sigma$ errors which we compute from the MCMC chains. The reduced $\chi^2$ value which we give in the last column is derived by comparing the SWML $\Phi$ values and their random and systematic errors to the best fit Schechter function parameters from STY.  

Note that blue galaxies with $\log (\mathrm{sSFR/}M_\odot \mathrm{yr^{-1}}) < -12$ and galaxies in underdense regions with $13.5 < \log (M_h/M_\odot) < 15$ are not listed in Tables \ref{tab:results_1} and \ref{tab:results_2} and are not shown in Figure \ref{fig:mass_fct_sSFR_color} and Figure \ref{fig:mass_fct_hm_dens} since these subsamples contain too few objects to constrain their stellar mass functions.

As discussed in detail in the previous sections, we have developed a robust method to generate stellar mass functions in the local Universe by combining three different techniques. We consider the M/L distribution of each subsample when determining the stellar mass completeness function and do not simply use magnitude dependent \zmax\ values. In comparison to previous work, we do not require a priori assumptions on which subsamples are better fit by a single and which subsamples are better described by a double Schechter function.

The extensive collection of stellar mass functions that we are presenting in this work only represents the first application of our method. An immediate follow-up project, which we are currently working on, is the generation of stellar mass functions based on the second Galaxy Zoo data release \citep{Willett:2013aa} which includes more detailed morphological information on, for example, spiral arms, bulges, galactic bars and pitch angles. In the future, we will also make our \textsc{python} code available online, so that users can generate mass functions of subsamples of their choice.  Ultimately, we want to use stellar mass functions to investigate the role of AGN feedback on the quenching of star formation for which the method and the stellar mass functions presented here will be very valuable.

\section{Summary}
\label{sec:summary}
We have used SDSS DR7 data to construct 135 stellar mass functions in the redshift range \zminsample\ to \zmaxsample. We not only determined the stellar mass function for the entire sample, but also split the main sample into various subsamples based on morphology \citep{Lintott:2008aa,Lintott:2011aa}, sSFR \citep{Brinchmann:2004aa,Kauffmann:2003ab}, colour (see Sec. \ref{sec:colour_def}), halo mass \citep{Yang:2007aa}, overdensity (see Sec. \ref{sec:overdensity}) and the classification into centrals and satellites \citep{Yang:2007aa}. 
Our analysis was based on three classical mass function estimators: the  1/\Vmax\ technique by \cite{Schmidt:1968aa}, the parametric maximum likelihood method STY by \cite{Sandage:1979aa} and the non-parametric maximum likelihood approach SWML by \cite{Efstathiou:1988aa}.

\begin{figure*}
	\includegraphics[scale = .3]{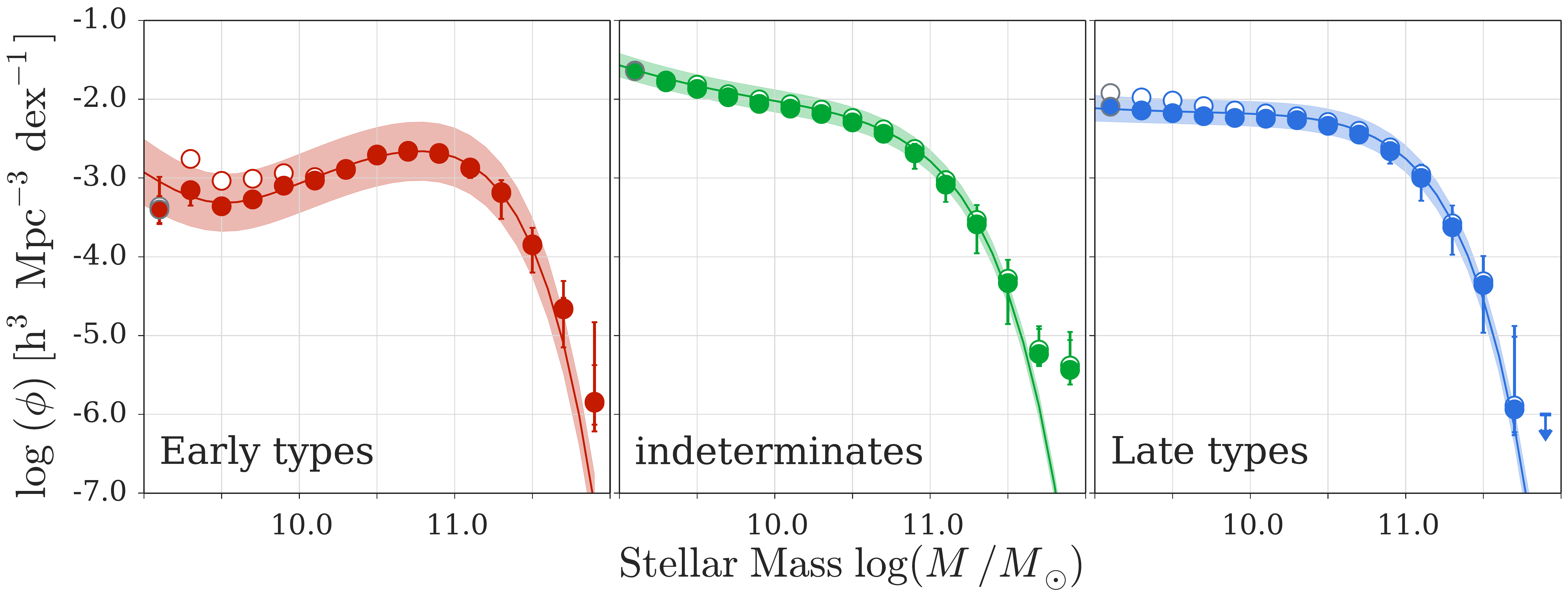}
	\caption{\label{fig:mass_fct_morphology} Entire sample split by morphology. The $\Phi$ values determined with $1/V_\mathrm{max}$ and SWML are shown with open and filled symbols, respectively. The best fit Schechter function based on STY is illustrated with a solid line. The shaded regions correspond to the STY $1\sigma$ uncertainties that we determine directly from the MCMC chain. For the $1/V_\mathrm{max}$ values we are showing random errors only. The errorbars on the SWML $\Phi$ values correspond to the combination of random errors and the systematic error due to stellar mass uncertainties.}
\end{figure*}

\begin{figure*}
	\includegraphics[scale = .3]{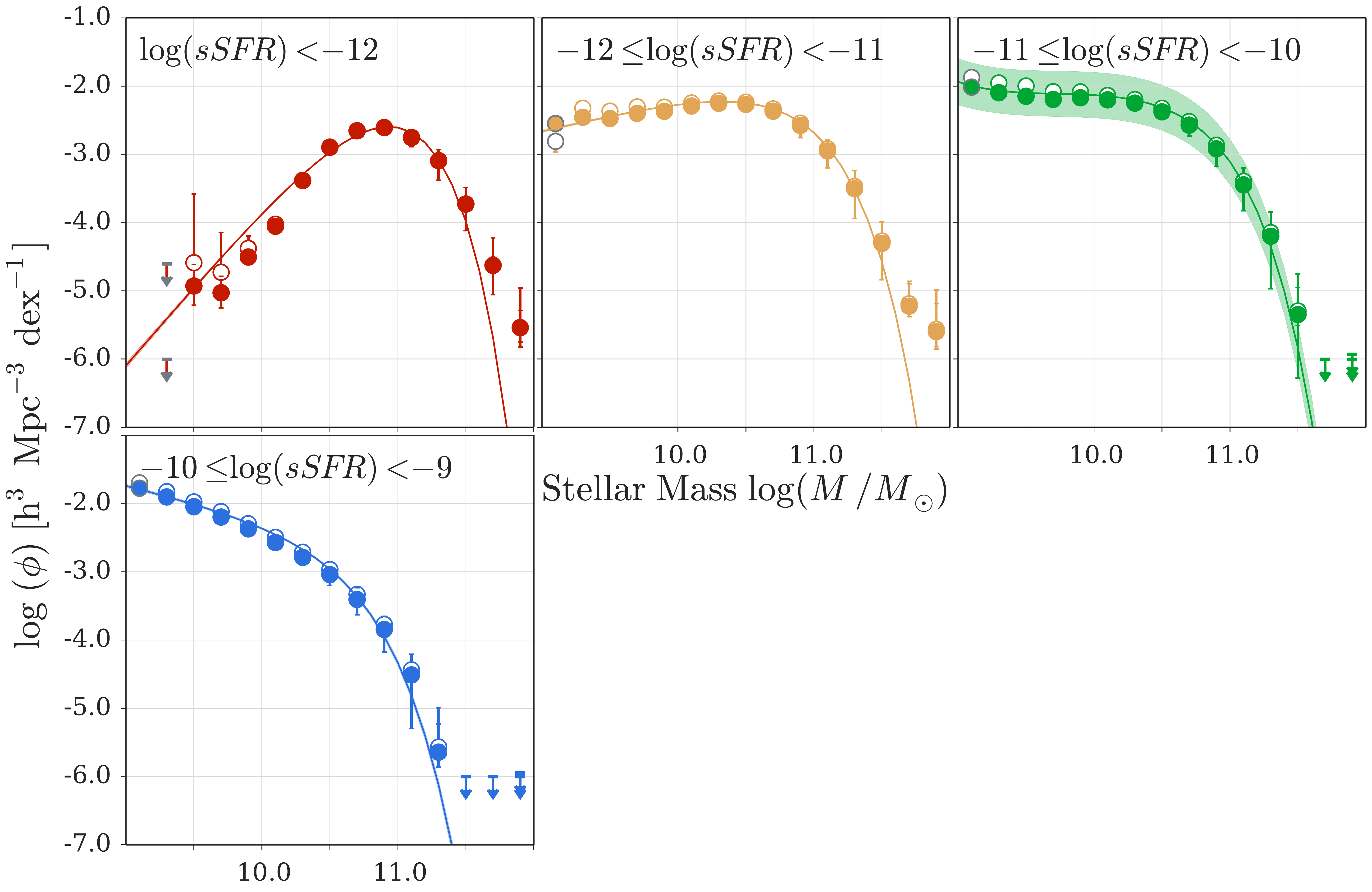}
	\caption{\label{fig:mass_fct_sSFR} Entire sample split by sSFR.The $\Phi$ values determined with $1/V_\mathrm{max}$ and SWML are shown with open and filled symbols, respectively. The best fit Schechter function based on STY is illustrated with a solid line. The shaded regions correspond to the STY $1\sigma$ uncertainties that we determine directly from the MCMC chain. For the $1/V_\mathrm{max}$ values we are showing random errors only. The errorbars on the SWML $\Phi$ values correspond to the combination of random errors and the systematic error due to stellar mass uncertainties.}
\end{figure*}

\begin{figure*}
	\includegraphics[scale = .3]{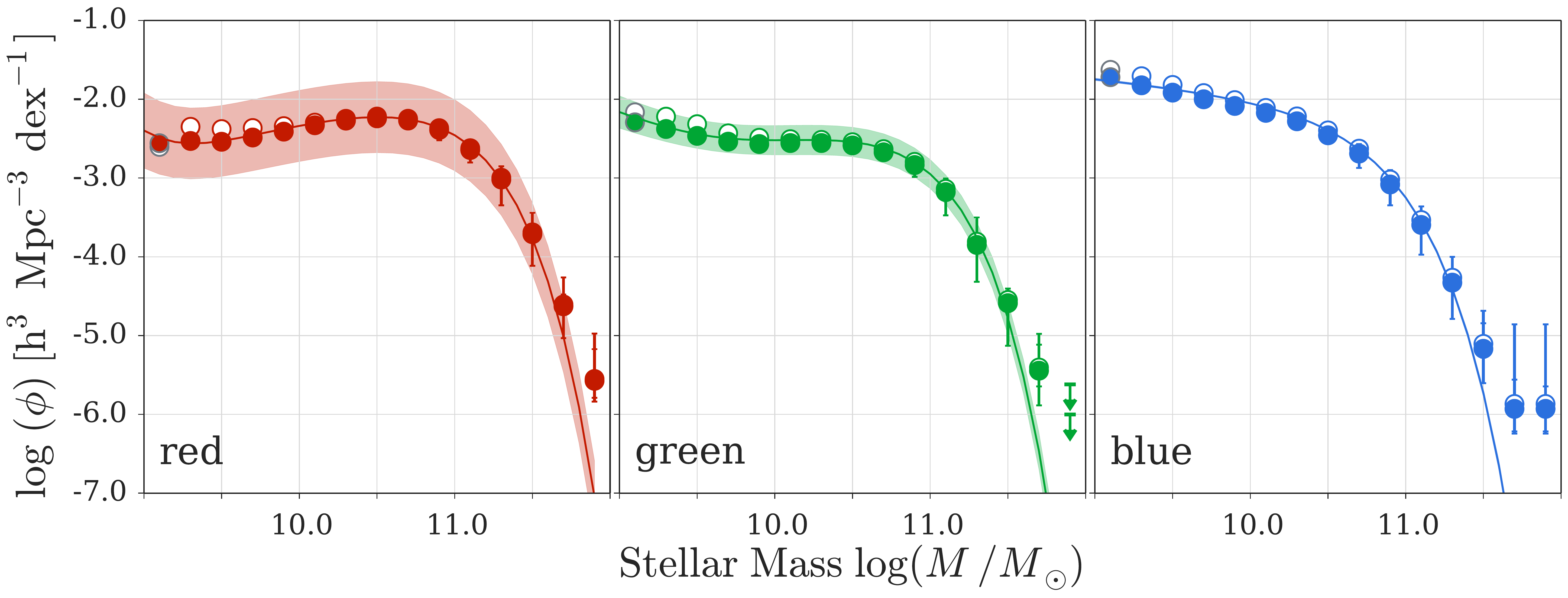}
	\caption{\label{fig:mass_fct_color} Entire sample split by colour. The $\Phi$ values determined with $1/V_\mathrm{max}$ and SWML are shown with open and filled symbols, respectively. The best fit Schechter function based on STY is illustrated with a solid line. The shaded regions correspond to the STY $1\sigma$ uncertainties that we determine directly from the MCMC chain. For the $1/V_\mathrm{max}$ values we are showing random errors only. The errorbars on the SWML $\Phi$ values correspond to the combination of random errors and the systematic error due to stellar mass uncertainties.}
\end{figure*}

\begin{figure*}
	\includegraphics[scale = .3]{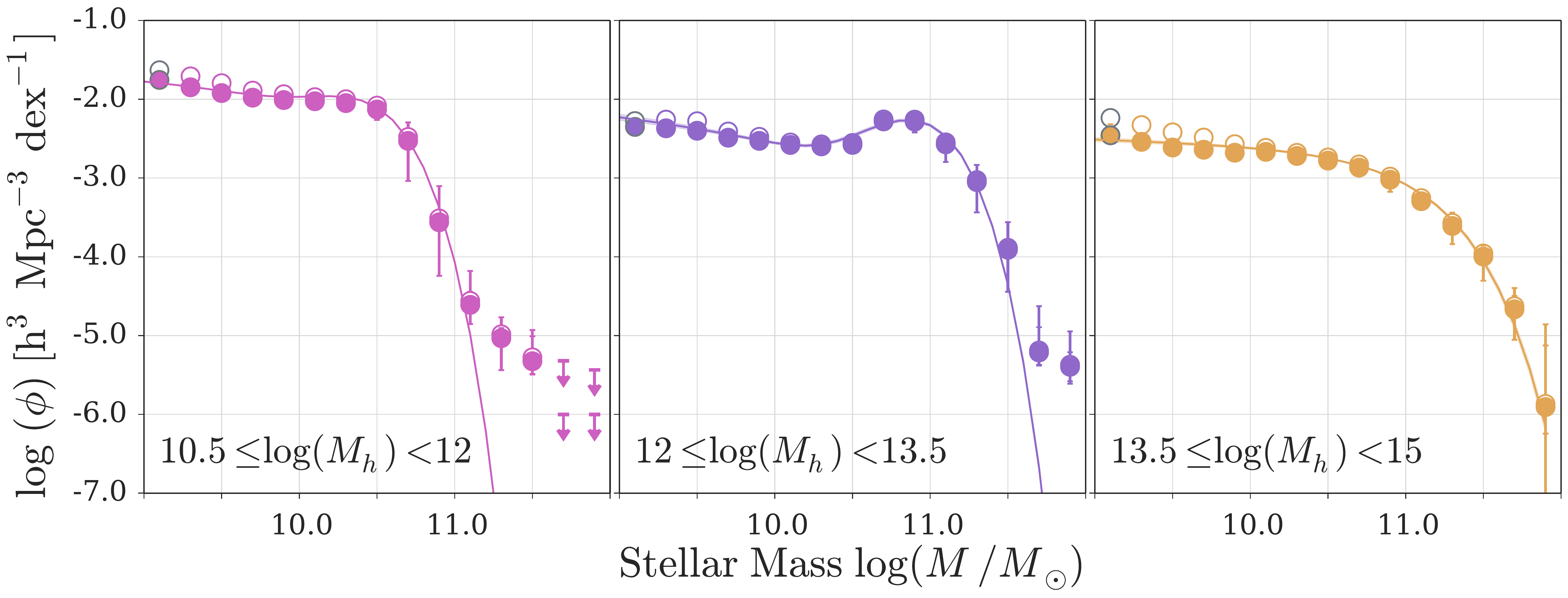}
	\caption{\label{fig:mass_fct_hm} Entire sample split by halo mass. The $\Phi$ values determined with $1/V_\mathrm{max}$ and SWML are shown with open and filled symbols, respectively. The best fit Schechter function based on STY is illustrated with a solid line. The shaded regions correspond to the STY $1\sigma$ uncertainties that we determine directly from the MCMC chain. For the $1/V_\mathrm{max}$ values we are showing random errors only. The errorbars on the SWML $\Phi$ values correspond to the combination of random errors and the systematic error due to stellar mass uncertainties.}
\end{figure*}

\begin{figure*}
	\includegraphics[scale = .3]{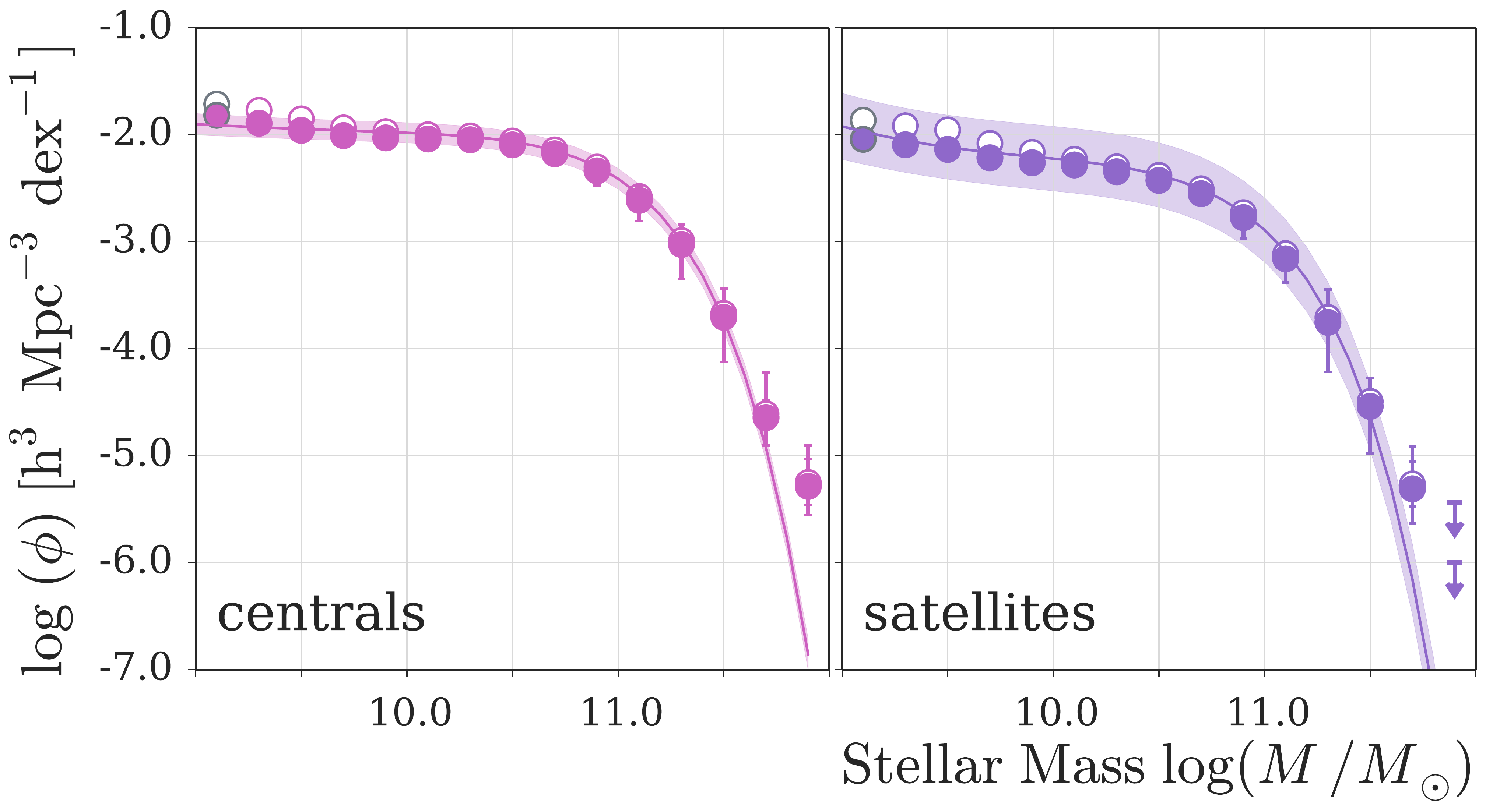}
	\caption{\label{fig:mass_fct_cent_sat} Entire sample split by centrals \& satellites. The $\Phi$ values determined with $1/V_\mathrm{max}$ and SWML are shown with open and filled symbols, respectively. The best fit Schechter function based on STY is illustrated with a solid line. The shaded regions correspond to the STY $1\sigma$ uncertainties that we determine directly from the MCMC chain. For the $1/V_\mathrm{max}$ values we are showing random errors only. The errorbars on the SWML $\Phi$ values correspond to the combination of random errors and the systematic error due to stellar mass uncertainties.}
\end{figure*}

\begin{figure*}
	\includegraphics[scale = .3]{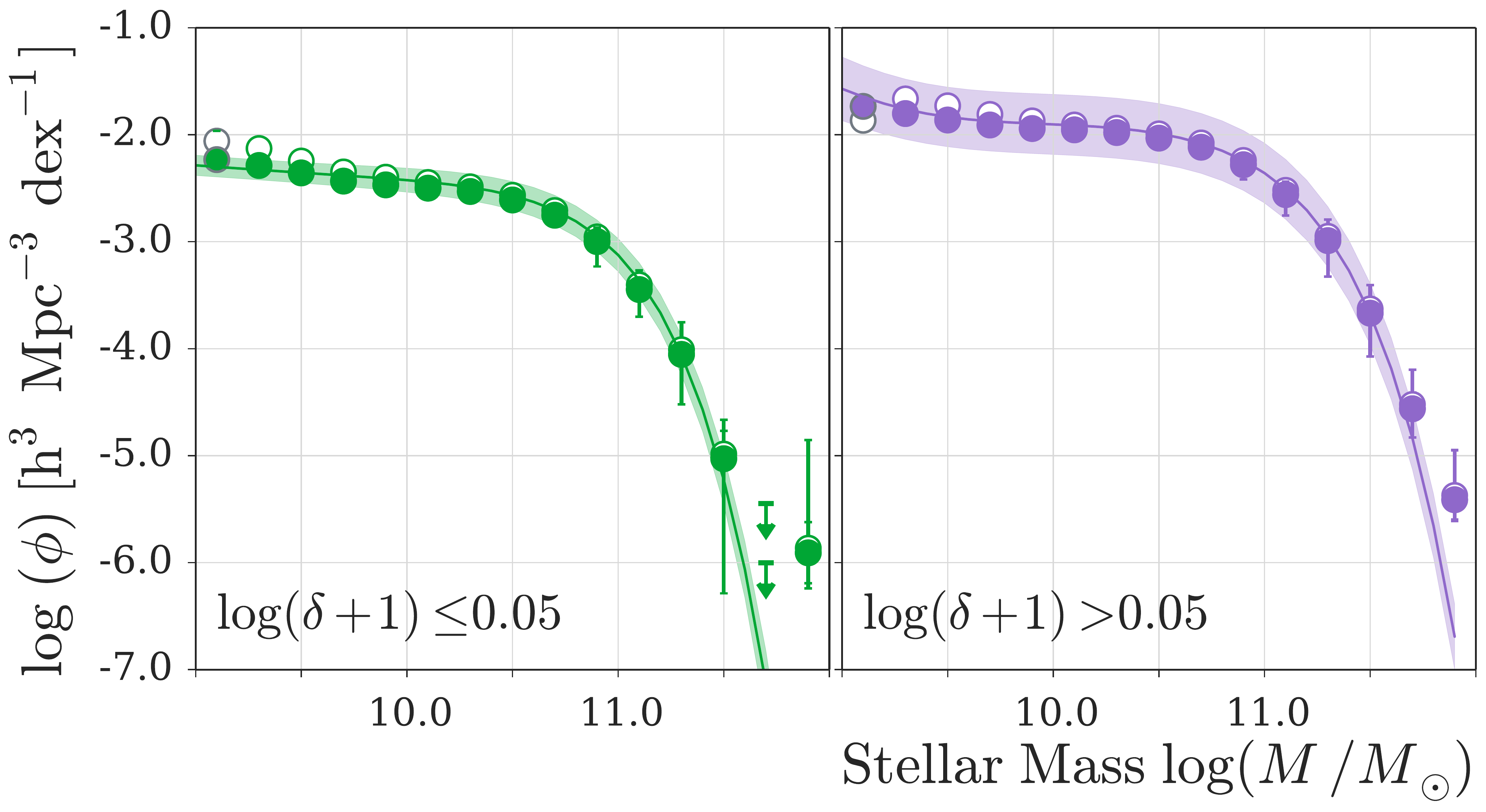}
	\caption{\label{fig:mass_fct_dens} Entire sample split by density.The $\Phi$ values determined with $1/V_\mathrm{max}$ and SWML are shown with open and filled symbols, respectively. The best fit Schechter function based on STY is illustrated with a solid line. The shaded regions correspond to the STY $1\sigma$ uncertainties that we determine directly from the MCMC chain. For the $1/V_\mathrm{max}$ values we are showing random errors only. The errorbars on the SWML $\Phi$ values correspond to the combination of random errors and the systematic error due to stellar mass uncertainties.}
\end{figure*}

\begin{figure*}
	\includegraphics[scale = .3]{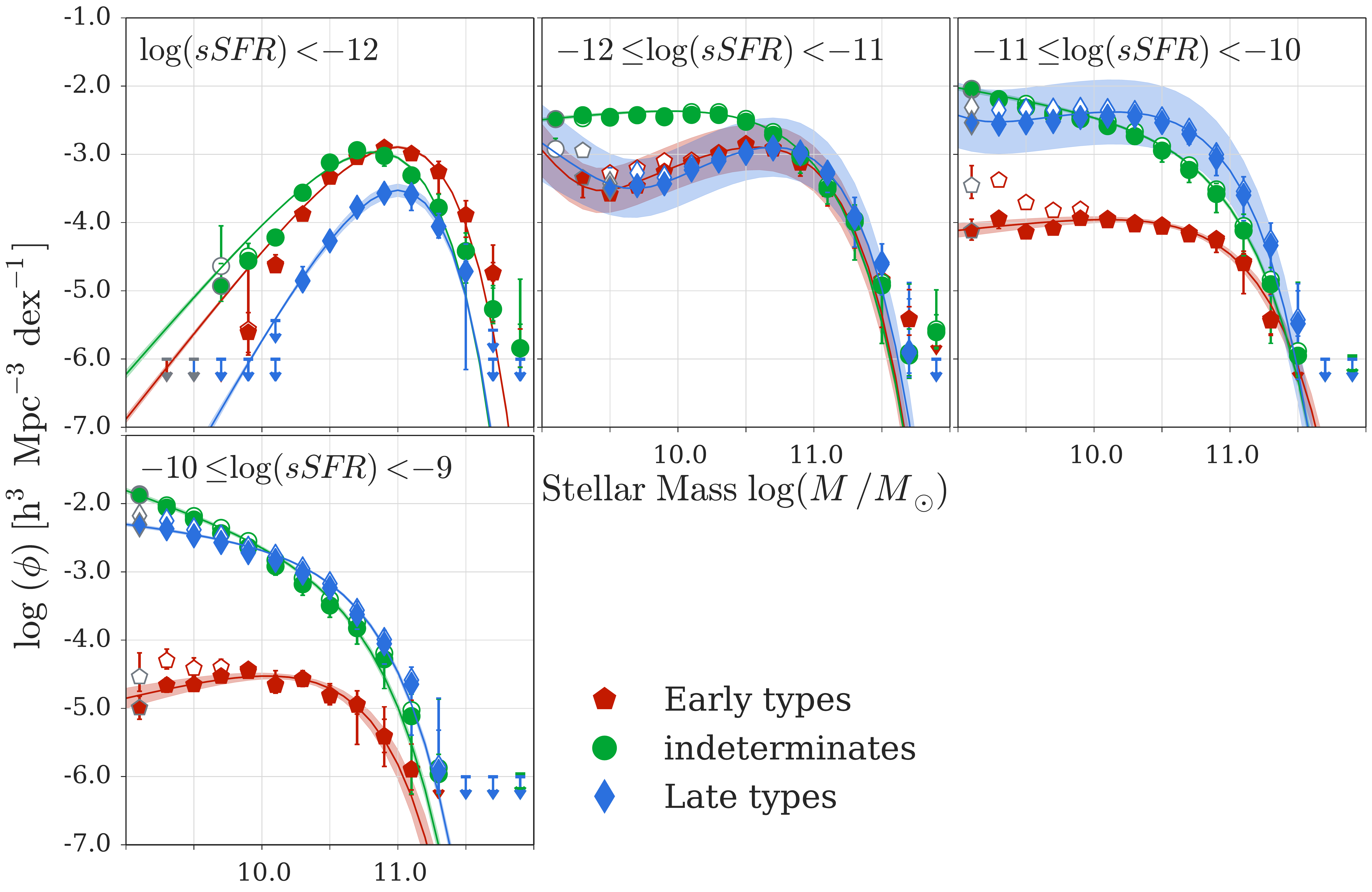}
	\caption{\label{fig:mass_fct_morph_sSFR} Entire sample split by morphology \& sSFR. The $\Phi$ values determined with $1/V_\mathrm{max}$ and SWML are shown with open and filled symbols, respectively. The best fit Schechter function based on STY is illustrated with a solid line. The shaded regions correspond to the STY $1\sigma$ uncertainties that we determine directly from the MCMC chain. For the $1/V_\mathrm{max}$ values we are showing random errors only. The errorbars on the SWML $\Phi$ values correspond to the combination of random errors and the systematic error due to stellar mass uncertainties.}
\end{figure*}

\begin{figure*}
	\includegraphics[scale = .3]{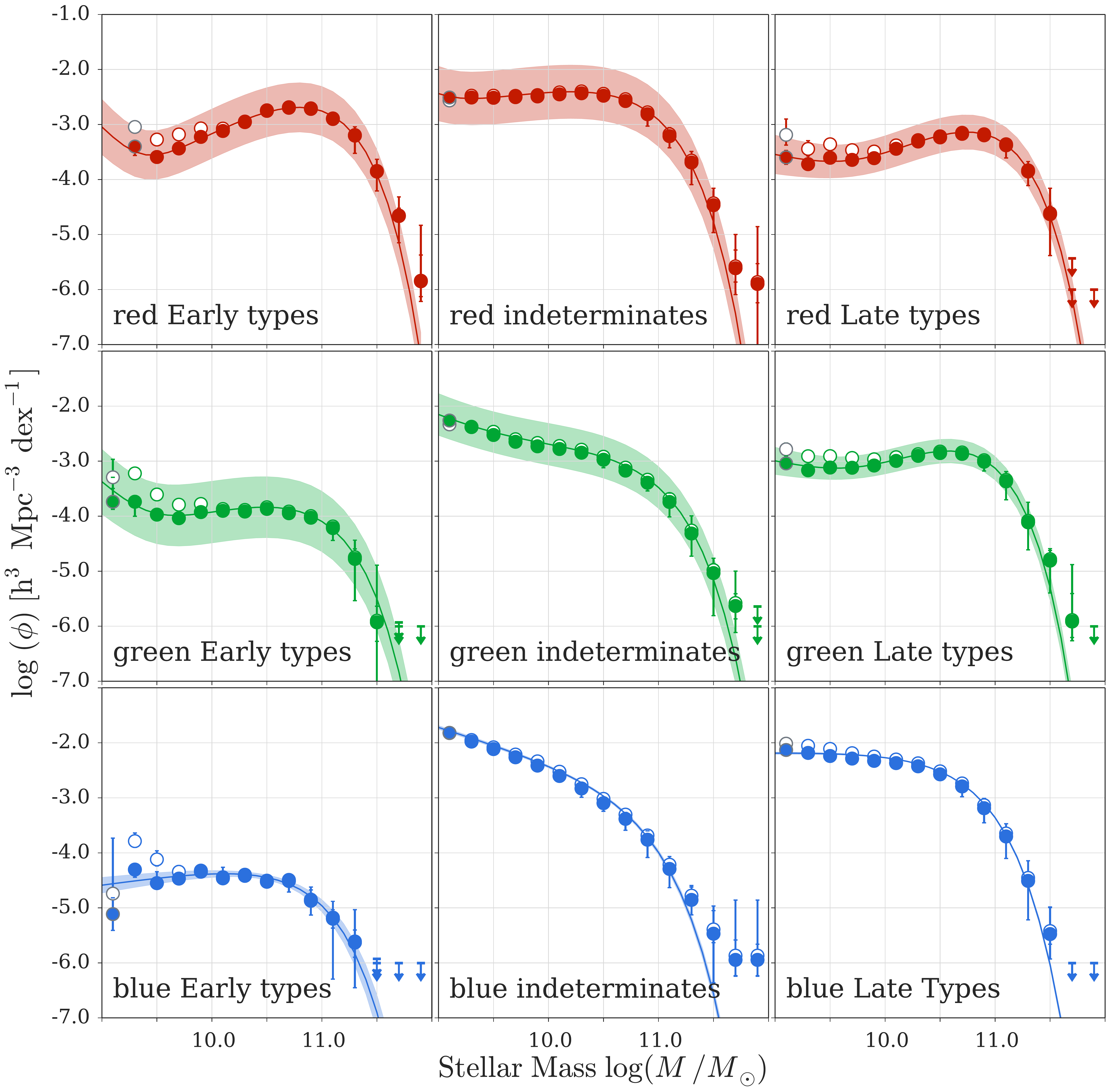}
	\caption{\label{fig:mass_fct_morph_color} Entire sample split by morphology \& colour.The $\Phi$ values determined with $1/V_\mathrm{max}$ and SWML are shown with open and filled symbols, respectively. The best fit Schechter function based on STY is illustrated with a solid line. The shaded regions correspond to the STY $1\sigma$ uncertainties that we determine directly from the MCMC chain. For the $1/V_\mathrm{max}$ values we are showing random errors only. The errorbars on the SWML $\Phi$ values correspond to the combination of random errors and the systematic error due to stellar mass uncertainties.}
\end{figure*}

\begin{figure*}
	\includegraphics[scale = .3]{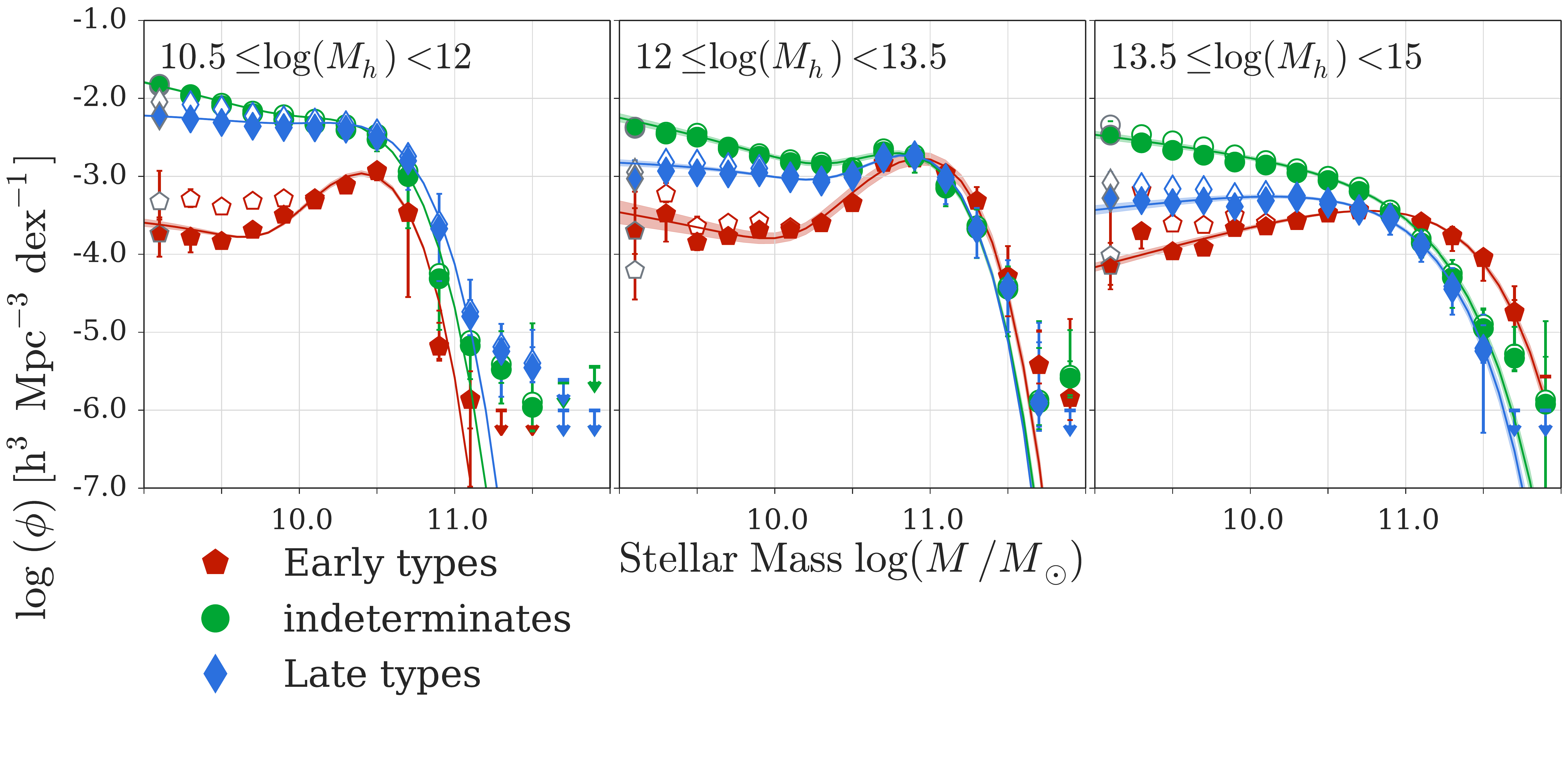}
	\caption{\label{fig:mass_fct_morph_hm} Entire sample split by morphology \& halo mass. The $\Phi$ values determined with $1/V_\mathrm{max}$ and SWML are shown with open and filled symbols, respectively. The best fit Schechter function based on STY is illustrated with a solid line. The shaded regions correspond to the STY $1\sigma$ uncertainties that we determine directly from the MCMC chain. For the $1/V_\mathrm{max}$ values we are showing random errors only. The errorbars on the SWML $\Phi$ values correspond to the combination of random errors and the systematic error due to stellar mass uncertainties.}
\end{figure*}

\begin{figure*}
	\includegraphics[scale = .3]{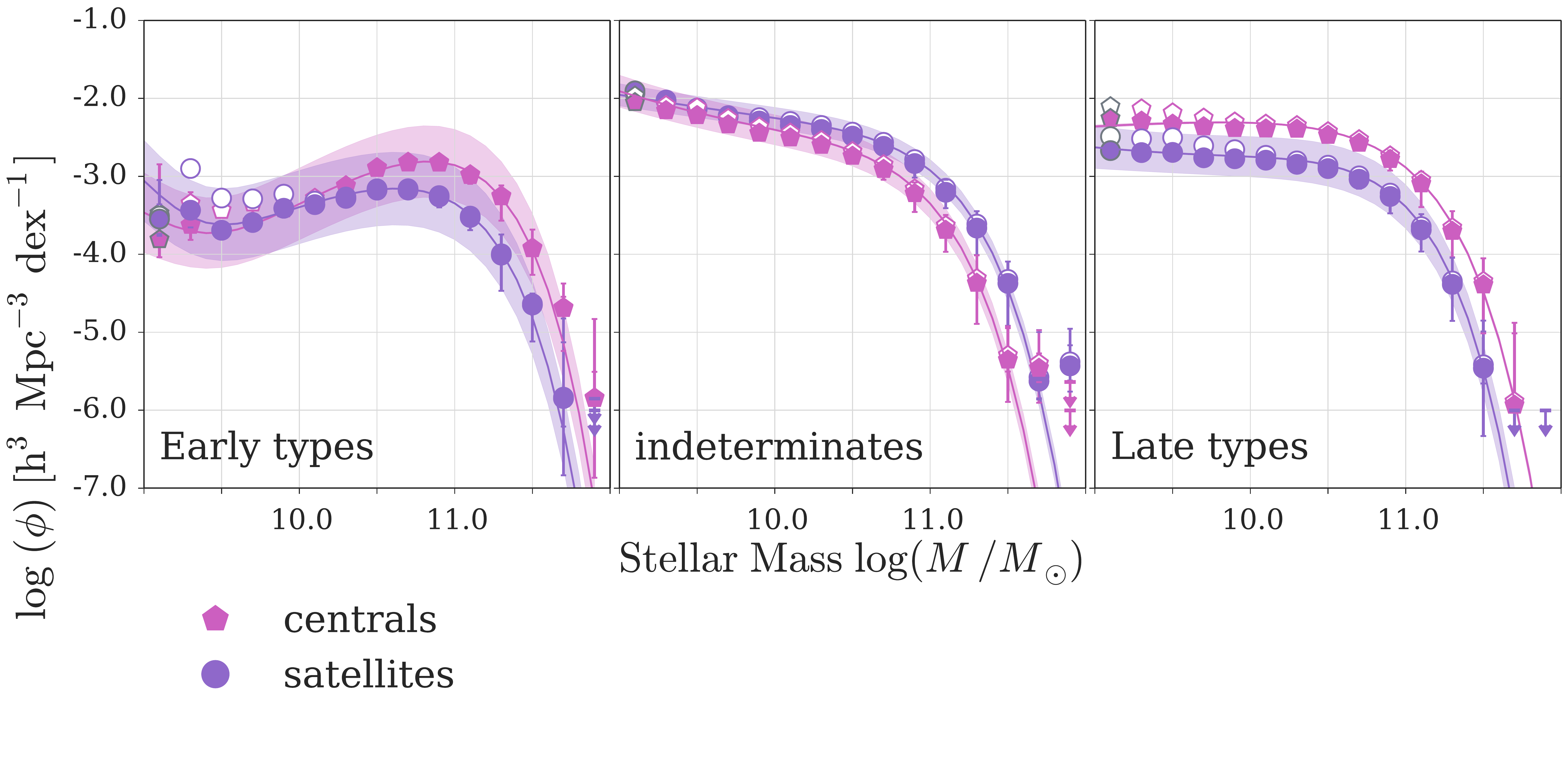}
	\caption{\label{fig:mass_fct_morph_cent_sat} Entire sample split by morphology \& centrals \& satellites. The $\Phi$ values determined with $1/V_\mathrm{max}$ and SWML are shown with open and filled symbols, respectively. The best fit Schechter function based on STY is illustrated with a solid line. The shaded regions correspond to the STY $1\sigma$ uncertainties that we determine directly from the MCMC chain. For the $1/V_\mathrm{max}$ values we are showing random errors only. The errorbars on the SWML $\Phi$ values correspond to the combination of random errors and the systematic error due to stellar mass uncertainties.}
\end{figure*}

\begin{figure*}
	\includegraphics[scale = .3]{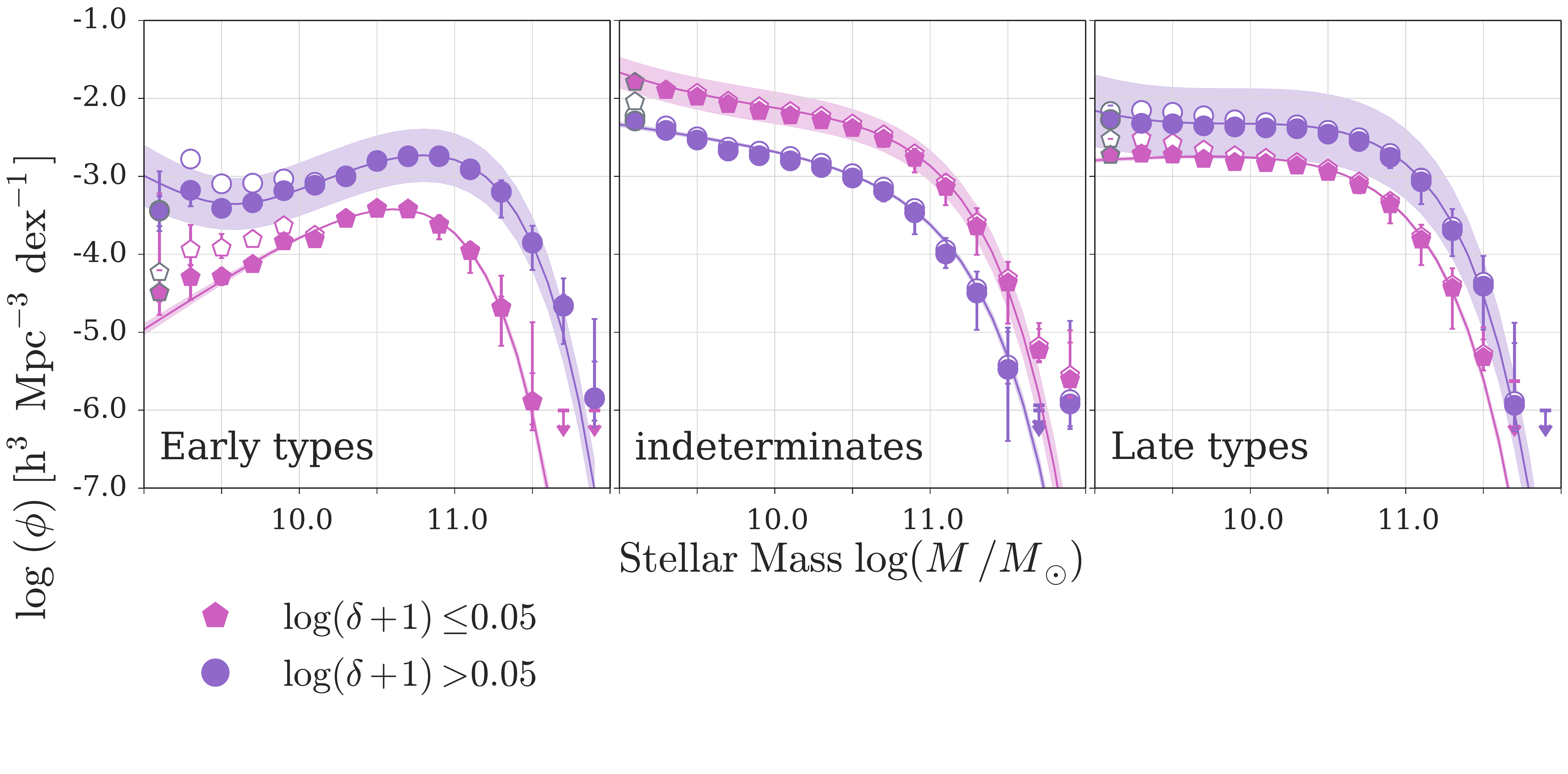}
	\caption{\label{fig:mass_fct_morph_dens} Entire sample split by morphology \& density. The $\Phi$ values determined with $1/V_\mathrm{max}$ and SWML are shown with open and filled symbols, respectively. The best fit Schechter function based on STY is illustrated with a solid line. The shaded regions correspond to the STY $1\sigma$ uncertainties that we determine directly from the MCMC chain. For the $1/V_\mathrm{max}$ values we are showing random errors only. The errorbars on the SWML $\Phi$ values correspond to the combination of random errors and the systematic error due to stellar mass uncertainties.}
\end{figure*}

\begin{figure*}
	\includegraphics[scale = .3]{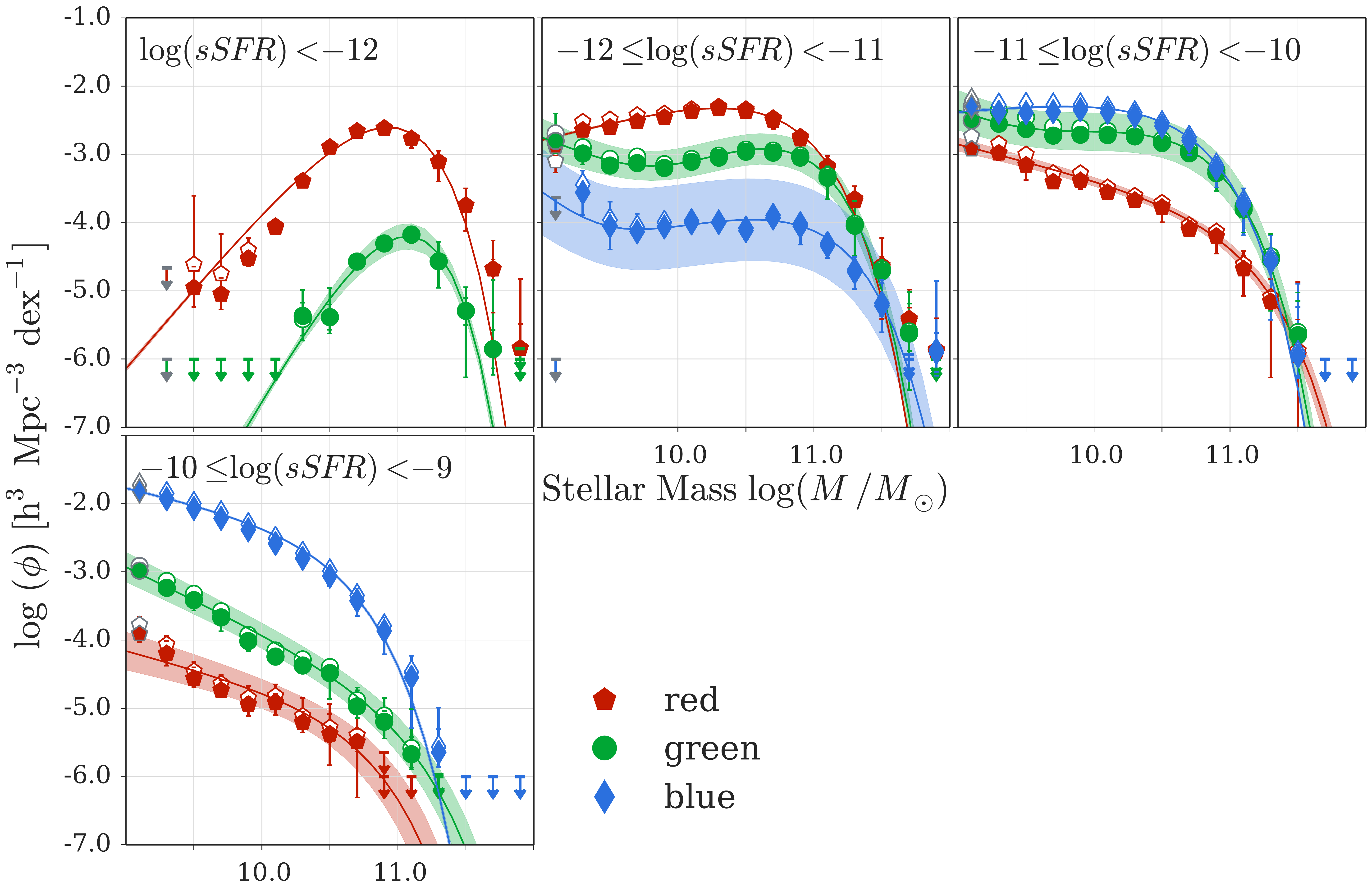}
	\caption{\label{fig:mass_fct_sSFR_color} Entire sample split by sSFR \& colour. The $\Phi$ values determined with $1/V_\mathrm{max}$ and SWML are shown with open and filled symbols, respectively. The best fit Schechter function based on STY is illustrated with a solid line. The shaded regions correspond to the STY $1\sigma$ uncertainties that we determine directly from the MCMC chain. For the $1/V_\mathrm{max}$ values we are showing random errors only. The errorbars on the SWML $\Phi$ values correspond to the combination of random errors and the systematic error due to stellar mass uncertainties.}
\end{figure*}

\begin{figure*}
	\includegraphics[scale = .3]{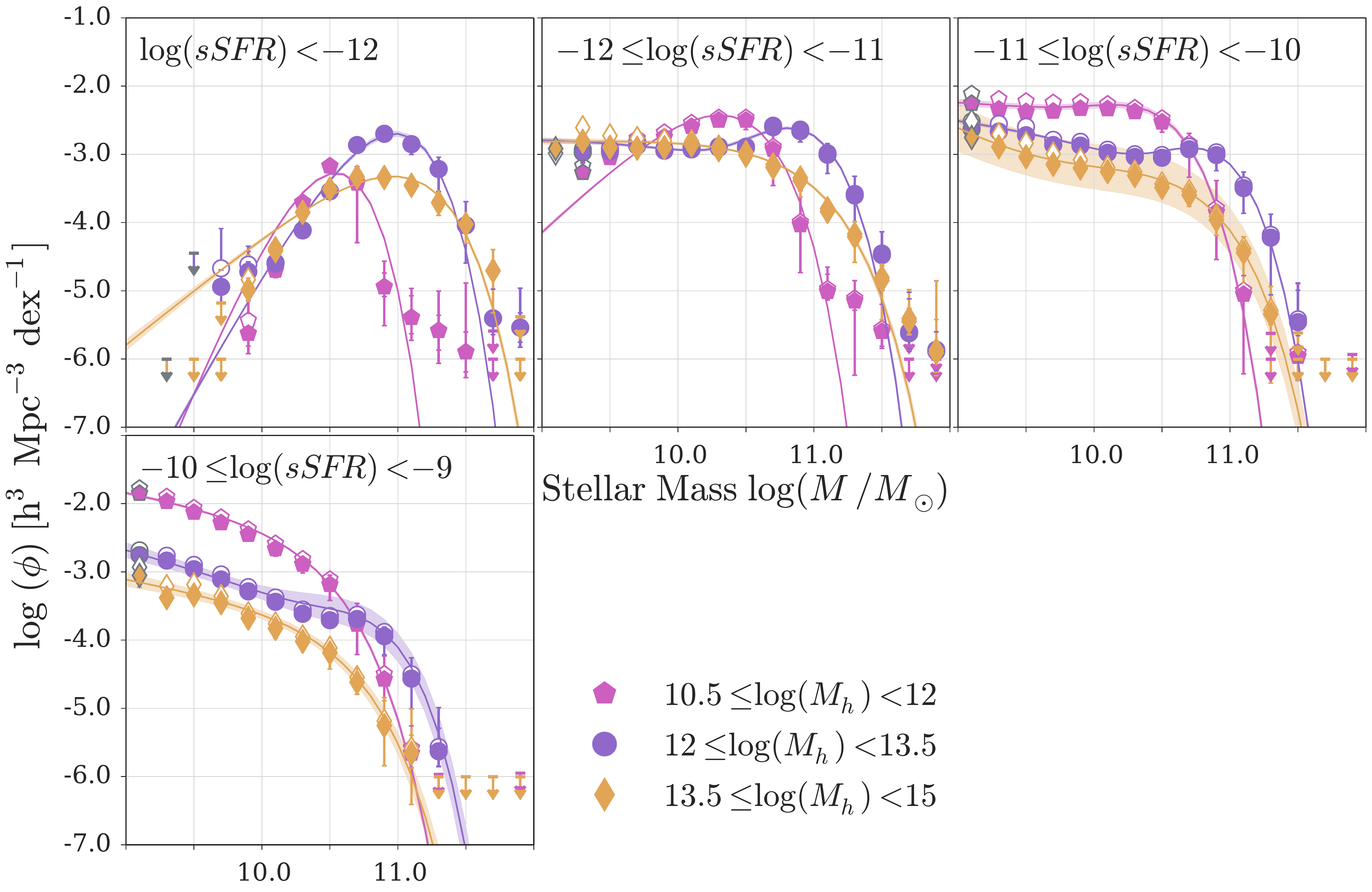}
	\caption{\label{fig:mass_fct_sSFR_hm} Entire sample split by sSFR \& halo mass. The $\Phi$ values determined with $1/V_\mathrm{max}$ and SWML are shown with open and filled symbols, respectively. The best fit Schechter function based on STY is illustrated with a solid line. The shaded regions correspond to the STY $1\sigma$ uncertainties that we determine directly from the MCMC chain. For the $1/V_\mathrm{max}$ values we are showing random errors only. The errorbars on the SWML $\Phi$ values correspond to the combination of random errors and the systematic error due to stellar mass uncertainties.}
\end{figure*}

\begin{figure*}
	\includegraphics[scale = .3]{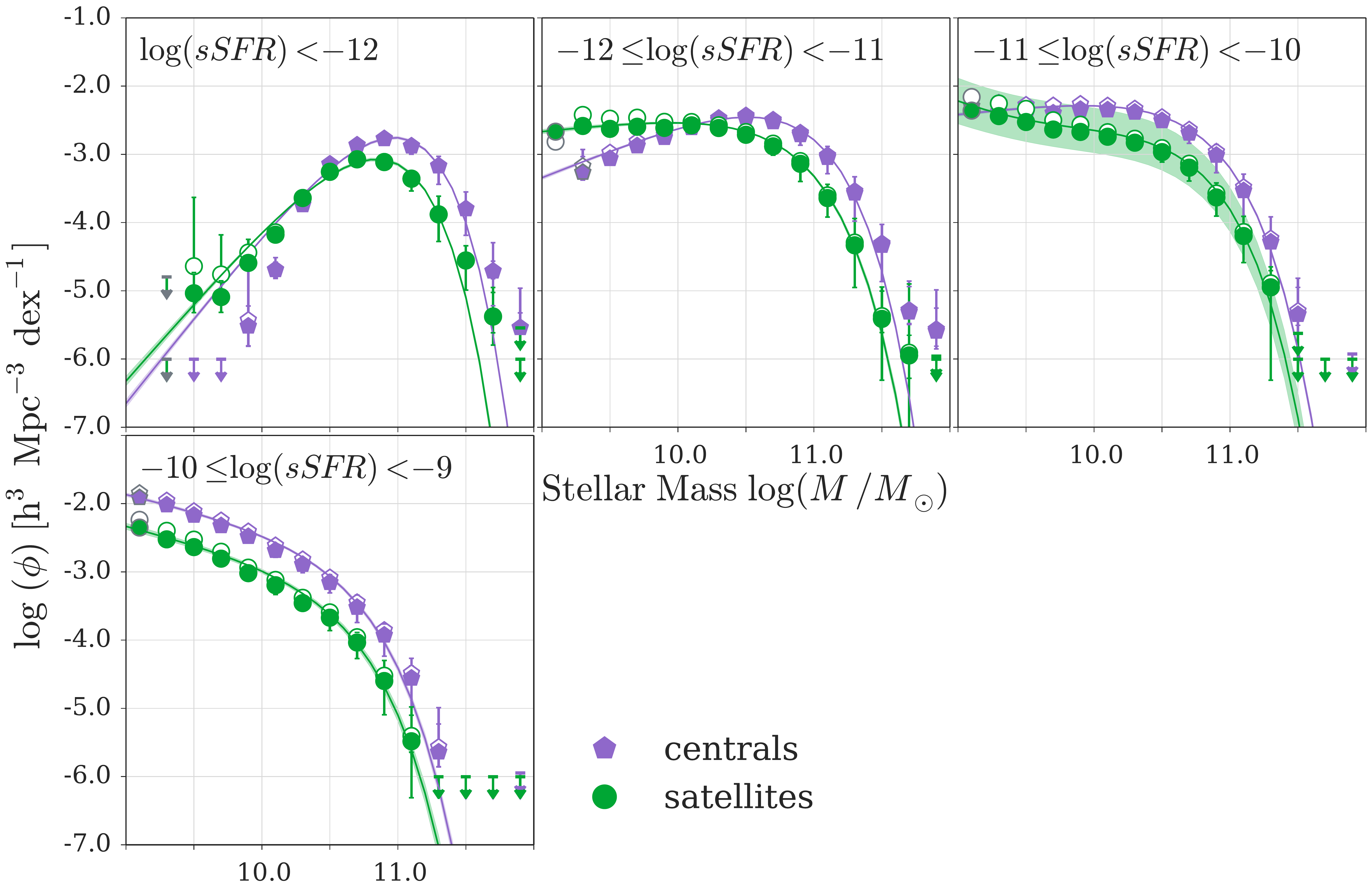}
	\caption{\label{fig:mass_fct_sSFR_cent_sat} Entire sample split by sSFR \& centrals \& satellites. The $\Phi$ values determined with $1/V_\mathrm{max}$ and SWML are shown with open and filled symbols, respectively. The best fit Schechter function based on STY is illustrated with a solid line. The shaded regions correspond to the STY $1\sigma$ uncertainties that we determine directly from the MCMC chain. For the $1/V_\mathrm{max}$ values we are showing random errors only. The errorbars on the SWML $\Phi$ values correspond to the combination of random errors and the systematic error due to stellar mass uncertainties.}
\end{figure*}

\begin{figure*}
	\includegraphics[scale = .3]{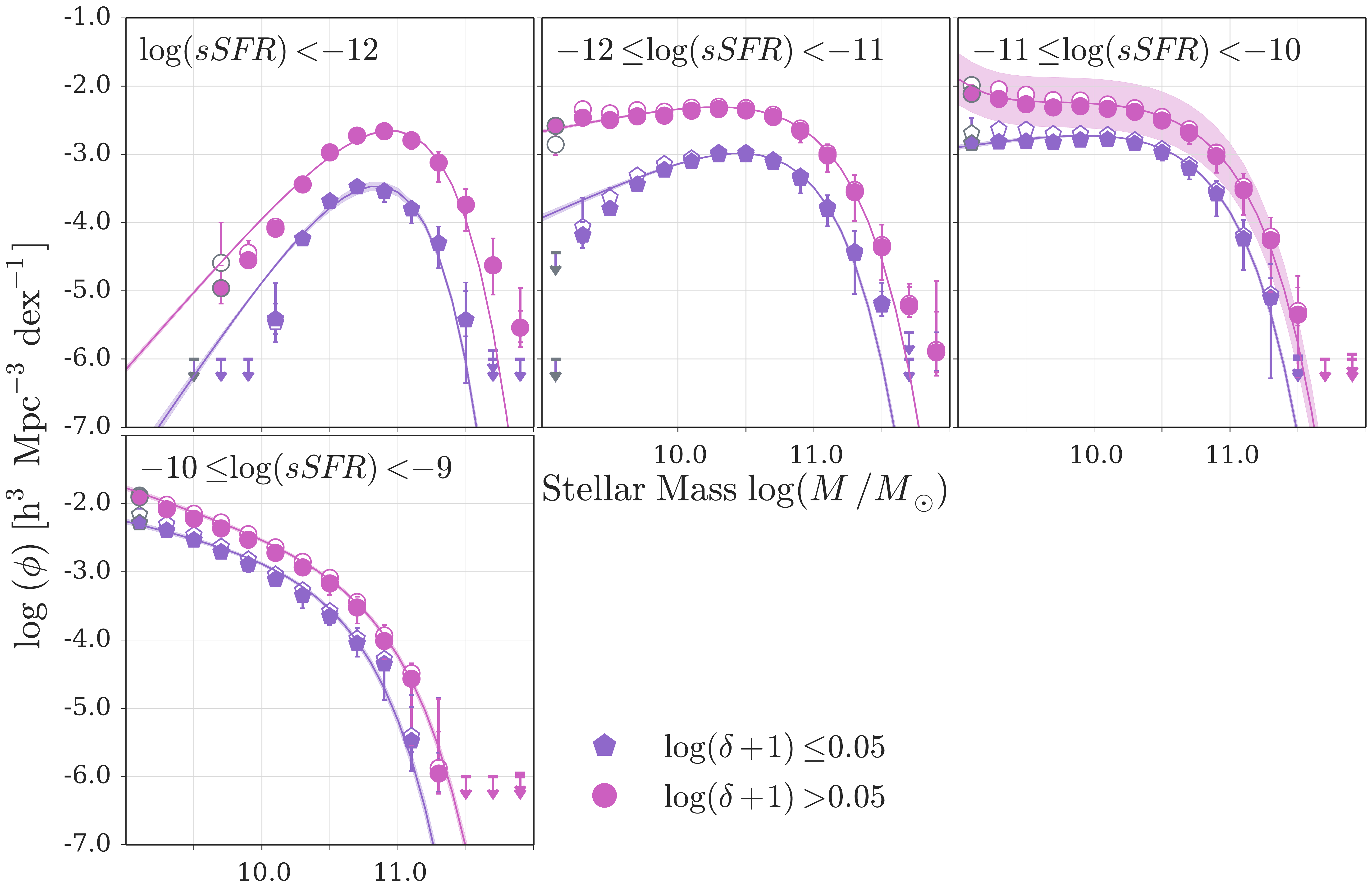}
	\caption{\label{fig:mass_fct_sSFR_dens} Entire sample split by sSFR \& density. The $\Phi$ values determined with $1/V_\mathrm{max}$ and SWML are shown with open and filled symbols, respectively. The best fit Schechter function based on STY is illustrated with a solid line. The shaded regions correspond to the STY $1\sigma$ uncertainties that we determine directly from the MCMC chain. For the $1/V_\mathrm{max}$ values we are showing random errors only. The errorbars on the SWML $\Phi$ values correspond to the combination of random errors and the systematic error due to stellar mass uncertainties.}
\end{figure*}

\begin{figure*}
	\includegraphics[scale = .3]{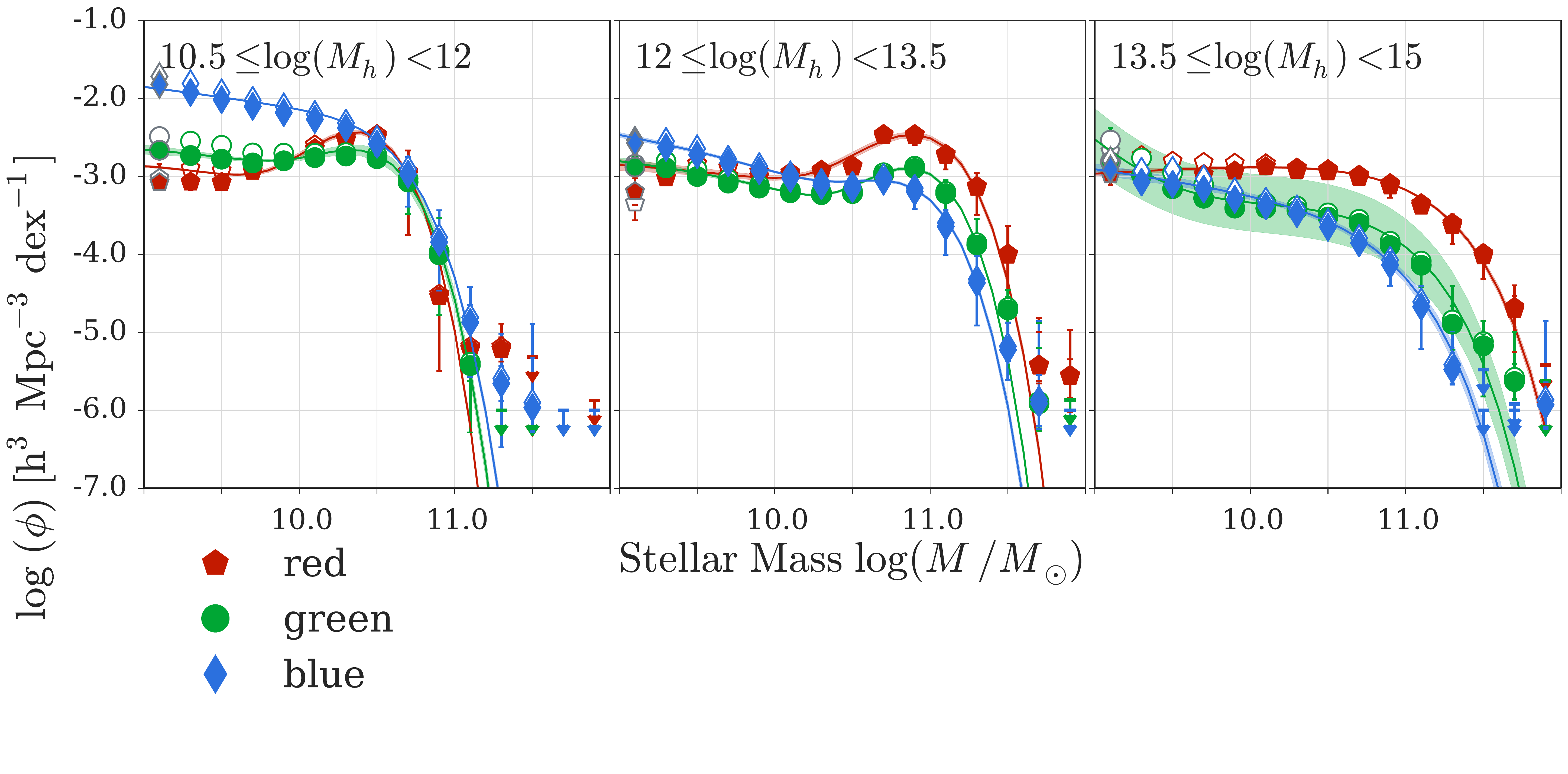}
	\caption{\label{fig:mass_fct_color_hm} Entire sample split by colour \& halo mass. The $\Phi$ values determined with $1/V_\mathrm{max}$ and SWML are shown with open and filled symbols, respectively. The best fit Schechter function based on STY is illustrated with a solid line. The shaded regions correspond to the STY $1\sigma$ uncertainties that we determine directly from the MCMC chain. For the $1/V_\mathrm{max}$ values we are showing random errors only. The errorbars on the SWML $\Phi$ values correspond to the combination of random errors and the systematic error due to stellar mass uncertainties. }
\end{figure*}

\begin{figure*}
	\includegraphics[scale = .3]{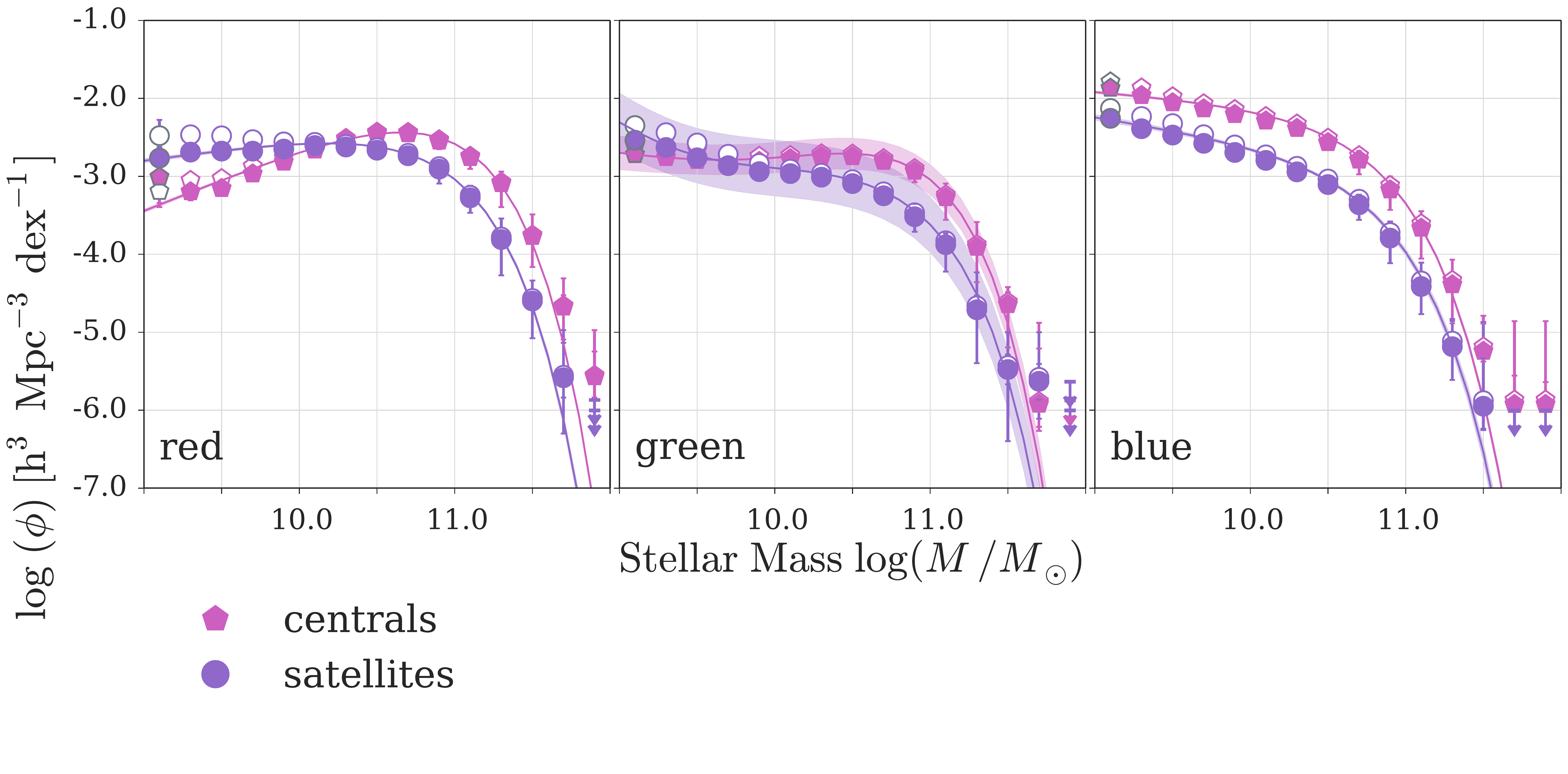}
	\caption{\label{fig:mass_fct_color_cent_sat} Entire sample split by colour \& centrals \& satellites. The $\Phi$ values determined with $1/V_\mathrm{max}$ and SWML are shown with open and filled symbols, respectively. The best fit Schechter function based on STY is illustrated with a solid line. The shaded regions correspond to the STY $1\sigma$ uncertainties that we determine directly from the MCMC chain. For the $1/V_\mathrm{max}$ values we are showing random errors only. The errorbars on the SWML $\Phi$ values correspond to the combination of random errors and the systematic error due to stellar mass uncertainties.}
\end{figure*}

\begin{figure*}
	\includegraphics[scale = .3]{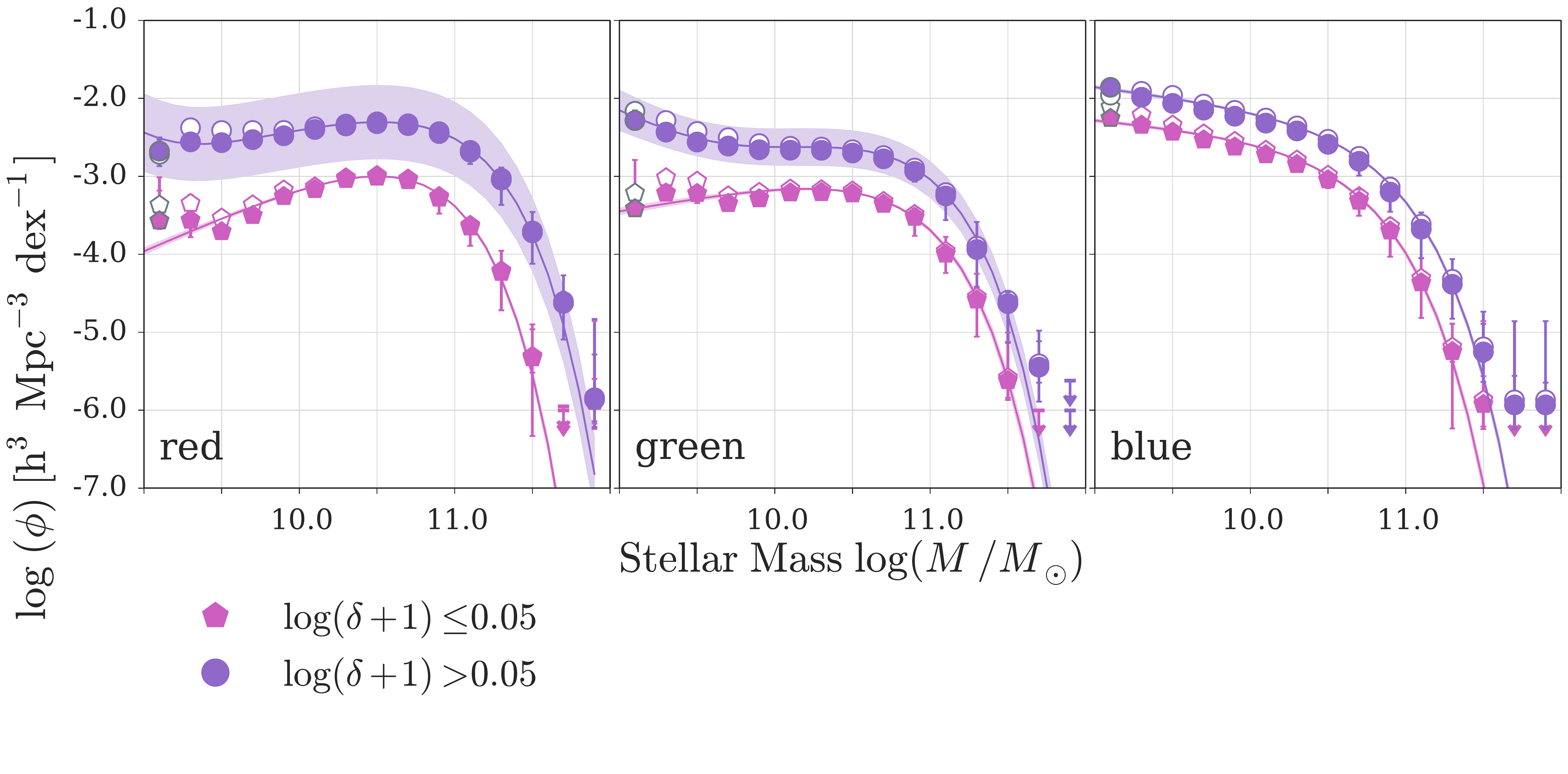}
	\caption{\label{fig:mass_fct_color_dens} Entire sample split by colour \& density. The $\Phi$ values determined with $1/V_\mathrm{max}$ and SWML are shown with open and filled symbols, respectively. The best fit Schechter function based on STY is illustrated with a solid line. The shaded regions correspond to the STY $1\sigma$ uncertainties that we determine directly from the MCMC chain. For the $1/V_\mathrm{max}$ values we are showing random errors only. The errorbars on the SWML $\Phi$ values correspond to the combination of random errors and the systematic error due to stellar mass uncertainties.}
\end{figure*}

\begin{figure*}
	\includegraphics[scale = .3]{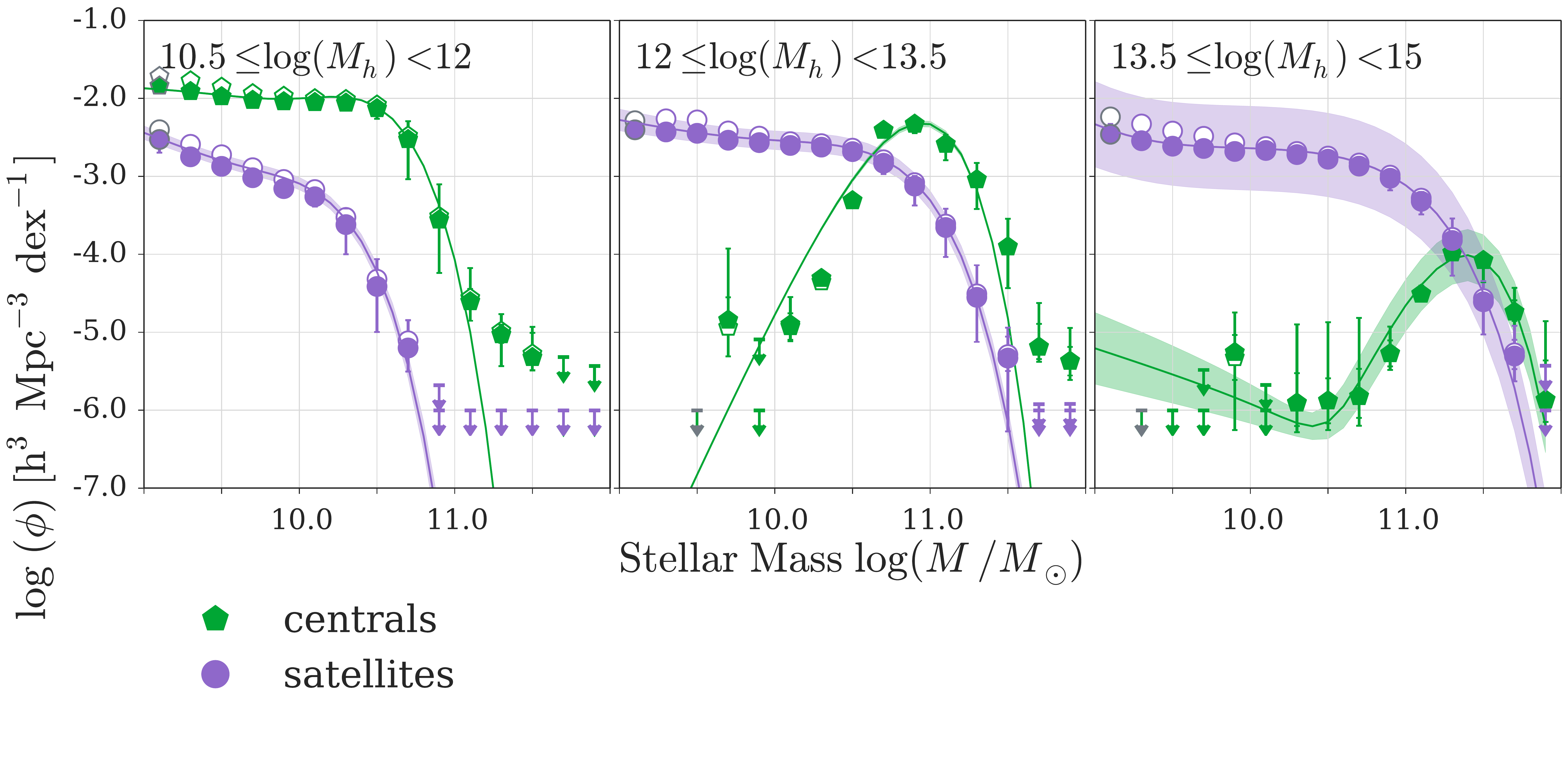}
	\caption{\label{fig:mass_fct_hm_cent_sat} Entire sample split by halo mass \& centrals \& satellites. The $\Phi$ values determined with $1/V_\mathrm{max}$ and SWML are shown with open and filled symbols, respectively. The best fit Schechter function based on STY is illustrated with a solid line. The shaded regions correspond to the STY $1\sigma$ uncertainties that we determine directly from the MCMC chain. For the $1/V_\mathrm{max}$ values we are showing random errors only. The errorbars on the SWML $\Phi$ values correspond to the combination of random errors and the systematic error due to stellar mass uncertainties.}
\end{figure*}

\begin{figure*}
	\includegraphics[scale = .3]{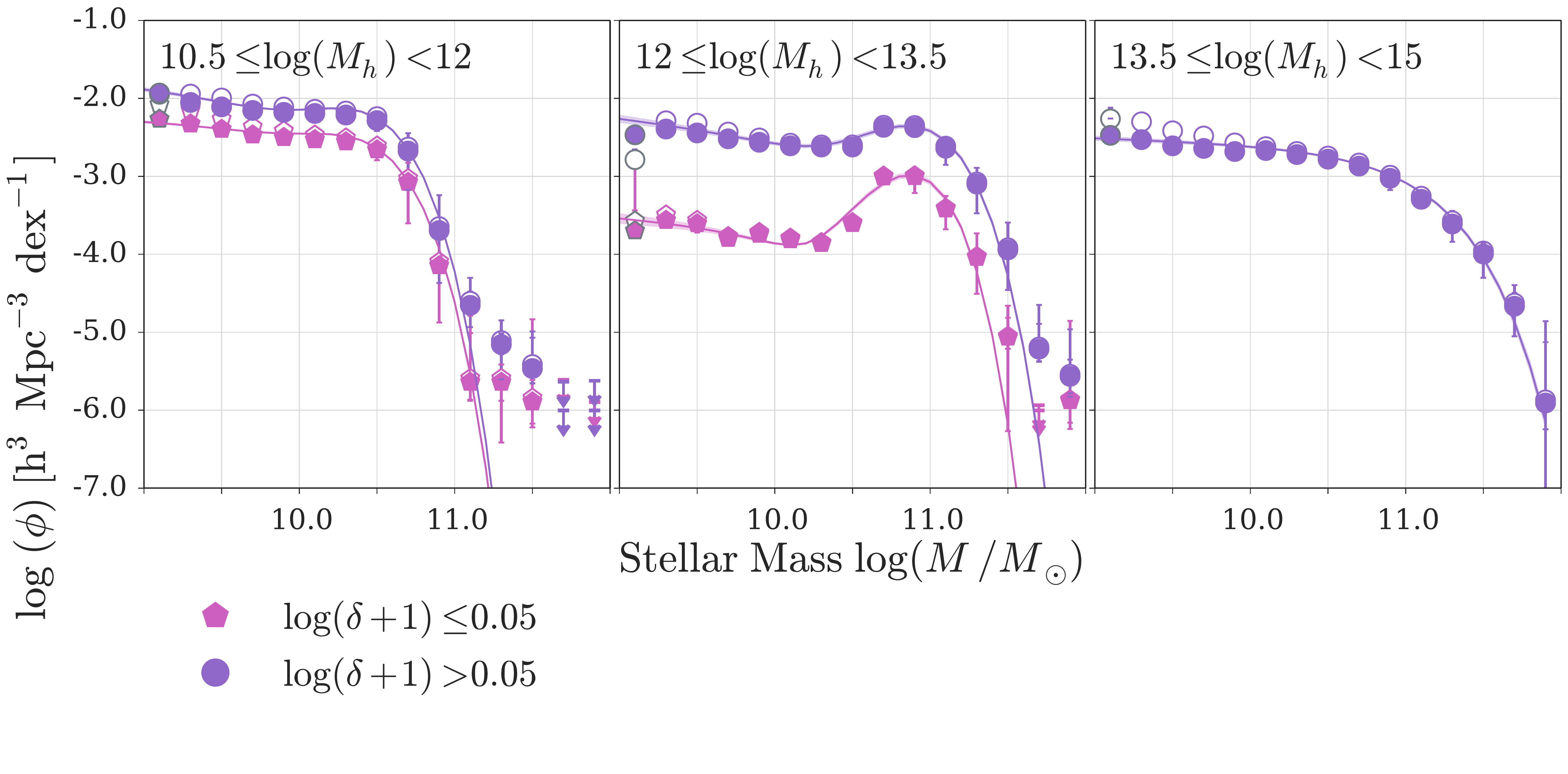}
	\caption{\label{fig:mass_fct_hm_dens} Entire sample split by halo mass \& density. The $\Phi$ values determined with $1/V_\mathrm{max}$ and SWML are shown with open and filled symbols, respectively. The best fit Schechter function based on STY is illustrated with a solid line. The shaded regions correspond to the STY $1\sigma$ uncertainties that we determine directly from the MCMC chain. For the $1/V_\mathrm{max}$ values we are showing random errors only. The errorbars on the SWML $\Phi$ values correspond to the combination of random errors and the systematic error due to stellar mass uncertainties.}
\end{figure*}

\begin{figure*}
	\includegraphics[scale = .3]{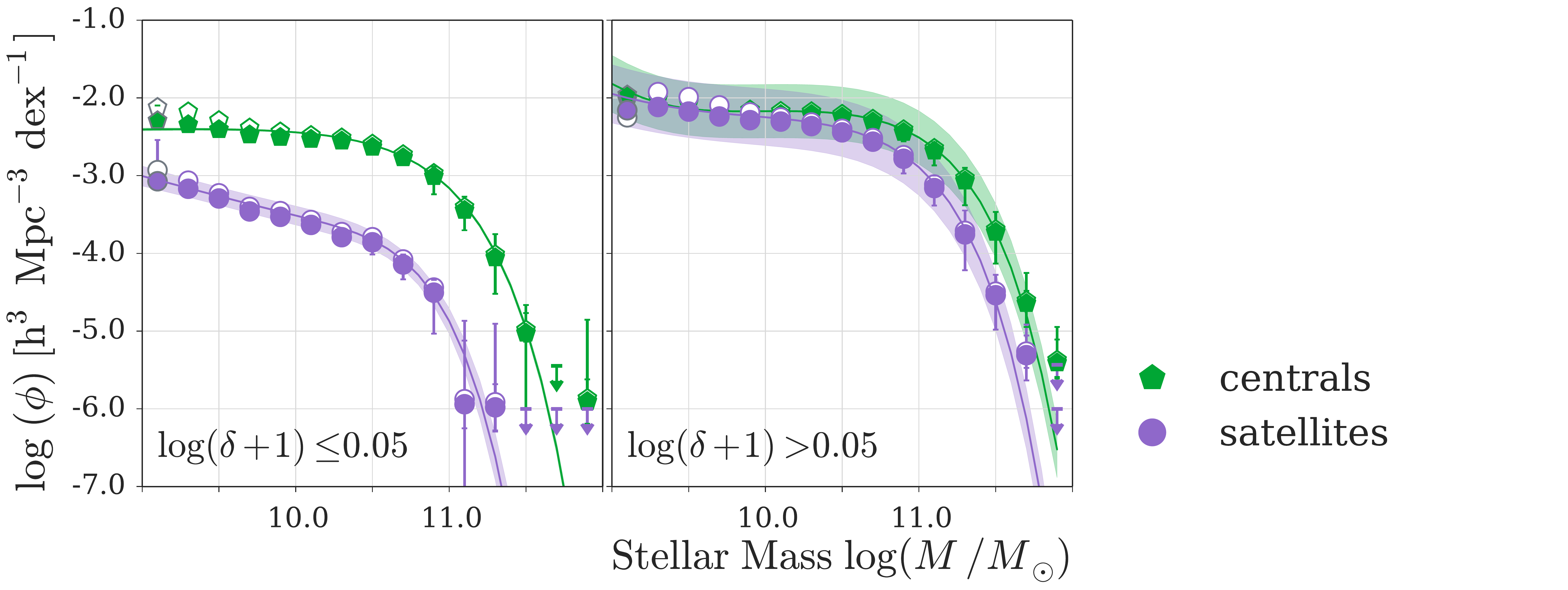}
	\caption{\label{fig:mass_fct_cent_sat_dens} Entire sample split by centrals \& satellites \& density. The $\Phi$ values determined with $1/V_\mathrm{max}$ and SWML are shown with open and filled symbols, respectively. The best fit Schechter function based on STY is illustrated with a solid line. The shaded regions correspond to the STY $1\sigma$ uncertainties that we determine directly from the MCMC chain. For the $1/V_\mathrm{max}$ values we are showing random errors only. The errorbars on the SWML $\Phi$ values correspond to the combination of random errors and the systematic error due to stellar mass uncertainties.}
\end{figure*}

For the 1/\Vmax\ technique we binned in stellar mass and weighed each galaxy by the maximum comoving volume that it could have been found in to correct for the Malmquist bias (see Sec. \ref{sec:Vmax_tech}). For STY we have to assume a functional form for the stellar mass function. We tested both a single and a double Schechter function (see Sec. \ref{sec:STY_tech}) by running two MCMC samplers for each subsample. For our third technique, the SWML approach, we again binned in stellar mass and estimated the number density $\Phi$ in each bin iteratively (see Sec. \ref{sec:SWML_tech}). For the entire sample which is shown in Figure \ref{fig:mass_fct_entire_sample} we measured the following best fit Schechter function parameters based on STY:  $\log (M^{*}/M_\odot) = $ \Mstarentiresample, $\log (\Phi^{*}_1/\mathrm{h^3\ Mpc^{-3}}) = $ \firstphistarentiresample, $\log (\Phi^{*}_2/\mathrm{h^3\ Mpc^{-3}}) = $ \secondphistarentiresample, $\alpha_1 = $ \firstalphaentiresample\ and $\alpha_2 = $ \secondalphaentiresample. The best fit parameters for the remaining subsamples are given in Tables \ref{tab:results_1} and \ref{tab:results_2}. Figure \ref{fig:single_vs_double} summarises which subsamples are well fit by a single, and which are well fit by a double Schechter function. The individual stellar mass functions are shown in Figures \ref{fig:mass_fct_morphology} to \ref{fig:mass_fct_cent_sat_dens}. An overview of all figures is given in Table \ref{tab:Fig_overview}.

A comparison to the mass functions by \cite{Peng:2010aa}, \cite{Baldry:2012aa} and \cite{Taylor:2015aa} is shown in Figure \ref{fig:comparison}. Note, that in contrast to this previous work, we did not make any a priori assumptions about the Schechter \citep{Schechter:1976aa} function parameters or if a sample is better described by a single or a double Schechter function. For each subsample, we estimated both, the likelihood of a single and the likelihood of a double Schechter description. We then used the likelihood ratio to determine which functional form provides a better fit to the subsample in question.  

To construct stellar mass functions we need to know the stellar mass completeness as a function of redshift for each of our subsamples. Determining $M_\mathrm{lim}(z)$ is not straightforward since the SDSS is a flux limited survey and the galaxies in our sample can show a wide range of M/L. We thus need to ensure that we do not falsely include sources with high M/L (high luminosity relative to their stellar mass) and wrongly exclude objects with low M/L (low luminosity relative to their stellar mass). We followed the approach by \cite{Pozzetti:2010aa} which relies on determining the stellar mass that each galaxy would have if its magnitude was equal to the magnitude limit while keeping its M/L and its redshift constant. The mass completeness function is then estimated by binning in redshift space and fitting a function to a certain mass completeness level (see Sec. \ref{sec:mass_completeness}).

To test our code we created mock catalogues based on input Schechter functions and constant M/L. We then used these simulated samples to check if our mass function code could retrieve the input Schechter function and which biases might affect the three techniques (see Sec. \ref{sec:simulation}). 

Our simulation results in Figure \ref{fig:sim_comp} showed a significant discrepancy between the 1/\Vmax\ $\Phi$ values and the input Schechter functions. This discrepancy might be due to us not correctly reproducing the underlying stellar mass - redshift distribution by using a constant M/L value. It could also be cause by an intrinsic bias in the 1/\Vmax\ method itself or a combination of the two effects. The maximum likelihood methods STY and SWML retrieved the input Schechter functions more reliably. SWML predicted marginally lower $\Phi$ values at the low mass end for the samples with $\alpha + 1 < 0$. STY was slightly underestimating the low mass end slope of two and overestimating the slope of one of the simulated samples. A possible bias causing 1/\Vmax\ to overestimate $\Phi$ at the low mass end and a tendency of STY to underpredict the low mass end slope have been discussed by \cite{Efstathiou:1988aa}  and \cite{Willmer:1997aa}. Figure \ref{fig:sim_comp_tests} illustrated that the 1/\Vmax\ results are strongly dependent on the shape and the normalisation of the stellar mass completeness function $M_\mathrm{lim}(z)$.  STY failed for a completeness function with an extreme shape, but in all other cases the results of both maximum likelihood methods matched the input Schechter function. 

We discussed the $M_\mathrm{lim}(z)$ dependency in more detail in Sec. \ref{sec:compl}. Compared to 1/\Vmax, the maximum likelihood methods have the advantage that the $\Phi$ values at low masses are dependent on the number density at higher masses. For the STY technique we are assuming that the mass function follows a Schechter function and we are maximizing the likelihood of all galaxies being part of the sample simultaneously. In SWML the $\Phi$ values in each bin also depend on each other since we are determining $\Phi$ iteratively. When determining the mass function with the 1/\Vmax\ method we are however assuming that the $\Phi$ values in each bin are completely independent from each other. This does give 1/\Vmax\ more freedom in comparison to STY, but it also implies that 1/\Vmax\ is more strongly affected by changes in the stellar mass completeness function. 

In summary our analysis of these classical estimators has shown how important it is to carefully determine the stellar mass completeness function when estimating stellar mass functions. At low masses, especially 1/\Vmax,  the simplest and most commonly used mass function technique is strongly dependent on the mass completeness function. For a true stellar mass function, one can not simply use the magnitude limit to measure the completeness but has to carefully treat the range in M/L to estimate $M_\mathrm{lim}(z)$. STY and SWML are clearly more robust towards changes of the mass completeness and are thus the more reliable mass function estimators. 

\begin{table*}
	\scriptsize
	\begin{center}
		\caption{\label{tab:results_1}Best fit Schechter function parameters 1/2. Given here are the best fit Schechter function parameters for our various subsamples. We determine these parameters with the parametric maximum likelihood approach by \protect\cite{Sandage:1979aa} (STY) (see Sec. \ref{sec:STY_tech}).  The errors on the parameters correspond to the $1\sigma$ random errors that we estimate from the MCMC chain. The third and the fourth column show if the subsample is better described by a single (S) or by a double (D) Schechter function, respectively. This conclusion is based on the likelihood ratio (see Sec. \ref{sec:single_double}). The reduced $\chi^2$, value which is given in the last column was determined by comparing the SWML $\Phi$ values to the STY best fit Schechter function. $\mathrm{sSFFR}_1$, $\mathrm{sSFFR}_2$, $\mathrm{sSFFR}_3$ and $\text{sSFR}_4$ refer to the following cuts in sSFR: $\log \mathrm{sSFR} < -12$, $-12 \leq \log \mathrm{sSFR} < -11$, $-11 \leq \log \mathrm{sSFR} < -10$ and $-10 \leq \log \mathrm{sSFR} < -9$. $M_{h, 1}$, $M_{h, 2}$ and $M_{h, 3}$  relate to the following halo mass bins: $10.5 \leq \log M_h < 12$, $12 \leq \log M_h < 13.5$, $13.5 \leq \log M_h < 15$.}
		\begin{tabular}{llllllllllll}
			\hline
			{sample} & {Fig.} & {S} & {D} & {$\log M^{*}$} & {$\log \Phi^{*}$} & {$\alpha$} & {$\log \Phi^{*}_2$} &  {$\alpha_2$} &{$\log(\Phi^{*}_{2}/\Phi^{*}_{1})$} & {$\chi^{2}_\mathrm{red}$} \\
			{} & {} & {} & {} & {$[M_\odot]$} & {$[\mathrm{h^3}\mathrm{Mpc}^{-3}]$} & {} & {$[\mathrm{h^3}\mathrm{Mpc}^{-3}]$} &  {} &{} & {} \\
			\hline
			{entire sample} & {\ref{fig:mass_fct_entire_sample}} & {} & {X} & {$10.79 \pm 0.01$} & {$-3.31 \pm 0.20$} & {$-1.69 \pm 0.10$} & {$-2.01 \pm 0.28$} & {$-0.79 \pm 0.04$} & {$1.30 \pm 0.20$} & {$9.78$}\\
			{Early types} & {\ref{fig:mass_fct_morphology}} & {} & {X} & {$10.75 \pm 0.01$} & {$-5.78 \pm 0.59$} & {$-2.41 \pm 0.41$} & {$-2.59 \pm 0.83$} & {$0.05 \pm 0.04$} & {$3.19 \pm 0.59$} & {$4.86$}\\
			{indeterminates} & {\ref{fig:mass_fct_morphology}} & {} & {X} & {$10.75 \pm 0.01$} & {$-3.50 \pm 0.17$} & {$-1.82 \pm 0.09$} & {$-2.40 \pm 0.25$} & {$-0.95 \pm 0.04$} & {$1.10 \pm 0.17$} & {$11.98$}\\
			{Late types} & {\ref{fig:mass_fct_morphology}} & {} & {X} & {$10.67 \pm 0.02$} & {$-2.80 \pm 0.20$} & {$-1.17 \pm 0.10$} & {$-2.47 \pm 0.34$} & {$-0.47 \pm 0.14$} & {$0.33 \pm 0.20$} & {$12.85$}\\
			{$\text{sSFR}_1: \log(\text{sSFR}) < -12$} & {\ref{fig:mass_fct_sSFR}} & {X} & {} & {$10.57 \pm 0.01$} & {$-2.80 \pm 0.01$} & {$1.33 \pm 0.04$} & {$$} & {$$} & {$$} & {$35.70$}\\
			{$\text{sSFR}_2: -12 \leq \log(\text{sSFR}) < -11$} & {\ref{fig:mass_fct_sSFR}} & {X} & {} & {$10.64 \pm 0.01$} & {$-2.23 \pm 0.00$} & {$-0.52 \pm 0.01$} & {$$} & {$$} & {$$} & {$9.75$}\\
			{$\text{sSFR}_3: -11 \leq \log(\text{sSFR}) < -10 $} & {\ref{fig:mass_fct_sSFR}} & {} & {X} & {$10.52 \pm 0.01$} & {$-4.77 \pm 0.50$} & {$-2.44 \pm 0.32$} & {$-2.26 \pm 0.71$} & {$-0.78 \pm 0.04$} & {$2.51 \pm 0.50$} & {$11.53$}\\
			{$\text{sSFR}_4: -10 \leq \log(\text{sSFR}) < -9$} & {\ref{fig:mass_fct_sSFR}} & {X} & {} & {$10.42 \pm 0.01$} & {$-2.76 \pm 0.02$} & {$-1.48 \pm 0.01$} & {$$} & {$$} & {$$} & {$0.84$}\\
			{red} & {\ref{fig:mass_fct_color}} & {} & {X} & {$10.77 \pm 0.01$} & {$-6.73 \pm 0.79$} & {$-3.12 \pm 0.51$} & {$-2.21 \pm 1.12$} & {$-0.46 \pm 0.02$} & {$4.52 \pm 0.79$} & {$6.67$}\\
			{green} & {\ref{fig:mass_fct_color}} & {} & {X} & {$10.65 \pm 0.02$} & {$-3.95 \pm 0.23$} & {$-1.84 \pm 0.15$} & {$-2.54 \pm 0.33$} & {$-0.44 \pm 0.07$} & {$1.41 \pm 0.23$} & {$5.94$}\\
			{blue} & {\ref{fig:mass_fct_color}} & {X} & {} & {$10.60 \pm 0.01$} & {$-2.43 \pm 0.01$} & {$-1.21 \pm 0.01$} & {$$} & {$$} & {$$} & {$4.70$}\\
			{$M_{h, 1}: 10.5 \leq \log(M_h) < 12$} & {\ref{fig:mass_fct_hm}} & {} & {X} & {$9.99 \pm 0.00$} & {$-2.20 \pm 0.00$} & {$-1.11 \pm 0.01$} & {$-2.20 \pm 0.01$} & {$1.16 \pm 0.02$} & {$0.00 \pm 0.00$} & {$31.30$}\\
			{$M_{h, 2}: 12 \leq \log(M_h) < 13.5 $} & {\ref{fig:mass_fct_hm}} & {} & {X} & {$10.45 \pm 0.01$} & {$-2.91 \pm 0.03$} & {$-1.23 \pm 0.03$} & {$-2.60 \pm 0.05$} & {$1.62 \pm 0.06$} & {$0.31 \pm 0.03$} & {$103.19$}\\
			{$M_{h, 3}: 13.5 \leq \log(M_h) < 15$} & {\ref{fig:mass_fct_hm}} & {X} & {} & {$11.00 \pm 0.01$} & {$-3.01 \pm 0.02$} & {$-1.07 \pm 0.02$} & {$$} & {$$} & {$$} & {$2.84$}\\
			{centrals} & {\ref{fig:mass_fct_cent_sat}} & {} & {X} & {$10.80 \pm 0.01$} & {$-2.60 \pm 0.11$} & {$-1.16 \pm 0.05$} & {$-2.33 \pm 0.19$} & {$-0.51 \pm 0.09$} & {$0.27 \pm 0.11$} & {$21.63$}\\
			{satellites} & {\ref{fig:mass_fct_cent_sat}} & {} & {X} & {$10.71 \pm 0.02$} & {$-3.87 \pm 0.43$} & {$-1.83 \pm 0.23$} & {$-2.47 \pm 0.62$} & {$-0.84 \pm 0.08$} & {$1.41 \pm 0.43$} & {$3.86$}\\
			{$\log(\delta + 1) \leq 0.05$} & {\ref{fig:mass_fct_dens}} & {} & {X} & {$10.60 \pm 0.04$} & {$-2.92 \pm 0.16$} & {$-1.16 \pm 0.05$} & {$-2.80 \pm 0.29$} & {$-0.30 \pm 0.22$} & {$0.13 \pm 0.16$} & {$6.91$}\\
			{$\log(\delta + 1) > 0.05$} & {\ref{fig:mass_fct_dens}} & {} & {X} & {$10.83 \pm 0.01$} & {$-4.51 \pm 0.39$} & {$-2.31 \pm 0.23$} & {$-2.10 \pm 0.55$} & {$-0.85 \pm 0.02$} & {$2.41 \pm 0.39$} & {$7.62$}\\
			{Early types \& $\text{sSFR}_1$} & {\ref{fig:mass_fct_morph_sSFR}} & {X} & {} & {$10.59 \pm 0.01$} & {$-3.18 \pm 0.02$} & {$1.55 \pm 0.05$} & {$$} & {$$} & {$$} & {$19.46$}\\
			{indeterminates \& $\text{sSFR}_1$} & {\ref{fig:mass_fct_morph_sSFR}} & {X} & {} & {$10.47 \pm 0.01$} & {$-3.17 \pm 0.02$} & {$1.32 \pm 0.06$} & {$$} & {$$} & {$$} & {$52.26$}\\
			{Late types \& $\text{sSFR}_1$} & {\ref{fig:mass_fct_morph_sSFR}} & {X} & {} & {$10.45 \pm 0.02$} & {$-4.33 \pm 0.06$} & {$2.61 \pm 0.19$} & {$$} & {$$} & {$$} & {$0.91$}\\
			{Early types \& $\text{sSFR}_2$} & {\ref{fig:mass_fct_morph_sSFR}} & {} & {X} & {$10.55 \pm 0.01$} & {$-7.18 \pm 0.50$} & {$-3.49 \pm 0.41$} & {$-2.82 \pm 0.71$} & {$0.07 \pm 0.04$} & {$4.35 \pm 0.50$} & {$9.39$}\\
			{indeterminates \& $\text{sSFR}_2$} & {\ref{fig:mass_fct_morph_sSFR}} & {X} & {} & {$10.59 \pm 0.01$} & {$-2.48 \pm 0.01$} & {$-0.77 \pm 0.02$} & {$$} & {$$} & {$$} & {$8.72$}\\
			{Late types \& $\text{sSFR}_2$} & {\ref{fig:mass_fct_morph_sSFR}} & {} & {X} & {$10.56 \pm 0.02$} & {$-5.43 \pm 0.74$} & {$-2.44 \pm 0.74$} & {$-2.85 \pm 1.05$} & {$0.37 \pm 0.08$} & {$2.58 \pm 0.74$} & {$2.66$}\\
			{Early types \& $\text{sSFR}_3$} & {\ref{fig:mass_fct_morph_sSFR}} & {X} & {} & {$10.73 \pm 0.05$} & {$-4.07 \pm 0.04$} & {$-0.77 \pm 0.07$} & {$$} & {$$} & {$$} & {$1.07$}\\
			{indeterminates \& $\text{sSFR}_3$} & {\ref{fig:mass_fct_morph_sSFR}} & {X} & {} & {$10.60 \pm 0.02$} & {$-2.92 \pm 0.02$} & {$-1.34 \pm 0.02$} & {$$} & {$$} & {$$} & {$3.49$}\\
			{Late types \& $\text{sSFR}_3$} & {\ref{fig:mass_fct_morph_sSFR}} & {} & {X} & {$10.47 \pm 0.01$} & {$-5.85 \pm 0.85$} & {$-2.88 \pm 0.61$} & {$-2.38 \pm 1.20$} & {$-0.53 \pm 0.03$} & {$3.47 \pm 0.85$} & {$12.81$}\\
			{Early types \& $\text{sSFR}_4$} & {\ref{fig:mass_fct_morph_sSFR}} & {X} & {} & {$10.34 \pm 0.09$} & {$-4.52 \pm 0.04$} & {$-0.50 \pm 0.16$} & {$$} & {$$} & {$$} & {$0.60$}\\
			{indeterminates \& $\text{sSFR}_4$} & {\ref{fig:mass_fct_morph_sSFR}} & {X} & {} & {$10.37 \pm 0.02$} & {$-3.08 \pm 0.03$} & {$-1.68 \pm 0.02$} & {$$} & {$$} & {$$} & {$0.84$}\\
			{Late types \& $\text{sSFR}_4$} & {\ref{fig:mass_fct_morph_sSFR}} & {X} & {} & {$10.40 \pm 0.02$} & {$-2.97 \pm 0.02$} & {$-1.23 \pm 0.02$} & {$$} & {$$} & {$$} & {$1.34$}\\
			{Early types \& red} & {\ref{fig:mass_fct_morph_color}} & {} & {X} & {$10.74 \pm 0.01$} & {$-7.09 \pm 0.79$} & {$-3.11 \pm 0.55$} & {$-2.62 \pm 1.11$} & {$0.13 \pm 0.03$} & {$4.47 \pm 0.79$} & {$4.40$}\\
			{indeterminates \& red} & {\ref{fig:mass_fct_morph_color}} & {} & {X} & {$10.67 \pm 0.01$} & {$-6.45 \pm 0.90$} & {$-2.98 \pm 0.64$} & {$-2.46 \pm 1.26$} & {$-0.65 \pm 0.02$} & {$3.99 \pm 0.90$} & {$6.98$}\\
			{Late types \& red} & {\ref{fig:mass_fct_morph_color}} & {} & {X} & {$10.67 \pm 0.02$} & {$-4.72 \pm 0.46$} & {$-1.48 \pm 0.32$} & {$-3.08 \pm 0.66$} & {$0.23 \pm 0.11$} & {$1.64 \pm 0.46$} & {$1.07$}\\
			{Early types \& green} & {\ref{fig:mass_fct_morph_color}} & {} & {X} & {$10.75 \pm 0.05$} & {$-7.05 \pm 1.11$} & {$-2.88 \pm 0.69$} & {$-3.82 \pm 1.57$} & {$-0.45 \pm 0.15$} & {$3.23 \pm 1.11$} & {$0.78$}\\
			{indeterminates \& green} & {\ref{fig:mass_fct_morph_color}} & {} & {X} & {$10.75 \pm 0.06$} & {$-4.43 \pm 0.61$} & {$-2.02 \pm 0.30$} & {$-3.08 \pm 0.86$} & {$-1.01 \pm 0.13$} & {$1.35 \pm 0.61$} & {$2.34$}\\
			{Late types \& green} & {\ref{fig:mass_fct_morph_color}} & {} & {X} & {$10.53 \pm 0.02$} & {$-4.08 \pm 0.28$} & {$-1.46 \pm 0.21$} & {$-2.77 \pm 0.41$} & {$0.16 \pm 0.10$} & {$1.32 \pm 0.28$} & {$1.65$}\\
			{Early types \& blue} & {\ref{fig:mass_fct_morph_color}} & {X} & {} & {$10.65 \pm 0.08$} & {$-4.46 \pm 0.06$} & {$-0.71 \pm 0.11$} & {$$} & {$$} & {$$} & {$0.79$}\\
			{indeterminates \& blue} & {\ref{fig:mass_fct_morph_color}} & {X} & {} & {$10.62 \pm 0.02$} & {$-3.07 \pm 0.03$} & {$-1.62 \pm 0.02$} & {$$} & {$$} & {$$} & {$7.41$}\\
			{Late types \& blue} & {\ref{fig:mass_fct_morph_color}} & {X} & {} & {$10.54 \pm 0.01$} & {$-2.49 \pm 0.01$} & {$-0.97 \pm 0.02$} & {$$} & {$$} & {$$} & {$5.61$}\\
			{Early types \& $M_{h, 1}$} & {\ref{fig:mass_fct_morph_hm}} & {} & {X} & {$9.82 \pm 0.01$} & {$-3.95 \pm 0.02$} & {$-1.07 \pm 0.07$} & {$-3.92 \pm 0.04$} & {$2.86 \pm 0.07$} & {$0.04 \pm 0.02$} & {$16.75$}\\
			{indeterminates \& $M_{h, 1}$} & {\ref{fig:mass_fct_morph_hm}} & {} & {X} & {$9.98 \pm 0.01$} & {$-2.46 \pm 0.00$} & {$-1.36 \pm 0.02$} & {$-2.45 \pm 0.01$} & {$0.94 \pm 0.03$} & {$0.01 \pm 0.00$} & {$16.47$}\\
			{Late types \& $M_{h, 1}$} & {\ref{fig:mass_fct_morph_hm}} & {} & {X} & {$10.05 \pm 0.01$} & {$-2.56 \pm 0.01$} & {$-1.01 \pm 0.01$} & {$-2.53 \pm 0.02$} & {$1.03 \pm 0.05$} & {$0.03 \pm 0.01$} & {$22.54$}\\
			{Early types \& $M_{h, 2}$} & {\ref{fig:mass_fct_morph_hm}} & {} & {X} & {$10.47 \pm 0.01$} & {$-4.29 \pm 0.10$} & {$-1.33 \pm 0.12$} & {$-3.23 \pm 0.16$} & {$1.94 \pm 0.09$} & {$1.06 \pm 0.10$} & {$42.77$}\\
			{indeterminates \& $M_{h, 2}$} & {\ref{fig:mass_fct_morph_hm}} & {} & {X} & {$10.44 \pm 0.02$} & {$-3.18 \pm 0.04$} & {$-1.41 \pm 0.04$} & {$-2.93 \pm 0.08$} & {$1.33 \pm 0.11$} & {$0.25 \pm 0.04$} & {$7.09$}\\
			{Late types \& $M_{h, 2}$} & {\ref{fig:mass_fct_morph_hm}} & {} & {X} & {$10.40 \pm 0.01$} & {$-3.25 \pm 0.02$} & {$-1.06 \pm 0.03$} & {$-3.20 \pm 0.04$} & {$1.94 \pm 0.06$} & {$0.04 \pm 0.02$} & {$6.48$}\\
			{Early types \& $M_{h, 3}$} & {\ref{fig:mass_fct_morph_hm}} & {X} & {} & {$11.02 \pm 0.02$} & {$-3.42 \pm 0.01$} & {$-0.46 \pm 0.03$} & {$$} & {$$} & {$$} & {$1.26$}\\
			{indeterminates \& $M_{h, 3}$} & {\ref{fig:mass_fct_morph_hm}} & {X} & {} & {$10.86 \pm 0.02$} & {$-3.28 \pm 0.03$} & {$-1.25 \pm 0.02$} & {$$} & {$$} & {$$} & {$3.41$}\\
			{Late types \& $M_{h, 3}$} & {\ref{fig:mass_fct_morph_hm}} & {X} & {} & {$10.77 \pm 0.02$} & {$-3.37 \pm 0.02$} & {$-0.77 \pm 0.04$} & {$$} & {$$} & {$$} & {$1.54$}\\
			{Early types \& centrals} & {\ref{fig:mass_fct_morph_cent_sat}} & {} & {X} & {$10.74 \pm 0.01$} & {$-6.04 \pm 0.82$} & {$-2.25 \pm 0.56$} & {$-2.75 \pm 1.16$} & {$0.22 \pm 0.04$} & {$3.30 \pm 0.82$} & {$3.06$}\\
			{indeterminates \& centrals} & {\ref{fig:mass_fct_morph_cent_sat}} & {} & {X} & {$10.77 \pm 0.02$} & {$-3.09 \pm 0.16$} & {$-1.40 \pm 0.07$} & {$-2.73 \pm 0.28$} & {$-0.75 \pm 0.12$} & {$0.36 \pm 0.16$} & {$9.70$}\\
			{Late types \& centrals} & {\ref{fig:mass_fct_morph_cent_sat}} & {X} & {} & {$10.76 \pm 0.01$} & {$-2.51 \pm 0.01$} & {$-0.89 \pm 0.01$} & {$$} & {$$} & {$$} & {$14.64$}\\
			{Early types \& satellites} & {\ref{fig:mass_fct_morph_cent_sat}} & {} & {X} & {$10.70 \pm 0.02$} & {$-7.08 \pm 0.84$} & {$-3.14 \pm 0.59$} & {$-3.09 \pm 1.18$} & {$-0.16 \pm 0.05$} & {$3.99 \pm 0.84$} & {$1.56$}\\
			{indeterminates \& satellites} & {\ref{fig:mass_fct_morph_cent_sat}} & {} & {X} & {$10.64 \pm 0.02$} & {$-3.61 \pm 0.24$} & {$-1.76 \pm 0.14$} & {$-2.79 \pm 0.37$} & {$-0.90 \pm 0.10$} & {$0.81 \pm 0.24$} & {$5.73$}\\
			{Late types \& satellites} & {\ref{fig:mass_fct_morph_cent_sat}} & {} & {X} & {$10.60 \pm 0.04$} & {$-3.45 \pm 0.38$} & {$-1.26 \pm 0.20$} & {$-2.96 \pm 0.61$} & {$-0.42 \pm 0.23$} & {$0.49 \pm 0.38$} & {$1.67$}\\
			{Early types \& $\log(\delta + 1) \leq 0.05$} & {\ref{fig:mass_fct_morph_dens}} & {X} & {} & {$10.50 \pm 0.02$} & {$-3.37 \pm 0.01$} & {$0.30 \pm 0.07$} & {$$} & {$$} & {$$} & {$1.79$}\\
			{indeterminates \& $\log(\delta + 1) \leq 0.05$} & {\ref{fig:mass_fct_morph_dens}} & {X} & {} & {$10.78 \pm 0.02$} & {$-3.20 \pm 0.02$} & {$-1.29 \pm 0.02$} & {$$} & {$$} & {$$} & {$3.40$}\\
			{Late types \& $\log(\delta + 1) \leq 0.05$} & {\ref{fig:mass_fct_morph_dens}} & {X} & {} & {$10.64 \pm 0.01$} & {$-2.94 \pm 0.01$} & {$-0.87 \pm 0.02$} & {$$} & {$$} & {$$} & {$3.76$}\\
			{Early types \& $\log(\delta + 1) > 0.05$} & {\ref{fig:mass_fct_morph_dens}} & {} & {X} & {$10.76 \pm 0.01$} & {$-5.36 \pm 0.51$} & {$-2.13 \pm 0.36$} & {$-2.66 \pm 0.73$} & {$0.10 \pm 0.05$} & {$2.70 \pm 0.51$} & {$4.15$}\\
			{indeterminates \& $\log(\delta + 1) > 0.05$} & {\ref{fig:mass_fct_morph_dens}} & {} & {X} & {$10.78 \pm 0.05$} & {$-4.05 \pm 0.25$} & {$-2.03 \pm 0.13$} & {$-2.50 \pm 0.34$} & {$-1.03 \pm 0.10$} & {$1.55 \pm 0.25$} & {$7.13$}\\
			{Late types \& $\log(\delta + 1) > 0.05$} & {\ref{fig:mass_fct_morph_dens}} & {} & {X} & {$10.71 \pm 0.02$} & {$-4.30 \pm 0.79$} & {$-1.92 \pm 0.45$} & {$-2.44 \pm 1.12$} & {$-0.72 \pm 0.08$} & {$1.86 \pm 0.79$} & {$7.12$}\\
			{red \& $\text{sSFR}_1$} & {\ref{fig:mass_fct_sSFR_color}} & {X} & {} & {$10.56 \pm 0.01$} & {$-2.82 \pm 0.01$} & {$1.36 \pm 0.04$} & {$$} & {$$} & {$$} & {$26.79$}\\
			{green \& $\text{sSFR}_1$} & {\ref{fig:mass_fct_sSFR_color}} & {X} & {} & {$10.52 \pm 0.06$} & {$-5.00 \pm 0.14$} & {$2.60 \pm 0.43$} & {$$} & {$$} & {$$} & {$4.20$}\\
			{red \& $\text{sSFR}_2$} & {\ref{fig:mass_fct_sSFR_color}} & {X} & {} & {$10.57 \pm 0.01$} & {$-2.31 \pm 0.00$} & {$-0.47 \pm 0.02$} & {$$} & {$$} & {$$} & {$14.37$}\\
			{green \& $\text{sSFR}_2$} & {\ref{fig:mass_fct_sSFR_color}} & {} & {X} & {$10.59 \pm 0.02$} & {$-4.39 \pm 0.30$} & {$-1.80 \pm 0.23$} & {$-2.86 \pm 0.43$} & {$0.15 \pm 0.10$} & {$1.53 \pm 0.30$} & {$2.28$}\\
			{blue \& $\text{sSFR}_2$} & {\ref{fig:mass_fct_sSFR_color}} & {} & {X} & {$10.86 \pm 0.06$} & {$-7.05 \pm 1.28$} & {$-2.67 \pm 0.76$} & {$-3.96 \pm 1.80$} & {$-0.50 \pm 0.20$} & {$3.10 \pm 1.28$} & {$1.83$}\\
			{red \& $\text{sSFR}_3$} & {\ref{fig:mass_fct_sSFR_color}} & {X} & {} & {$10.90 \pm 0.06$} & {$-4.16 \pm 0.09$} & {$-1.50 \pm 0.05$} & {$$} & {$$} & {$$} & {$1.94$}\\
			{green \& $\text{sSFR}_3$} & {\ref{fig:mass_fct_sSFR_color}} & {} & {X} & {$10.52 \pm 0.02$} & {$-4.28 \pm 0.38$} & {$-1.98 \pm 0.25$} & {$-2.72 \pm 0.54$} & {$-0.54 \pm 0.09$} & {$1.56 \pm 0.38$} & {$4.31$}\\
			{blue \& $\text{sSFR}_3$} & {\ref{fig:mass_fct_sSFR_color}} & {X} & {} & {$10.48 \pm 0.01$} & {$-2.44 \pm 0.01$} & {$-0.80 \pm 0.02$} & {$$} & {$$} & {$$} & {$6.22$}\\
		\end{tabular}
	\end{center}
\end{table*}

\begin{table*}
	\scriptsize
	\begin{center}
		\caption{\label{tab:results_2}Best fit Schechter function parameters 2/2. Given here are the best fit Schechter function parameters for our various subsamples. We determine these parameters with the parametric maximum likelihood approach by \protect\cite{Sandage:1979aa} (STY) (see Sec. \ref{sec:STY_tech}).  The errors on the parameters correspond to the $1\sigma$ random errors that we estimate from the MCMC chain. The third and the fourth column show if the subsample is better described by a single (S) or by a double (D) Schechter function, respectively. This conclusion is based on the likelihood ratio (see Sec. \ref{sec:single_double}). The reduced $\chi^2$, value which is given in the last column was determined by comparing the SWML $\Phi$ values to the STY best fit Schechter function. $\mathrm{sSFFR}_1$, $\mathrm{sSFFR}_2$, $\mathrm{sSFFR}_3$ and $\text{sSFR}_4$ refer to the following cuts in sSFR: $\log \mathrm{sSFR} < -12$, $-12 \leq \log \mathrm{sSFR} < -11$, $-11 \leq \log \mathrm{sSFR} < -10$ and $-10 \leq \log \mathrm{sSFR} < -9$. $M_{h, 1}$, $M_{h, 2}$ and $M_{h, 3}$  relate to the following halo mass bins: $10.5 \leq \log M_h < 12$, $12 \leq \log M_h < 13.5$, $13.5 \leq \log M_h < 15$.}
		\begin{tabular}{llllllllllll}
			\hline
			{sample} & {Fig.} & {S} & {D} & {$\log M^{*}$} & {$\log \Phi^{*}$} & {$\alpha$} & {$\log \Phi^{*}_2$} &  {$\alpha_2$} &{$\log(\Phi^{*}_{2}/\Phi^{*}_{1})$} & {$\chi^{2}_\mathrm{red}$} \\
			{} & {} & {} & {} & {$[M_\odot]$} & {$[\mathrm{h^3}\mathrm{Mpc}^{-3}]$} & {} & {$[\mathrm{h^3}\mathrm{Mpc}^{-3}]$} &  {} &{} & {} \\
			\hline
			{red \& $\text{sSFR}_4$} & {\ref{fig:mass_fct_sSFR_color}} & {X} & {} & {$10.58 \pm 0.24$} & {$-5.34 \pm 0.37$} & {$-1.53 \pm 0.18$} & {$$} & {$$} & {$$} & {$2.83$}\\
			{green \& $\text{sSFR}_4$} & {\ref{fig:mass_fct_sSFR_color}} & {X} & {} & {$10.90 \pm 0.17$} & {$-5.10 \pm 0.33$} & {$-1.96 \pm 0.07$} & {$$} & {$$} & {$$} & {$1.33$}\\
			{blue \& $\text{sSFR}_4$} & {\ref{fig:mass_fct_sSFR_color}} & {X} & {} & {$10.40 \pm 0.01$} & {$-2.74 \pm 0.02$} & {$-1.45 \pm 0.01$} & {$$} & {$$} & {$$} & {$0.92$}\\
			{$\text{sSFR}_1$ \& $M_{h, 1}$} & {\ref{fig:mass_fct_sSFR_hm}} & {X} & {} & {$9.85 \pm 0.01$} & {$-4.94 \pm 0.03$} & {$3.97 \pm 0.03$} & {$$} & {$$} & {$$} & {$68.77$}\\
			{$\text{sSFR}_2$ \& $M_{h, 1}$} & {\ref{fig:mass_fct_sSFR_hm}} & {X} & {} & {$10.02 \pm 0.01$} & {$-2.50 \pm 0.01$} & {$0.92 \pm 0.05$} & {$$} & {$$} & {$$} & {$29.65$}\\
			{$\text{sSFR}_3$ \& $M_{h, 1}$} & {\ref{fig:mass_fct_sSFR_hm}} & {} & {X} & {$10.02 \pm 0.01$} & {$-2.63 \pm 0.05$} & {$-1.06 \pm 0.05$} & {$-2.42 \pm 0.09$} & {$0.81 \pm 0.10$} & {$0.22 \pm 0.05$} & {$12.63$}\\
			{$\text{sSFR}_4$ \& $M_{h, 1}$} & {\ref{fig:mass_fct_sSFR_hm}} & {X} & {} & {$10.22 \pm 0.01$} & {$-2.62 \pm 0.02$} & {$-1.36 \pm 0.02$} & {$$} & {$$} & {$$} & {$1.02$}\\
			{$\text{sSFR}_1$ \& $M_{h, 2}$} & {\ref{fig:mass_fct_sSFR_hm}} & {X} & {} & {$10.43 \pm 0.01$} & {$-3.50 \pm 0.02$} & {$2.60 \pm 0.07$} & {$$} & {$$} & {$$} & {$20.04$}\\
			{$\text{sSFR}_2$ \& $M_{h, 2}$} & {\ref{fig:mass_fct_sSFR_hm}} & {} & {X} & {$10.40 \pm 0.01$} & {$-3.19 \pm 0.03$} & {$-1.03 \pm 0.03$} & {$-3.00 \pm 0.06$} & {$1.73 \pm 0.09$} & {$0.19 \pm 0.03$} & {$14.93$}\\
			{$\text{sSFR}_3$ \& $M_{h, 2}$} & {\ref{fig:mass_fct_sSFR_hm}} & {} & {X} & {$10.36 \pm 0.01$} & {$-3.23 \pm 0.01$} & {$-1.28 \pm 0.03$} & {$-3.21 \pm 0.02$} & {$1.49 \pm 0.05$} & {$0.02 \pm 0.01$} & {$6.54$}\\
			{$\text{sSFR}_4$ \& $M_{h, 2}$} & {\ref{fig:mass_fct_sSFR_hm}} & {} & {X} & {$10.40 \pm 0.06$} & {$-3.78 \pm 0.21$} & {$-1.54 \pm 0.09$} & {$-3.69 \pm 0.41$} & {$0.54 \pm 0.37$} & {$0.10 \pm 0.21$} & {$5.60$}\\
			{$\text{sSFR}_1$ \& $M_{h, 3}$} & {\ref{fig:mass_fct_sSFR_hm}} & {X} & {} & {$10.75 \pm 0.02$} & {$-3.32 \pm 0.01$} & {$0.61 \pm 0.07$} & {$$} & {$$} & {$$} & {$10.58$}\\
			{$\text{sSFR}_2$ \& $M_{h, 3}$} & {\ref{fig:mass_fct_sSFR_hm}} & {X} & {} & {$10.76 \pm 0.02$} & {$-3.10 \pm 0.02$} & {$-0.96 \pm 0.03$} & {$$} & {$$} & {$$} & {$5.66$}\\
			{$\text{sSFR}_3$ \& $M_{h, 3}$} & {\ref{fig:mass_fct_sSFR_hm}} & {} & {X} & {$10.53 \pm 0.05$} & {$-4.60 \pm 0.49$} & {$-2.03 \pm 0.30$} & {$-3.35 \pm 0.70$} & {$-0.66 \pm 0.19$} & {$1.25 \pm 0.49$} & {$1.38$}\\
			{$\text{sSFR}_4$ \& $M_{h, 3}$} & {\ref{fig:mass_fct_sSFR_hm}} & {X} & {} & {$10.41 \pm 0.05$} & {$-3.98 \pm 0.07$} & {$-1.37 \pm 0.06$} & {$$} & {$$} & {$$} & {$2.38$}\\
			{centrals \& $\text{sSFR}_1$} & {\ref{fig:mass_fct_sSFR_cent_sat}} & {X} & {} & {$10.57 \pm 0.01$} & {$-3.03 \pm 0.02$} & {$1.52 \pm 0.05$} & {$$} & {$$} & {$$} & {$45.77$}\\
			{satellites \& $\text{sSFR}_1$} & {\ref{fig:mass_fct_sSFR_cent_sat}} & {X} & {} & {$10.48 \pm 0.01$} & {$-3.27 \pm 0.02$} & {$1.30 \pm 0.07$} & {$$} & {$$} & {$$} & {$7.68$}\\
			{centrals \& $\text{sSFR}_2$} & {\ref{fig:mass_fct_sSFR_cent_sat}} & {X} & {} & {$10.60 \pm 0.01$} & {$-2.39 \pm 0.00$} & {$-0.19 \pm 0.02$} & {$$} & {$$} & {$$} & {$5.49$}\\
			{satellites \& $\text{sSFR}_2$} & {\ref{fig:mass_fct_sSFR_cent_sat}} & {X} & {} & {$10.59 \pm 0.01$} & {$-2.66 \pm 0.01$} & {$-0.77 \pm 0.02$} & {$$} & {$$} & {$$} & {$4.05$}\\
			{centrals \& $\text{sSFR}_3$} & {\ref{fig:mass_fct_sSFR_cent_sat}} & {X} & {} & {$10.53 \pm 0.01$} & {$-2.39 \pm 0.01$} & {$-0.76 \pm 0.02$} & {$$} & {$$} & {$$} & {$12.76$}\\
			{satellites \& $\text{sSFR}_3$} & {\ref{fig:mass_fct_sSFR_cent_sat}} & {} & {X} & {$10.47 \pm 0.03$} & {$-4.08 \pm 0.49$} & {$-1.95 \pm 0.30$} & {$-2.84 \pm 0.69$} & {$-0.77 \pm 0.15$} & {$1.25 \pm 0.49$} & {$3.50$}\\
			{centrals \& $\text{sSFR}_4$} & {\ref{fig:mass_fct_sSFR_cent_sat}} & {X} & {} & {$10.43 \pm 0.02$} & {$-2.89 \pm 0.02$} & {$-1.47 \pm 0.02$} & {$$} & {$$} & {$$} & {$0.84$}\\
			{satellites \& $\text{sSFR}_4$} & {\ref{fig:mass_fct_sSFR_cent_sat}} & {X} & {} & {$10.39 \pm 0.03$} & {$-3.37 \pm 0.04$} & {$-1.50 \pm 0.03$} & {$$} & {$$} & {$$} & {$0.90$}\\
			{$\log(\delta + 1) \leq 0.05$ \& $\text{sSFR}_1$} & {\ref{fig:mass_fct_sSFR_dens}} & {X} & {} & {$10.37 \pm 0.02$} & {$-3.95 \pm 0.05$} & {$1.99 \pm 0.15$} & {$$} & {$$} & {$$} & {$5.12$}\\
			{$\log(\delta + 1) > 0.05$ \& $\text{sSFR}_1$} & {\ref{fig:mass_fct_sSFR_dens}} & {X} & {} & {$10.58 \pm 0.01$} & {$-2.85 \pm 0.01$} & {$1.31 \pm 0.04$} & {$$} & {$$} & {$$} & {$32.18$}\\
			{$\log(\delta + 1) \leq 0.05$ \& $\text{sSFR}_2$} & {\ref{fig:mass_fct_sSFR_dens}} & {X} & {} & {$10.49 \pm 0.02$} & {$-2.92 \pm 0.00$} & {$-0.09 \pm 0.05$} & {$$} & {$$} & {$$} & {$3.06$}\\
			{$\log(\delta + 1) > 0.05$ \& $\text{sSFR}_2$} & {\ref{fig:mass_fct_sSFR_dens}} & {X} & {} & {$10.67 \pm 0.01$} & {$-2.33 \pm 0.01$} & {$-0.59 \pm 0.02$} & {$$} & {$$} & {$$} & {$9.51$}\\
			{$\log(\delta + 1) \leq 0.05$ \& $\text{sSFR}_3$} & {\ref{fig:mass_fct_sSFR_dens}} & {X} & {} & {$10.44 \pm 0.01$} & {$-2.80 \pm 0.01$} & {$-0.69 \pm 0.03$} & {$$} & {$$} & {$$} & {$4.99$}\\
			{$\log(\delta + 1) > 0.05$ \& $\text{sSFR}_3$} & {\ref{fig:mass_fct_sSFR_dens}} & {} & {X} & {$10.55 \pm 0.01$} & {$-5.93 \pm 0.57$} & {$-3.25 \pm 0.39$} & {$-2.41 \pm 0.80$} & {$-0.83 \pm 0.03$} & {$3.53 \pm 0.57$} & {$9.20$}\\
			{$\log(\delta + 1) \leq 0.05$ \& $\text{sSFR}_4$} & {\ref{fig:mass_fct_sSFR_dens}} & {X} & {} & {$10.33 \pm 0.02$} & {$-3.18 \pm 0.03$} & {$-1.44 \pm 0.02$} & {$$} & {$$} & {$$} & {$0.94$}\\
			{$\log(\delta + 1) > 0.05$ \& $\text{sSFR}_4$} & {\ref{fig:mass_fct_sSFR_dens}} & {X} & {} & {$10.58 \pm 0.02$} & {$-3.17 \pm 0.03$} & {$-1.66 \pm 0.02$} & {$$} & {$$} & {$$} & {$1.94$}\\
			{red \& $M_{h, 1}$} & {\ref{fig:mass_fct_color_hm}} & {} & {X} & {$9.86 \pm 0.01$} & {$-3.19 \pm 0.03$} & {$-1.02 \pm 0.02$} & {$-3.03 \pm 0.08$} & {$2.21 \pm 0.09$} & {$0.16 \pm 0.03$} & {$33.54$}\\
			{green \& $M_{h, 1}$} & {\ref{fig:mass_fct_color_hm}} & {} & {X} & {$9.98 \pm 0.01$} & {$-3.03 \pm 0.03$} & {$-1.05 \pm 0.06$} & {$-2.98 \pm 0.07$} & {$1.51 \pm 0.09$} & {$0.05 \pm 0.03$} & {$8.95$}\\
			{blue \& $M_{h, 1}$} & {\ref{fig:mass_fct_color_hm}} & {} & {X} & {$10.11 \pm 0.01$} & {$-2.42 \pm 0.01$} & {$-1.21 \pm 0.01$} & {$-2.42 \pm 0.01$} & {$0.24 \pm 0.06$} & {$0.01 \pm 0.01$} & {$3.93$}\\
			{red \& $M_{h, 2}$} & {\ref{fig:mass_fct_color_hm}} & {} & {X} & {$10.48 \pm 0.01$} & {$-3.37 \pm 0.04$} & {$-1.11 \pm 0.05$} & {$-2.78 \pm 0.08$} & {$1.59 \pm 0.07$} & {$0.59 \pm 0.04$} & {$45.49$}\\
			{green \& $M_{h, 2}$} & {\ref{fig:mass_fct_color_hm}} & {} & {X} & {$10.38 \pm 0.01$} & {$-3.46 \pm 0.02$} & {$-1.22 \pm 0.04$} & {$-3.42 \pm 0.04$} & {$2.05 \pm 0.05$} & {$0.03 \pm 0.02$} & {$3.87$}\\
			{blue \& $M_{h, 2}$} & {\ref{fig:mass_fct_color_hm}} & {} & {X} & {$10.40 \pm 0.01$} & {$-3.32 \pm 0.02$} & {$-1.36 \pm 0.02$} & {$-3.28 \pm 0.04$} & {$1.22 \pm 0.07$} & {$0.04 \pm 0.02$} & {$12.58$}\\
			{red \& $M_{h, 3}$} & {\ref{fig:mass_fct_color_hm}} & {X} & {} & {$10.98 \pm 0.02$} & {$-3.08 \pm 0.01$} & {$-0.88 \pm 0.02$} & {$$} & {$$} & {$$} & {$2.71$}\\
			{green \& $M_{h, 3}$} & {\ref{fig:mass_fct_color_hm}} & {} & {X} & {$10.78 \pm 0.04$} & {$-5.99 \pm 0.57$} & {$-2.72 \pm 0.35$} & {$-3.57 \pm 0.80$} & {$-0.88 \pm 0.09$} & {$2.42 \pm 0.57$} & {$3.91$}\\
			{blue \& $M_{h, 3}$} & {\ref{fig:mass_fct_color_hm}} & {X} & {} & {$10.70 \pm 0.04$} & {$-3.73 \pm 0.05$} & {$-1.27 \pm 0.04$} & {$$} & {$$} & {$$} & {$1.36$}\\
			{red \& centrals} & {\ref{fig:mass_fct_color_cent_sat}} & {X} & {} & {$10.75 \pm 0.01$} & {$-2.37 \pm 0.00$} & {$-0.18 \pm 0.02$} & {$$} & {$$} & {$$} & {$3.75$}\\
			{green \& centrals} & {\ref{fig:mass_fct_color_cent_sat}} & {} & {X} & {$10.62 \pm 0.02$} & {$-3.59 \pm 0.25$} & {$-1.31 \pm 0.16$} & {$-2.71 \pm 0.38$} & {$-0.12 \pm 0.14$} & {$0.89 \pm 0.25$} & {$2.94$}\\
			{blue \& centrals} & {\ref{fig:mass_fct_color_cent_sat}} & {X} & {} & {$10.59 \pm 0.01$} & {$-2.52 \pm 0.01$} & {$-1.15 \pm 0.01$} & {$$} & {$$} & {$$} & {$5.51$}\\
			{red \& satellites} & {\ref{fig:mass_fct_color_cent_sat}} & {X} & {} & {$10.72 \pm 0.01$} & {$-2.66 \pm 0.01$} & {$-0.71 \pm 0.02$} & {$$} & {$$} & {$$} & {$3.30$}\\
			{green \& satellites} & {\ref{fig:mass_fct_color_cent_sat}} & {} & {X} & {$10.66 \pm 0.03$} & {$-5.09 \pm 0.55$} & {$-2.41 \pm 0.34$} & {$-3.09 \pm 0.77$} & {$-0.82 \pm 0.09$} & {$2.00 \pm 0.55$} & {$4.38$}\\
			{blue \& satellites} & {\ref{fig:mass_fct_color_cent_sat}} & {X} & {} & {$10.59 \pm 0.02$} & {$-3.09 \pm 0.02$} & {$-1.31 \pm 0.02$} & {$$} & {$$} & {$$} & {$1.59$}\\
			{red \& $\log(\delta + 1) \leq 0.05$} & {\ref{fig:mass_fct_color_dens}} & {X} & {} & {$10.56 \pm 0.01$} & {$-2.93 \pm 0.00$} & {$-0.11 \pm 0.04$} & {$$} & {$$} & {$$} & {$2.03$}\\
			{green \& $\log(\delta + 1) \leq 0.05$} & {\ref{fig:mass_fct_color_dens}} & {X} & {} & {$10.65 \pm 0.02$} & {$-3.21 \pm 0.01$} & {$-0.64 \pm 0.03$} & {$$} & {$$} & {$$} & {$5.56$}\\
			{blue \& $\log(\delta + 1) \leq 0.05$} & {\ref{fig:mass_fct_color_dens}} & {X} & {} & {$10.51 \pm 0.01$} & {$-2.92 \pm 0.02$} & {$-1.19 \pm 0.02$} & {$$} & {$$} & {$$} & {$1.18$}\\
			{red \& $\log(\delta + 1) > 0.05$} & {\ref{fig:mass_fct_color_dens}} & {} & {X} & {$10.80 \pm 0.01$} & {$-6.02 \pm 0.86$} & {$-2.68 \pm 0.55$} & {$-2.29 \pm 1.21$} & {$-0.48 \pm 0.02$} & {$3.73 \pm 0.86$} & {$4.23$}\\
			{green \& $\log(\delta + 1) > 0.05$} & {\ref{fig:mass_fct_color_dens}} & {} & {X} & {$10.67 \pm 0.02$} & {$-4.24 \pm 0.31$} & {$-2.01 \pm 0.20$} & {$-2.65 \pm 0.44$} & {$-0.48 \pm 0.08$} & {$1.59 \pm 0.31$} & {$4.40$}\\
			{blue \& $\log(\delta + 1) > 0.05$} & {\ref{fig:mass_fct_color_dens}} & {X} & {} & {$10.64 \pm 0.01$} & {$-2.62 \pm 0.02$} & {$-1.25 \pm 0.02$} & {$$} & {$$} & {$$} & {$3.97$}\\
			{centrals \& $M_{h, 1}$} & {\ref{fig:mass_fct_hm_cent_sat}} & {} & {X} & {$9.99 \pm 0.00$} & {$-2.23 \pm 0.00$} & {$-1.04 \pm 0.01$} & {$-2.23 \pm 0.00$} & {$1.22 \pm 0.02$} & {$0.00 \pm 0.00$} & {$42.94$}\\
			{satellites \& $M_{h, 1}$} & {\ref{fig:mass_fct_hm_cent_sat}} & {} & {X} & {$9.74 \pm 0.03$} & {$-3.23 \pm 0.09$} & {$-1.66 \pm 0.11$} & {$-3.09 \pm 0.16$} & {$0.32 \pm 0.20$} & {$0.14 \pm 0.09$} & {$3.97$}\\
			{centrals \& $M_{h, 2}$} & {\ref{fig:mass_fct_hm_cent_sat}} & {X} & {} & {$10.30 \pm 0.01$} & {$-3.60 \pm 0.02$} & {$3.40 \pm 0.06$} & {$$} & {$$} & {$$} & {$409.87$}\\
			{satellites \& $M_{h, 2}$} & {\ref{fig:mass_fct_hm_cent_sat}} & {} & {X} & {$10.46 \pm 0.02$} & {$-3.20 \pm 0.14$} & {$-1.38 \pm 0.10$} & {$-2.66 \pm 0.22$} & {$-0.11 \pm 0.15$} & {$0.54 \pm 0.14$} & {$5.35$}\\
			{centrals \& $M_{h, 3}$} & {\ref{fig:mass_fct_hm_cent_sat}} & {} & {X} & {$10.71 \pm 0.02$} & {$-6.65 \pm 0.47$} & {$-1.64 \pm 0.48$} & {$-5.57 \pm 0.70$} & {$3.82 \pm 0.16$} & {$1.08 \pm 0.47$} & {$10.20$}\\
			{satellites \& $M_{h, 3}$} & {\ref{fig:mass_fct_hm_cent_sat}} & {} & {X} & {$10.81 \pm 0.02$} & {$-5.40 \pm 1.05$} & {$-2.38 \pm 0.58$} & {$-2.83 \pm 1.48$} & {$-0.85 \pm 0.08$} & {$2.57 \pm 1.05$} & {$2.83$}\\
			{$\log(\delta + 1) \leq 0.05$ \& $M_{h, 1}$} & {\ref{fig:mass_fct_hm_dens}} & {} & {X} & {$10.00 \pm 0.01$} & {$-2.68 \pm 0.00$} & {$-1.06 \pm 0.02$} & {$-2.68 \pm 0.01$} & {$1.02 \pm 0.04$} & {$0.01 \pm 0.00$} & {$14.27$}\\
			{$\log(\delta + 1) > 0.05$ \& $M_{h, 1}$} & {\ref{fig:mass_fct_hm_dens}} & {} & {X} & {$9.99 \pm 0.01$} & {$-2.39 \pm 0.01$} & {$-1.19 \pm 0.03$} & {$-2.36 \pm 0.03$} & {$1.21 \pm 0.04$} & {$0.03 \pm 0.01$} & {$26.04$}\\
			{$\log(\delta + 1) \leq 0.05$ \& $M_{h, 2}$} & {\ref{fig:mass_fct_hm_dens}} & {} & {X} & {$10.27 \pm 0.01$} & {$-4.06 \pm 0.02$} & {$-1.14 \pm 0.06$} & {$-4.02 \pm 0.05$} & {$3.00 \pm 0.07$} & {$0.05 \pm 0.02$} & {$12.32$}\\
			{$\log(\delta + 1) > 0.05$ \& $M_{h, 2}$} & {\ref{fig:mass_fct_hm_dens}} & {} & {X} & {$10.49 \pm 0.01$} & {$-2.96 \pm 0.03$} & {$-1.24 \pm 0.03$} & {$-2.61 \pm 0.06$} & {$1.41 \pm 0.07$} & {$0.35 \pm 0.03$} & {$63.57$}\\
			{$\log(\delta + 1) > 0.05$ \& $M_{h, 3}$} & {\ref{fig:mass_fct_hm_dens}} & {X} & {} & {$11.00 \pm 0.01$} & {$-3.01 \pm 0.01$} & {$-1.07 \pm 0.02$} & {$$} & {$$} & {$$} & {$2.45$}\\
			{centrals \& $\log(\delta + 1) \leq 0.05$} & {\ref{fig:mass_fct_cent_sat_dens}} & {X} & {} & {$10.71 \pm 0.01$} & {$-2.69 \pm 0.01$} & {$-0.96 \pm 0.01$} & {$$} & {$$} & {$$} & {$10.29$}\\
			{satellites \& $\log(\delta + 1) \leq 0.05$} & {\ref{fig:mass_fct_cent_sat_dens}} & {} & {X} & {$10.34 \pm 0.08$} & {$-4.03 \pm 0.14$} & {$-1.50 \pm 0.09$} & {$-3.85 \pm 0.22$} & {$-0.13 \pm 0.37$} & {$0.18 \pm 0.14$} & {$1.90$}\\
			{centrals \& $\log(\delta + 1) > 0.05$} & {\ref{fig:mass_fct_cent_sat_dens}} & {} & {X} & {$10.85 \pm 0.01$} & {$-5.20 \pm 0.52$} & {$-2.55 \pm 0.31$} & {$-2.30 \pm 0.74$} & {$-0.77 \pm 0.03$} & {$2.90 \pm 0.52$} & {$9.40$}\\
			{satellites \& $\log(\delta + 1) > 0.05$} & {\ref{fig:mass_fct_cent_sat_dens}} & {} & {X} & {$10.72 \pm 0.02$} & {$-4.04 \pm 0.57$} & {$-1.90 \pm 0.31$} & {$-2.47 \pm 0.80$} & {$-0.84 \pm 0.08$} & {$1.56 \pm 0.57$} & {$4.34$}\\
		\end{tabular}
	\end{center}
\end{table*}

\clearpage

\clearpage

\section*{Acknowledgements}

We thank D. Marchesini, O. Ilbert, K. Kova\v{c}, A. Nicola, N. Caplar and S. Seehars for helpful discussions. 

We also thank the anonymous referee for helpul comments.

AKW, KS gratefully acknowledge support from Swiss National Science Foundation Grant PP00P2\_138979/1. CB gratefully acknowledges support from Swiss National Science Foundation Grants $200021\_14944$ and $200021\_143906$. 

This research made use of NASA's ADS Service. This publication made use of Astropy, a community-developed core Python package for Astronomy (Astropy Collaboration, 2013). This publication made extensive use of the Tool for OPerations on Catalogues And Tables (TOPCAT), which can be found at \url{http://www.starlink.ac.uk/topcat/}.

Funding for the Sloan Digital Sky Survey (SDSS) has been provided by the Alfred P. Sloan Foundation, the Participating Institutions, the National Aeronautics and Space Administration, the National Science Foundation, the U.S. Department of Energy, the Japanese Monbukagakusho, and the Max Planck Society. The SDSS Web site is http://www.sdss.org/.

The SDSS is managed by the Astrophysical Research Consortium (ARC) for the Participating Institutions. The Participating Institutions are The University of Chicago, Fermilab, the Institute for Advanced Study, the Japan Participation Group, The Johns Hopkins University, Los Alamos National Laboratory, the Max-Planck-Institute for Astronomy (MPIA), the Max-Planck-Institute for Astrophysics (MPA), New Mexico State University, University of Pittsburgh, Princeton University, the United States Naval Observatory, and the University of Washington.

\bibliographystyle{mnras}
\bibliography{Bib_Lib.bib}
\appendix
\section{Mass-redshift distribution of the mock catalogues}
\begin{figure*}
	\includegraphics[width=\textwidth]{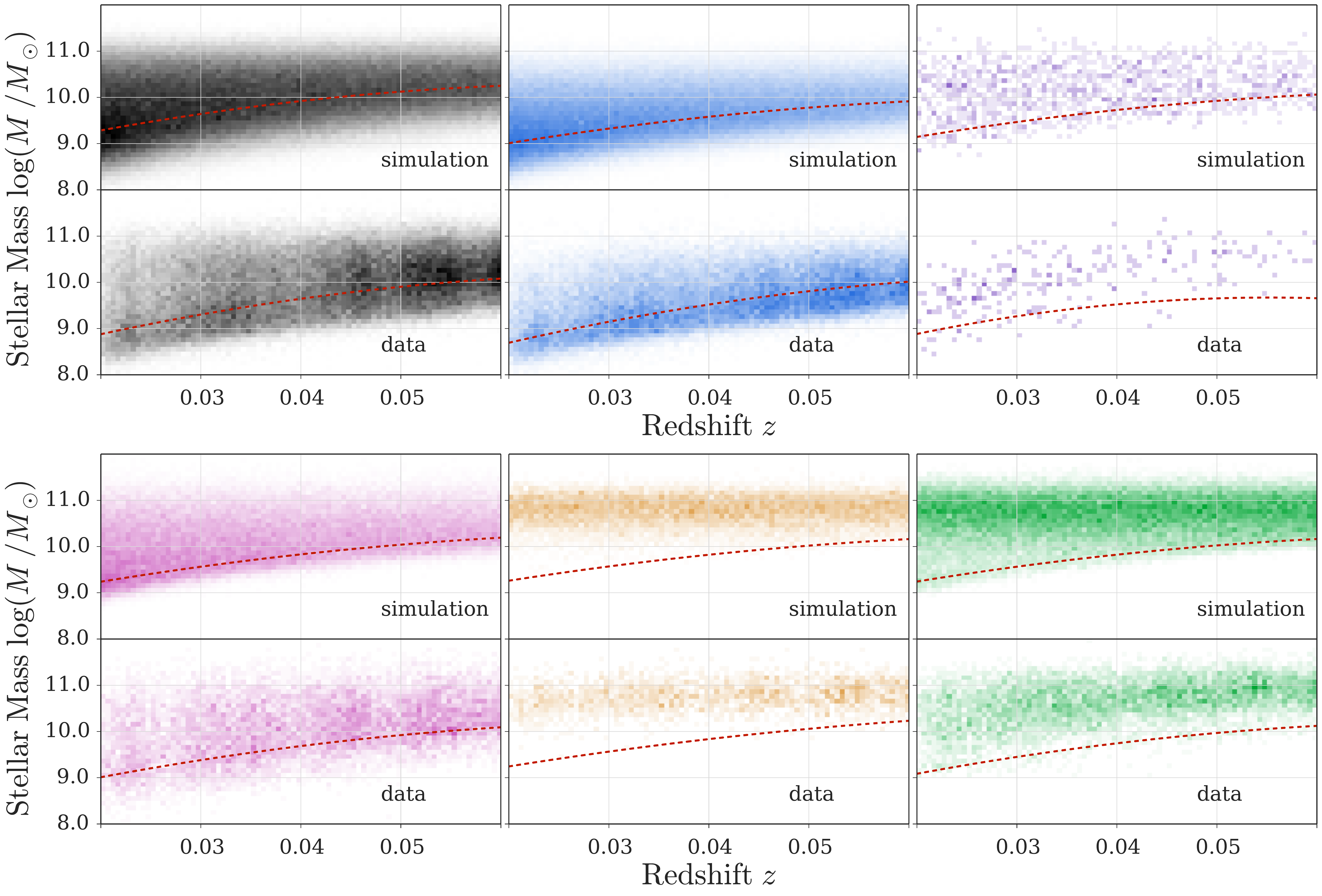}
	\caption{\label{fig:mass_z_dis_mock}Mass-redshift distributions of the mock catalogues discussed in Sec. \ref{sec:simulation}. To test the reliability of our method and investigate possible systematics we create mock catalogues by drawing from stellar mass functions. We test how well the input Schechter function can be recovered by treating these mock catalogues as if they were equivalent to  observed data and generating their stellar mass functions. The panels shown here correspond to the panels in Fig. \ref{fig:sim_comp}. From top left to bottom right, the top panels show the stellar mass-redshift distributions of the mock catalgues which are based on the M/L ratio of the entire sample, blue galaxies, blue Early type galaxies, galaxies in halos with $13.5 \leq \log M_h < 15$, satellite galaxies with $\log \mathrm{sSFR} < -12$ and elliptical galaxies in overdense regions. The bottom panels illustrate the mass-redshift distribution of the corresponding SDSS subsamples which we used to calibrate our simulations. The red, dashed line shows the mass completeness limit. Note, that in our simulations we are modelling the entire sky. The plots in the upper panels thus contain about four times as many objects as the lower panels which are based on SDSS data.}
\end{figure*}

Figure \ref{fig:mass_z_dis_mock} illustrates the stellar mass-redshift distributions of the mock catalogues discussed in Sec. \ref{sec:simulation} in comparison to the corresponding SDSS subsamples which were used to calibrate the simulations.

\bsp	
\label{lastpage}
\end{document}